\documentclass[%
reprint,
aps,
superscriptaddress,
amsmath,
amssymb,
twocolumn,
showpacs,
pra,
longbibliography
]{revtex4-2}
\usepackage{tikz}
\usepackage{graphicx}
\usepackage{dcolumn}
\usepackage{bm}
\usepackage{braket}
\usepackage{hyperref}
\usepackage[caption=false]{subfig} 
\usepackage{tabularx}
\usepackage{booktabs}
\usepackage{amsmath}
\usepackage{graphicx}
\usepackage{fancyhdr}
\usepackage{float}
\usepackage[separate-uncertainty=true]{siunitx}
\usepackage{xcolor}
\usepackage{xspace}
\usepackage{eucal}
\usepackage{multirow}
\usepackage{lineno}
\usepackage{comment}
\usepackage{pgfplots}
\pgfplotsset{compat=1.14}
\usepackage{lineno}
\newcommand{\mf}{m_F\xspace}

\newcommand{\DOne}{D\textsubscript{1}\xspace}
\newcommand{\DTwo}{D\textsubscript{2}\xspace}

\newcommand{\Rb}{\textsuperscript{87}Rb\xspace}
\newcommand{\K}{\textsuperscript{39}K\xspace}

\DeclareSIUnit{\gauss}{G}
\begin{document}
\title{
Matter-wave collimation to picokelvin energies with scattering length and potential shape control
}
\author{Alexander~Herbst}
\affiliation{Leibniz Universit\"at Hannover, Institut f\"ur Quantenoptik,\\ Welfengarten 1, 30167 Hannover, Germany}
\author{Timoth{\'e}~Estrampes}
\affiliation{Leibniz Universit\"at Hannover, Institut f\"ur Quantenoptik,\\ Welfengarten 1, 30167 Hannover, Germany}
\affiliation{Universit\'e Paris-Saclay, CNRS, Institut des Sciences Mol\'eculaires d'Orsay, 91405 Orsay, France}
\author{Henning~Albers}
\affiliation{Leibniz Universit\"at Hannover, Institut f\"ur Quantenoptik,\\ Welfengarten 1, 30167 Hannover, Germany}
\author{Robin~Corgier}
\affiliation{LNE-SYRTE, Observatoire de Paris, Universit\'e PSL, CNRS, Sorbonne Universit\'e 61 avenue de l’Observatoire, 75014 Paris, France}
\author{Knut~Stolzenberg}
\affiliation{Leibniz Universit\"at Hannover, Institut f\"ur Quantenoptik,\\ Welfengarten 1, 30167 Hannover, Germany}
\author{Sebastian~Bode}
\affiliation{Leibniz Universit\"at Hannover, Institut f\"ur Quantenoptik,\\ Welfengarten 1, 30167 Hannover, Germany}
\author{Eric~Charron}
\affiliation{Universit\'e Paris-Saclay, CNRS, Institut des Sciences Mol\'eculaires d'Orsay, 91405 Orsay, France}
\author{Ernst~M.~Rasel}
\affiliation{Leibniz Universit\"at Hannover, Institut f\"ur Quantenoptik,\\ Welfengarten 1, 30167 Hannover, Germany}
\author{Naceur~Gaaloul}
\affiliation{Leibniz Universit\"at Hannover, Institut f\"ur Quantenoptik,\\ Welfengarten 1, 30167 Hannover, Germany}
\author{Dennis~Schlippert}\email{Email: schlippert@iqo.uni-hannover.de}
\affiliation{Leibniz Universit\"at Hannover, Institut f\"ur Quantenoptik,\\ Welfengarten 1, 30167 Hannover, Germany}
\begin{abstract}
\section{Abstract}
The sensitivity of atom interferometers depends on their ability to realize long pulse separation times and prevent loss of contrast by limiting the expansion of the atomic ensemble within the interferometer beam through matter-wave collimation.
Here we investigate the impact of atomic interactions on collimation by applying a lensing protocol to a \K Bose-Einstein condensate at different scattering lengths.
Tailoring interactions, we measure energies corresponding to \SI{340 \pm 12}{\pico\kelvin} in one direction.
Our results are supported by an accurate simulation, which allows us to extrapolate a 2D ballistic expansion energy of \SI{438 \pm 77}{\pico\kelvin}.
Based on our findings we propose an advanced scenario, which enables 3D expansion energies below \SI{16}{\pico\kelvin} by implementing an additional pulsed delta-kick.
Our results pave the way to realize ensembles with more than \num{1e5} atoms and 3D energies in the two-digit \si{\pico\kelvin} range in typical dipole trap setups without the need for micro-gravity or long baseline environments. 
\end{abstract}
\maketitle
\section{Introduction}
Cooling quantum gases to sub-nanokelvin temperatures has enabled breakthroughs in the fields of quantum sensing~\cite{Degen2017RMP}, quantum information~\cite{Mandel2003Nature} and quantum simulation~\cite{Georgescu2014RMP}. 
Especially in precision sensing and metrology, atom interferometers~\cite{Kasevich1991PRL,Kasevich1992APB,Riehle1991PRL,Cronin09RMP} have become a state-of-the-art solution and are used for probing general relativity~\cite{Schlippert14PRL,Tarallo2014PRL,Albers2020EPJD,Asenbaum2020PRL}, quantum mechanics~\cite{Carlesso2022NatPhys,Kovachy15Nature,Bassi2013RMP,Schrinski23PRA}, determining fundamental constants~\cite{Rosi14Nature,Parker2018Science,Morel2020Nature} and measuring inertial effects~\cite{Gustavson97PRL,Canuel06PRL,Dickerson13PRL,Dutta16PRL,Savoie2018SciAdv}. 
Interferometers utilizing molasses-cooled atoms, characterized by expansion energies in the range of several microkelvin, offer short experimental cycle times and a high sensor bandwidth~\cite{LeGouet2008APB, Hu2013PRA, Menoret2018SR}. 
Despite these advantages, their velocity spread limits the accessible free fall distance and their systematic uncertainty is typically restrained at a few $10^{-8}$ \si{\meter\per\second\squared} due to wave-front distortions, when the ensemble is expanding within the interferometer beam~\cite{Louchet2011NJP,Schkolnik2015APB}. 
In contrast, Bose-Einstein condensates (BECs)~\cite{Anderson95Science,Davis95PRL} offer significant advantages with respect to controlling systematic errors and their dynamic behavior~\cite{Schlippert2020,Hensel2021}.
In optical dipole traps (ODTs), BECs of various atomic species readily achieve expansion energies in the range of a few tens of nanokelvin~\cite{Weber2003Science, Hardman2016PRL, Gochnauer2021Atoms}, enhancing coherence time and signal-to-noise ratio.
However, to meet the demands of future precision experiments, further collimation into the picokelvin regime is required to achieve the long pulse separation times necessary and to avoid loss of contrast~\cite{Aguilera2014CQG,Trimeche2019CQG,Loriani2019NJP,Corgier2020NJP,Struckmann2024PRD}. 
Expansion energies of a few hundred picokelvin have been achieved by direct evaporative cooling~\cite{Leanhardt2003Science} and spin gradient cooling~\cite{Medley2011PRL}.
Additionally, advancements using different types of matter-wave lenses have further reduced expansion energies by an order of magnitude~\cite{Ammann1997PRL,Kalnins2005PRA, Muntinga2013PRL}.
In this regime, extended free fall times prior to applying the lens are crucial to minimize atomic interactions, which would otherwise drive the expansion post-lensing~\cite{Ketterle96AAMOP,Kovachy2015PRL}.
Hence, recent records of a few tens of picokelvin have been realized in unique experimental settings utilizing micro-gravity~\cite{Deppner2021PRL,Gaaloul2022NatComm} or long-baseline devices~\cite{Kovachy2015PRL} which both allow for an initial prolonged expansion of the ensemble.
In this paper we demonstrate an alternative approach to resolve this issue by use of a Feshbach resonance~\cite{Inouye1998Nature, Masi2021PRR} to tailor interactions during the lens and upon release from the trapping potential.
Using a \K BEC in the weak interaction regime, we observe expansion energies below \SI{400}{\pico\kelvin} in one dimension.
Through dedicated theory simulations, we extrapolate this result to two dimensions, yielding a 2D energy below \SI{500}{\pico\kelvin}. 
We hence demonstrate a substantial improvement over previous results achieved with the same method and setup using \Rb~\cite{Albers2022Commun}.
Furthermore, our systematic analysis reveals that the careful adjustment of trapping frequencies and interactions will allow to reach 3D expansion energies below \SI{16}{\pico\kelvin}, when implementing an additional delta-kick collimation (DKC) pulse~\cite{Ammann1997PRL} after a few milliseconds of free fall.  
Hence, our method allows for state-of-the-art collimation in typical or even compact quantum optics experiments, without excessive hardware or environmental requirements. 

\begin{figure*}[tbh!]
    \centering
    \includegraphics[width = 0.99\textwidth]{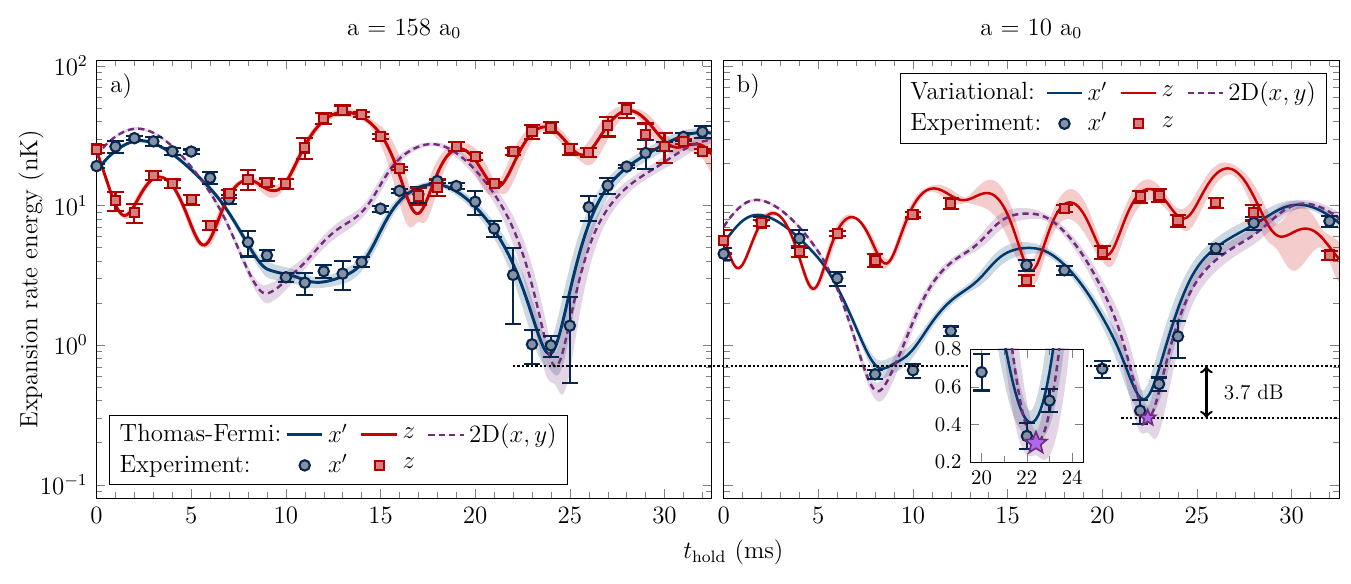}
    \caption{
    \textbf{Measured and simulated expansion rate energies in one and two dimensions.}
    All measurements (blue circles and red squares) are performed within the camera frame and based on time-of-flight (TOF) series with a total length of \SI{25}{\milli\second}.
    The error bars arise from the fit uncertainty of the expansion in the individual TOF series (c.f. data acquisition and analysis section).
    The dynamics of the ensemble are simulated (lines) simultaneously for all directions within the trap frame for \SI{25}{\milli\second} TOF and are subsequently transformed into the camera frame.
    Uncertainty bands are obtained by a Monte Carlo method based on the detection angle and trap frequency errors matching the oscillations of the ensemble size.
    Panel a shows the results obtained in the strong interaction regime for a scattering length of \SI{158}{a_0}, using the Thomas-Fermi approximation in the theoretical description.
    Panel b shows the results obtained in the weak interaction regime at \SI{10}{a_0} scattering length. 
    Here we simulate the dynamics based on a variational approach (c.f. theoretical model section). 
    For the data we choose a lower sampling rate, allowing to increase the number of points per TOF measurement to resolve the lower expansion energies, effectively. 
    In both interaction regimes the measurements agree well with the simulation and the coupling of the dynamics in all dimensions allows to extrapolate the behavior in the entire horizontal plane as shown by the purple dashed lines.
    We find an overall improvement by 3.7 dB in the extrapolated 2D expansion rate energy, when reducing the scattering length.
    The purple star highlights the minimum 2D energy at \SI{10}{a_0}, as prominently featured in the inset. 
    }
    \label{fig:Lens_temperature}
\end{figure*}

\section{Results and Discussion}\label{sec:results}
\subsection{Lensing protocol}
We apply the matter-wave lensing protocol as described by Albers et al.~\cite{Albers2022Commun}.
A detailed overview of the setup is provided in the experimental apparatus section. 
The atoms are held in a crossed ODT with recycled beam configuration crossing under an angle of $\SI{70}{\degree}$.
In the following, the $\{x,y,z\}$-coordinate system refers to the trap frame as defined by the principal axes of the confining potential and used to specify all trap frequencies. 
We image the $\{x',z\}$-plane, obtaining the camera frame $\{x',y',z\}$ by rotating around the vertical $z$-axis by approximately $\SI{30}{\degree}$.
To implement time-averaged optical potentials~\cite{Roy2016PRA} we perform a center-position modulation (CPM) along the horizontal axis of the trapping beams, using an acousto-optical modulator (AOM). 
This approach allows to create harmonic traps with variable width and depth in the horizontal $\{x,y\}$-plane, but does not feature independent control of the trap frequencies in $x$- and $y$-direction or changing the potential shape in $z$-direction.  
By rapidly relaxing the trap within \SI{50}{\micro\second} we cause a sudden reduction in trap frequencies from initial frequencies $\omega^\text{I}_i$ to final frequencies $\omega^{\text{F}}_i<\omega^\text{I}_i$ for $i \in \{x,y\}$, inducing collective mode excitations~\cite{Jin1996PRL,Mewes1996PRL}. 
Subsequently, the ensemble is collimated by turning off the trapping potential at the turning point of the resulting oscillations of the ensemble size.

We apply this method at two different scattering lengths \SI{158}{a_0} and \SI{10}{a_0} at which the interaction and kinetic energy terms respectively dominate~(c.f. theoretical model section).
In the following we differentiate between expansion energies along a singular axis in $i$-direction (E$^\text{1D}_i$), 2D energies in the horizontal plane in which the matter-wave lens is applied (E$^\text{2D}$) and the full three dimensional expansion energy (E$^\text{3D}$). 
For both measurements at the two different scattering lengths, we use the same initial and final trap configurations, with small variations of the parameters resulting only from pointing instabilities of the ODT beams which we relate to the time passed between the two measurement campaigns.
In both cases the initial trap is realized without any CPM.
We find initial trapping frequencies of $2\pi\times\{72, 144, 115\}$ Hz for \SI{158}{a_0} and $2\pi\times \{62, 149, 96\}$ Hz for \SI{10}{a_0}.
After relaxation our final trap frequencies are $2\pi\times \{23, 36, 126\}$ Hz for \SI{158}{a_0} and $2\pi\times \{24, 38, 129\}$ Hz for \SI{10}{a_0}.
In parallel the trap depth is maintained by increasing the laser intensity, suppressing atom number loss.
Based on time-of-flight (TOF) measurements of the ensemble’s expansion, we determine the expansion energies along the horizontal (collimated) $x'$- and vertical (not collimated) $z$-direction within the camera frame for different holding times $t_\text{hold}$, after relaxing the trap.
\subsection{Obtained energies}
At a scattering length of \SI{158}{a_0} (Fig.~1a) the minimal value in the collimated direction yields E$^\text{1D}_{x'}=\SI{1\pm 0.17}{\nano\kelvin}$ and is achieved after a holding time of \SI{24}{\milli\second}.
For \SI{10}{a_0} (Fig.~1b) we find the minimum for a holding time of \SI{22}{\milli\second} after decompression, resulting in a minimal value of E$^\text{1D}_{x'}=\SI{340 \pm 12}{\pico\kelvin}$ after up to \SI{25}{\milli\second} TOF.
While the behavior derived from simulations (c.f. data acquisition and analysis section) agrees with these findings, for the points below \SI{1}{\nano\kelvin} a portion of interaction energy remains and the ensemble has not yet reached the linear expansion regime at that point.
When correcting for this effect, by simulating for a TOF of \SI{250}{\milli\second} the asymptotic behavior yields a minimum of E$^\text{1D}_{x'}=\SI{429 \pm 56}{\pico\kelvin}$ after a holding time of \SI{22.1}{\milli\second}.
The excellent agreement between experiment and simulation allows to understand the ensemble's dynamics in the entire horizontal plane, as both theoretical approaches feature coupling of ensemble oscillations in all directions.
Including the axis which cannot be directly observed, we extrapolate the resulting 2D expansion energies as depicted by the dashed purple lines in Fig~1.
At \SI{10}{a_0} we find a minimal value of E$^\text{2D} = \SI{301 \pm 65}{\pico\kelvin}$ for a TOF of \SI{25}{\milli\second}, which corresponds to an improvement by \SI{3.7}{dB} over the \SI{158}{a_0} case. 
Extending the simulation to the ballistic regime as before, yields a final value of E$^\text{2D} = \SI{438 \pm 77}{\pico\kelvin}$ for \SI{250}{\milli\second} TOF~(Fig.~2a).
\subsection{Comparison to previous results}\label{sec:Evaluation}
In this work we applied our matter-wave collimation protocol previously developed for \Rb to a \K BEC and prove the ease of application to another atomic species, demonstrating a reduction of the expansion energy by \SI{13}{dB} compared to the non-collimated case, as given for vanishing holding time.
Considering the mass ratio of both elements, the obtained energy of \SI{1 \pm 0.17}{\nano\kelvin} for \K at \SI{158}{a_0}, corresponding to an expansion velocity of \SI{0.46 \pm 0.04}{\milli\meter\per\second}, is comparable to the previously achieved result of $\SI{3.2\pm 0.6}{\nano\kelvin}\equiv \SI{0.55 \pm 0.05}{\milli\meter\per\second}$ with \Rb~\cite{Albers2022Commun} at its natural background scattering length of $\sim\SI{100}{a_0}$~\cite{EgorovPRA2013}.
The remaining difference can be attributed to variations in the trap frequency ratios between the two experiments, rather than to the difference in scattering length, since changing the latter by less than a few multiples does not significantly affect the expansion rate when staying within the strong interaction regime~\cite{Kraemer2004APB, Roati2007PRL}.
Hence, the observed outcome aligns with expectations as the technique only depends on the ensemble's dynamics governed by interactions and trap frequencies and accurately described through the Gross-Pitaevskii equation.
More importantly, we show that the final expansion energy after the lens can be further reduced by transitioning into the weak interaction regime, as done here through minimizing the scattering length by means of a magnetic Feshbach resonance.
By reducing the repulsive forces driving the expansion after release from the trap, we achieve expansion energies well below \SI{1}{\nano\kelvin}, which is necessary to match the requirements of proposed experiments, e.g. for, but not limited to, gravitational wave detection~\cite{Hogan2011GRG,Canuel2018SciRep,Zhan2019quq,schubert_scalable_2019,Canuel2020CQG,Badurina2020JCA}, test of the Weak Equivalence Principle~\cite{Schlippert14PRL,Asenbaum2020PRL,Ahlers2022} or the search for dark matter~\cite{ElNeaj2020EPJQ,Du2022PRD,Badurina2023PRD}.
While the energies realized here are still an order of magnitude larger than in previous demonstrations in two~\cite{Kovachy2015PRL} and three dimensions~\cite{Deppner2021PRL}, our method can be applied directly in the ODT. 
Hence it is suitable for setups and applications which do not allow for an extended pre-expansion time before applying the lens due to constraints regarding experimental cycle time or spatial dimensions. 
Lower expansion energies are currently limited by the achievable maximum CPM amplitude of \SI{200}{\micro\meter} which in turn restricts the range of accessible trapping frequencies to the values given in lensing protocol section.
\subsection{Scattering length and trap frequency dependencies}\label{sec:Scat_dependency}
\begin{figure*}[tbh!]
    \centering
    \includegraphics[width = 0.99\textwidth]{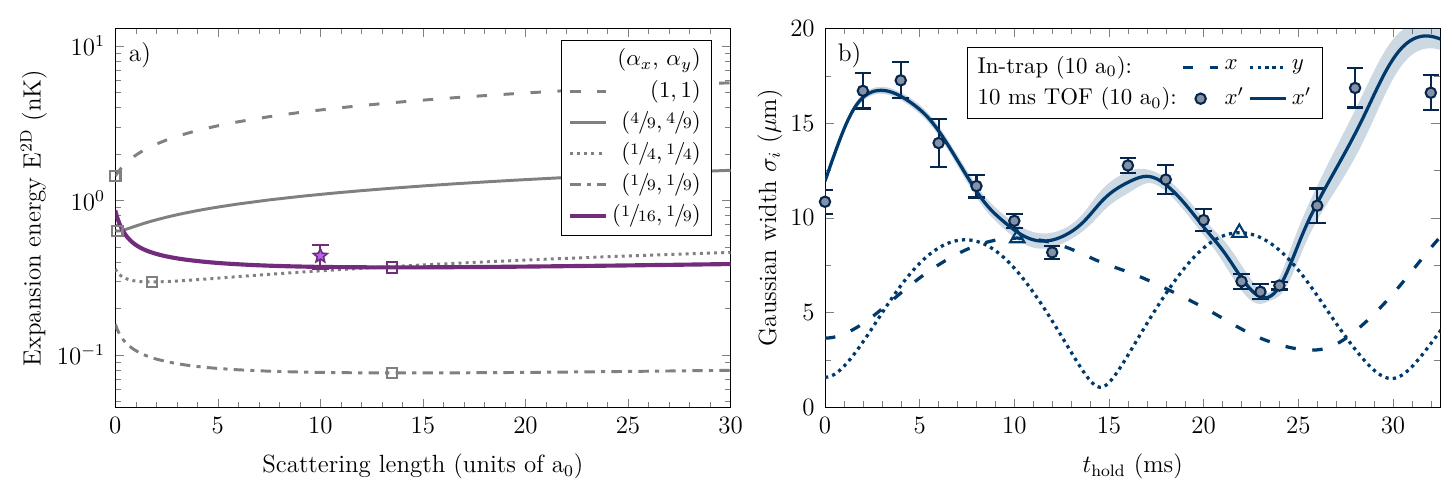}
    \caption{
    \textbf{Ensemble dynamics in the collimated plane.}
    Panel a shows the simulated 2D expansion energy E$^\text{2D}$ in the collimated plane for different scaling factors $\alpha_i$ and for the configuration used in the experiment (see lensing protocol section) for \SI{250}{\milli\second} time-of-flight (TOF).
    For a common frequency reduction along both lensed directions with $\alpha_{x}=\alpha_{y}$ taking the values $\{1, 4/9, 1/4, 1/9\}$, the minimal energies are obtained for a scattering length $a$ being respectively $\{0, 0.1, 1.8, 13.5\}$~\si{a_0} and identified by the squares for each case. 
    The sequence becomes more robust against changes of the scattering length with larger frequency reduction, as the minima become more shallow.
    The purple star resembles the lowest 2D experimental expansion energy presented in Fig.~1b (obtained here for 250 ms TOF). 
    While the curves are simulated for a fixed set of parameters, this point is obtained within a Monte-Carlo simulation including all experimental uncertainties. 
    The error bar denotes 2-$\sigma$ deviation while the central point stands for the mean value (see data acquisition and analysis section).
    Qualitatively, the experimental configuration closely resembles the case of $\alpha_{x}=\alpha_{y} = 1/9$. 
    The resulting expansion energies are globally shifted towards higher values, since $\alpha_{x}\approx 1/16$ and $\alpha_{y}\approx 1/9$.
    This causes the optimal release points to differ for each axis as marked by the triangle symbols in panel b, highlighting the importance of a symmetric choice of $\alpha$-values. 
    Here, the measured ensemble width in $\mathbf{x'}$-direction is shown as blue circles for \SI{10}{\milli\second} TOF at \SI{10}{a_0} and the error bars represent the standard deviation of at least four measurements, 
    The simulated size after 10 ms TOF is shown as solid blue line, while the corresponding oscillations of the ensemble widths in $\mathbf{x}$- and $\mathbf{y}$-direction within the trapping potential are shown with blue dashed and dotted lines, respectively.
    }
    \label{fig:Scattering_study}
\end{figure*}
To gain insight into the impact of the scattering length onto the collimation, we analyze the ensemble's behavior in the weak interaction regime by simulating the dynamics with an adapted theoretical model for two different scenarios (Fig.~2a). 
Starting from trap frequencies of $2\pi\times\SI{60}{\hertz}$ in all directions, we apply a common reduction in the horizontal plane while maintaining the frequency along the $z$-axis, as shown by the theoretical grey lines. 
For all scenarios we evaluate the optimal holding time after relaxing the trap for minimizing E$^\text{2D}$ and study the behavior for different squared trap frequency ratios $\alpha_{i} = (\omega_i^\text{F} / \omega_i^\text{I})^2$, which would provide the energy scaling in the ideal gas regime~\cite{Chu86OL}. 
As previously discussed by Kovachy et al.~\cite{Kovachy2015PRL} and also observed here, the energy reduction for a BEC is significantly higher due to an interplay of interactions and coupling of the oscillations of the ensemble widths along each axis. 
From the theory simulations we find a reduction of the expansion energy towards smaller scattering length for $\alpha_{x}=\alpha_{y} > 4/9$ as shown by the dashed grey line. 
This result matches the expected dynamics of an ensemble during free-fall expansion without any additional collimation and is explained by repulsive interactions after removing the trapping potential~\cite{Weber2003Science, Kraemer2004APB, Roati2007PRL}.
We identify a minimal expansion energy at non-zero interactions for $\alpha_{x}=\alpha_{y}\lessapprox 4/9$, as depicted by the continuous, dotted and dash-dotted grey lines.
These curves clearly show that for smaller values of $\alpha_{i}$, reaching optimal energies requires to move towards higher scattering length values. 
While in these cases minimizing the interaction energy still reduces the corresponding forces upon release, it also reduces the final ensemble width within the trap.
Hence, the kinetic energy from the resulting fundamental momentum spread as given by the uncertainty principle increases and the optimal scattering length must be found considering both contributions.
To achieve a minimum momentum spread in free fall one hence wants to increase the interactions during the lens to maximize the cloud size and cancel them at the release time.
However, given that the interactions are controlled via magnetic fields such an optimization is technically not feasible, as it typically takes tens of millisecond to change the magnetic field~\cite{Masi2021PRR}.
Therefore the optimal scattering length has to be found as a trade-off between the maximum size achievable within the trap and minimal repulsive interactions in free fall along the horizontal direction. 
Such optimized configurations are highlighted by the empty squares in Fig.~2a. 
Regarding the experimental setup, as shown by the purple line, where the change in aspect ratio is not the same for both directions, $\alpha_{x}\approx 1/16$ and $\alpha_{y}\approx 1/9$, we recover the same behavior as described above for the case $\alpha_{x}=\alpha_{y}\lessapprox 4/9$, but with globally higher energies
The similarity can be explained by the fact that $\sqrt{\alpha_{x}\alpha_{y}}=1/12$.
Note here that the purple star is identical to the one already shown in Fig.~1b.
\begin{figure*}[tbh!]
    \centering
    \includegraphics[width = \textwidth]{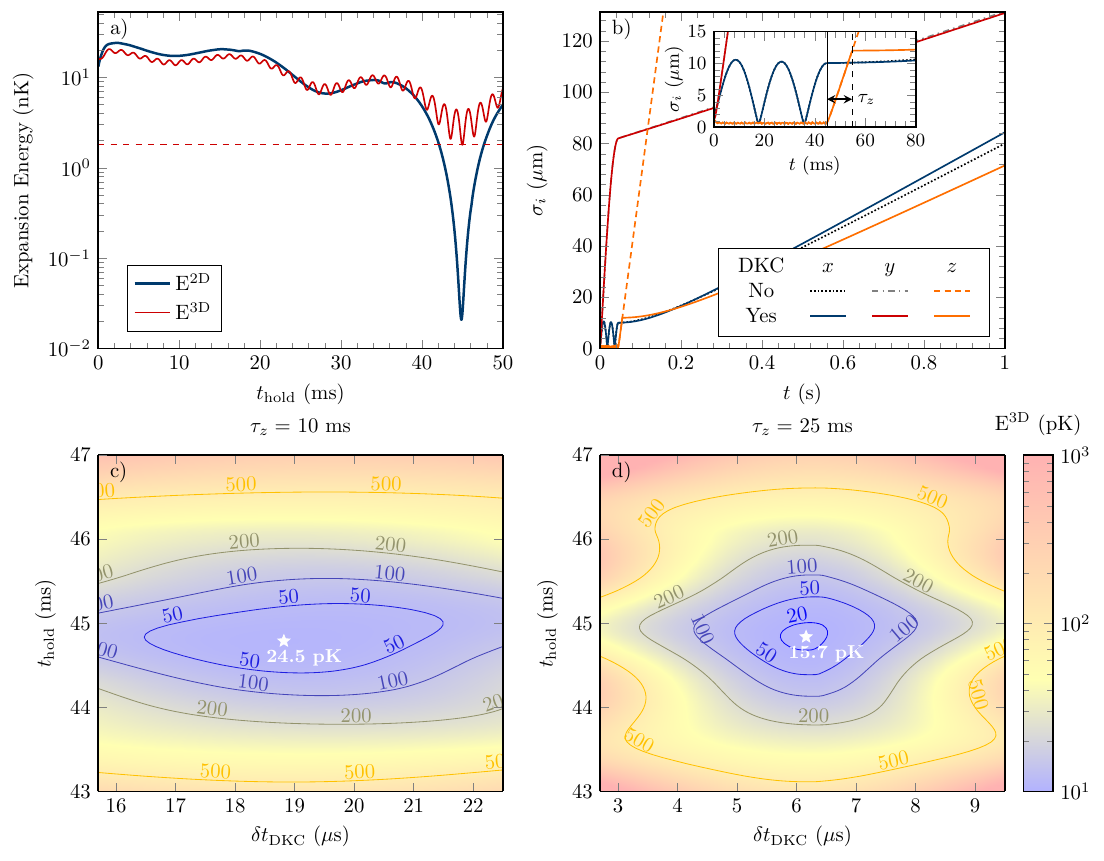}
    \caption{\textbf{Generating a delta-kick collimated Bose-Einstein condensate in the regime of tens of pK.}
    We take advantage of the holding process after trap relaxation to minimize the energy in the $\{x,y\}$-plane (a) (red line).
    Subsequently, a short free-fall time (pre-TOF) $\tau_z$ allows the ensemble to expand, followed by a delta-kick collimation (DKC) to collimate the third direction (b).
    We show the width evolution in all three directions after the optimal holding period without (non-solid lines) and with a DKC (solid lines). 
    The inset shows the dynamics in the trapping potential, highlighting release (solid black line) and DKC (black dashed line) timings.
    This process leads to a reduced 3D expansion energy expressed as a function of the lensing time and the DKC duration after a pre-TOF of 10 ms (c) and 25 ms (d), leading to respectively \SI{24.5}{\pico\kelvin} and \SI{15.7}{\pico\kelvin}.
    \label{fig:Holding_DKC}
    } 
\end{figure*}

To get further insight of the complex behavior of the matter wave for different aspect ratios we now compare in Fig~2b the simulated in-trap oscillations in $x$- and $y$-direction, respectively presented with dashed and dotted lines, with the observed ensemble width after \SI{10}{\milli\second} TOF at $a=\SI{10}{a_0}$. 
From the experimental measurements (blue points) we find two distinct minima, each of which can be assigned to the maxima of the underlying ensemble widths within the trap along a different axis (empty blue triangles).
In this specific case the lowest expansion energy E$^\text{2D}$ is found near the optimal collimation of the $y$-direction, closely resembling the case $\alpha_{x} = \alpha_{y} = 1/9$ in Fig.~2a 
Compared to the symmetric case the globally higher energies are hence explained by the energy contribution of the other direction, which always exhibits a non-vanishing expansion at release, as long as the aspect ratios are not integer multiples of each other.
When performing additional simulations at $a = \SI{30}{a_0}$, we further note that the optimal release timing is extremely robust with respect to changes of the scattering length, giving only \SI{0.3}{\milli\second} offset in this particular case.
While in practice such changes might arise from technical limitations, e.g. due to imperfect control of the Feshbach field, the offset is fundamentally expected, since changing the repulsive interactions within the trap alters the frequency of the excited oscillations.

Overall, our analysis yields the choice of trapping frequencies to be more important due to their enhanced scaling and the effects of asymmetric expansion compared to the exact scattering length knowledge which is more difficult to pinpoint in practice.
Besides allowing to extract the 2D expansion energy from the measurements, the adequacy between the experiment and theory model in Fig.~1 and Fig.~2 allow us in the following to identify advanced collimation scenarios.
Consecutively, we discuss two 3D collimation sequences based on the combination of a 2D in-trapped collimation combined with a pulse delta-kick method collimating the third axis to reach the pK regime~\cite{Corgier2018NJP,Deppner2021PRL}.
\subsection{Advanced Scenario}\label{sec:Adv_scenario}
Since neither the demonstrated method, nor the experimental apparatus is designed to collimate the remaining vertical axis, the achievable 3D expansion energies are limited to the nanokelvin regime, regardless of the performance in the horizontal plane, as shown by the horizontal dashed line in Fig.~3a.
To overcome this limitation we consider a short free-fall time (pre-TOF) $\tau_\text{z}$ at the end of the holding process followed by a pulsed DKC protocol~\cite{Ammann1997PRL,Kovachy2015PRL,Deppner2021PRL,Gaaloul2022NatComm}.
We study the theoretically achievable expansion energies for this sequence in an advanced scenario which is specifically tailored towards the capabilities of an improved apparatus~\cite{Herbst2024PRR} and highlight the crucial requirements for the implementation.
Instead of a recycled ODT, the setup features two individually controllable beams, each with up to \SI{16}{\watt} of optical power at a wavelength of \SI{1064}{\nano\meter}.
This configuration allows to realize a variety of possible trap geometries and especially to design common turning points for the oscillations of the ensemble widths along both principal axes. 
Furthermore, 2D acousto-optical deflectors (AODs) [AA Opto-Electronic DTSXY-400-1064] are used to create time-averaged optical potentials instead of the previously used AOM.
In combination with the implemented lens-system, their superior bandwidth allows for CPM amplitudes of at least \SI{1.5}{\milli\meter} and consequently to access lower final trap frequencies and expansion energies.
Finally, for the DKC, the second AOD axis is needed to shift the ODT beams vertically and match the position of the atomic cloud for a maximum pre-TOF of $\tau_z = \SI{25}{\milli\second}$, corresponding to a free-fall distance of \SI{3}{\milli\meter}. 

For E$^\text{2D}$ we numerically find minimal values below \SI{20}{\pico\kelvin} for a holding time of \SI{42.5}{\milli\second} switching the trap frequencies from $2\pi\times \{152.7, 310.7, 342.6\}$ Hz to $2\pi\times \{28.1, 5.6, 340.0\}$ Hz at \SI{10}{a_0} scattering length (Fig.~3a).
For the final trap configuration \SI{150}{\milli\watt} of optical power at a CPM amplitude of \SI{175}{\micro\meter} for one and \SI{450}{\milli\watt} with \SI{800}{\micro\meter} modulation stroke for the other beam is required.
Since the frequency along the vertical axis is much higher than the two others, the DKC (black dashed line in the inset of Fig.~3b) will not significantly affect the other direction as shown by the black dotted and dash-dotted curves in Fig.~3b.
For an easy configuration with only $\tau_z =\SI{10}{\milli\second}$ pre-TOF, corresponding to a free fall distance of \SI{490}{\micro\meter}, experimentally accessible in practice, we obtain a minimal 3D expansion energy of E$^{\text{3D}}=\SI{24.5}{\pico\kelvin}$ with a $\delta t_{\text{DKC}}=\SI{18.8}{\micro\second}$ long delta-kick pulse and \SI{44.8}{\milli\second} of lensing as shown in Fig.~3c and optimised using a simulated annealing algorithm~\cite{Kirkpatrick1983Science}.
Moreover, the implementation is expected to be robust against variations of the experimental parameters as it allows to achieve energies below \SI{50}{\pico\kelvin} for a wide range of holding and delta-kick durations.
Even better performance can be obtained by increasing the pre-TOF duration at the expense of the overall robustness with respect to the delta-kick timing~\cite{Ammann1997PRL}. 
For $\tau_z = $ \SI{25}{\milli\second} of pre-TOF, we find final energies as low as E$^\text{3D}=\SI{15.7}{\pico\kelvin}$, but requiring a DKC of only $\delta t_{\mathrm{DKC}}=\SI{6.2}{\micro\second}$ (Fig.~3d).
The simulation explicitly takes the AOD's response time of \SI{3}{\micro\second} into account, as it is on the same order of magnitude as $\delta t_\mathrm{DKC}$.
While other experimental limitation e.g. due to the bandwidth of the different control loops may apply, the AOD is the slowest component involved and therefore poses the relevant limitation for the advanced scenario, contrary to the measurements in Fig.~1.
Nevertheless, such timings can be experimentally challenging when being limited to the center-position modulation frequency below \SI{100}{\kilo\hertz} as relevant time scale or using rf-switches with switching times of several \si{\micro\second}.
For such short signals arbitrary waveform generators based on direct digital synthesizers (DDS) with a high sampling rate offer a convincing solution.
In this case we use a software defined radio [Ettus USRP X310], whose DDS allows to interrupt the waveform at any given sample and match the pulse length with a resolution of \SI{50}{\nano\second} for a typical sampling rate of \SI{20}{\mega\hertz}.
As before using a low scattering length assists in the overall collimation.
However, simulating the same sequence for a scattering length of \SI{158}{a_0} in particular, still leads to expansion energies of E$^\text{3D}=\SI{97.5}{\pico\kelvin}$ for $\tau_z=\SI{10}{\milli\second}$ of pre-TOF and E$^\text{3D}=\SI{81.9}{\pico\kelvin}$ for \SI{25}{\milli\second} pre-TOF with \SI{17.6}{\micro\second} and \SI{7.1}{\micro\second} long delta-kick pulses, respectively.

Hence, the analyzed two-step process opens up the path to approach ($a = \SI{158}{a_0}$) and even exceed ($a = \SI{10}{a_0}$) the results that were obtained in a drop-tower~\cite{Deppner2021PRL}, on the International Space Station~\cite{Gaaloul2022NatComm} and within a long-baseline device~\cite{Kovachy2015PRL}.
Combining these results with a strategy for rapid evaporation~\cite{Herbst2024PRR} and a bright source for fast magneto-optical trap (MOT) loading~\cite{Catani2006PRA}, compact or even field-deployable devices can reach experimental repetition rates higher than \SI{0.5}{\hertz} with BECs consisting of \num{3e5} atoms and state-of-the-art collimation. 
\begin{figure}[tbh!]
    \centering
    \includegraphics[width = 0.99\linewidth]{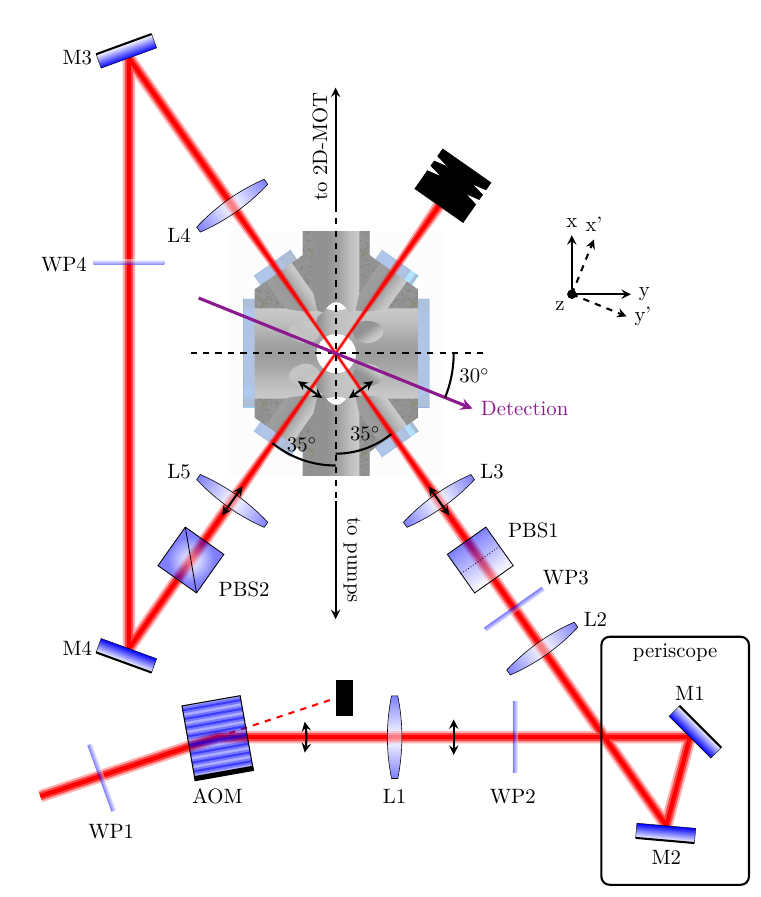}
    \caption{\textbf{Optical dipole trap setup.}
    Time-averaged potentials are implemented with an acousto-optical modulator (AOM). 
    A three-lens system (L1, L2, L3) with focal lengths $f_1=\SI{100}{\milli\meter}$, $f_2=\SI{300}{\milli\meter}$ and $f_3=\SI{150}{\milli\meter}$,
    translates the change in AOM deflection angle into a parallel displacement,
    while simultaneously focusing the beam to a waist of \num{30} (\num{45}) \si{\micro\meter} in horizontal (vertical) direction.
    The lenses L4 and L5 ($f_{4,5}=\SI{150}{\milli\meter}$) are used to re-collimate the beam after passing the experimental chamber and to re-focus it on the atoms. 
    At each point optimal polarization is ensured by wave plates (WP1 - WP4) with additional orthogonal oriented polarizing beam splitters (PBS1 and PBS2) in front of the chamber for polarization cleaning.
    Dielectric mirrors (M1 - M4) are used to guide the beam through the setup. 
    For perfect alignment and equal beam power the trap frame $\{x,y,z\}$, as given by the principal axes of the optical potential, resembles the symmetry axes of the vacuum chamber. 
    The camera frame $\{x',y',z\}$ is obtained by a rotation around the $z$-axis, with the exact detection angle depending on the beam configuration and alignment. 
    The figure is taken from Albers et al.~\cite{Albers2022Commun} and openly licensed via \href{https://creativecommons.org/licenses/by/4.0/}{CC BY 4.0}. 
    Here the orientation of the detection arrow and the naming of the coordinate systems was altered to account for changes of the apparatus compared to the source material.
    }
    \label{fig:Setup}
\end{figure}
\section{Methods}
\subsection{Experimental apparatus}\label{sec:apparatus}
We use the same setup as for the previous matter-wave lens study with \Rb~\cite{Albers2022Commun}, featuring a crossed ODT in recycled beam configuration (Fig.~4). 
A detailed description of the vacuum, laser and coil systems used can be found in previous publications~\cite{Schlippert14PRL, Albers2020EPJD, Herbst2022PRA}.
The ODT is based on a \SI{1960}{\nano\meter} fiber laser [IPG TLR-50-1960-LP] which is intensity stabilized by a feedback loop, controlling a linearized Pockels cell.  
The crossed trap is realized by recycling the same beam, passing the atoms again under an angle of \SI{70}{\degree}.
We ensure orthogonal beam polarization to avoid running lattice formation.
Due to the elliptical beam shape of the fiber laser output we obtain different beam waists of \SI{30}{\micro\meter} in horizontal and \SI{45}{\micro\meter} in vertical direction. 
Taking losses at all optical elements into account, the maximal power which can be delivered to the atoms is limited to \SI{8}{\watt} for the initial and \SI{6}{\watt} for the recycled beam.
A custom-made AOM [Polytec ATM-1002FA53.24] is used to deflect the beam, thereby creating time-averaged optical potentials of harmonic shape along one beam axis~\cite{Roy2016PRA,Albers2020phd}.
It is further utilized to control the beam power at the lower end of the intensity stabilization. 
By focusing the beam onto the atomic cloud, the change in deflection angle of the AOM is translated into a parallel displacement of the beam.
For the initial beam, the bandwidth of the AOM allows for a maximum CPM amplitude of \SI{200}{\micro\meter}.
For the recycled beam, the same configuration corresponds to a CPM amplitude of \SI{300}{\micro\meter} due to the additional re-collimation and re-focusing, and the increased path length in-between.
Due to the experimental configuration, the recycled beam is fully determined by the state of the initial beam and hence the setup does not allow to choose the trap frequencies in $x$- and $y$-direction independently.
The required frequency modulation of the rf-signal driving the AOM is generated with a combination of voltage controlled oscillator [Mini-Circuits ZOS-150+] to provide the actual signal and an arbitrary waveform generator [Rigol DG1022Z], which provides the waveform.
For the whole experimental sequence a constant modulation frequency of \SI{20}{\kilo\hertz} is used, which is sufficiently large compared to any occurring trap frequency. 
We define the trap frame $\{x,y,z\}$ as the principal axes of the trapping potential, given by the eigenvectors of the curvature in the critical point. 
In the case of equal ODT beams in terms of power and waist it resembles the symmetry axis of the experimental setup as shown in Fig.~4, while any deviations from the ideal beam configuration result in rotations of the coordinate system around the $z$-axis.
Finally, detection is performed by absorption imaging with unity magnification.
The camera is situated in the $\{x,y\}$-plane and the related camera frame $\{x',y',z\}$ can be obtained from the trap frame $\{x,y,z\}$ by rotating clockwise around the $z$-axis by an angle of $\sim\SI{30}{\degree}$. 
Note, that the exact angle depends on the beam configuration prior to detection due to the resulting rotation of the trap frame.
\subsection{Ensemble preparation}\label{sec:preparation}
We apply a trap loading and state preparation sequence optimized for \K as described earlier~\cite{Herbst2022PRA}.
We load a 3D-magneto optical trap on the \DTwo-line from a 2D-MOT, trapping \num{1e9} atoms within \SI{4}{\second}.
Subsequently, we apply a hybrid \DOne-\DTwo compression MOT to increase the ensemble's density and grey molasses cooling on the \DOne-line for cooling the ensemble to sub-Doppler temperatures~\cite{Salomon13EPL}.
In this manner we prepare \num{4e8} atoms at a temperature of \SI{12}{\micro\kelvin} within \SI{56}{\milli\second} after turning off the 2D-MOT.
For loading the ODT we use a center-position modulation amplitude of \SI{160}{\micro\meter} to improve the mode matching of the crossing region with the cloud, transferring \num{12e6} atoms into the \SI{54}{\micro\kelvin} deep ODT, with a temperature of \SI{8.5}{\micro\kelvin}. 
Afterwards, we prepare the ensemble in $\ket{F = 1, \mf=-1}$ with a multi-loop state preparation scheme based on microwave adiabatic rapid passages.
This allows to use the broad Feshbach resonance at \SI{32.6}{\gauss}~\cite{DErrico07NJP} to adjust the scattering length to positive values, necessary for direct evaporative cooling~\cite{Landini12PRA}.
We use the evaporation sequence optimized for the largest number of condensed particle, rather than the shortest experimental cycle time with the highest atomic flux, realizing a quasi-pure BEC of up to \num{2e5} atoms after \SI{3.9}{\second} of evaporative cooling at a scattering length of \SI{158}{a_0}. 
For the measurement at \SI{158}{a_0} we perform the matter-wave lens \SI{100}{\milli\second} after creating the BEC by increasing the center-position modulation amplitude to the achievable maximum, as stated before.
For the measurement at \SI{10}{a_0}, we additionally adiabatically sweep the magnetic field towards the broad minimum between the resonance at \SI{32.6}{\gauss} and the next higher one at \SI{162.8}{\gauss} after creating the BEC and before performing the matter-wave lens.
\subsection{Data acquisition and analysis}\label{sec:measurement}
We perform TOF measurements for different holding times after relaxing the trap with a total experimental cycle time of \SI{12}{\second}.
Subsequently we describe the obtained density profile either by a Gaussian or a Thomas-Fermi distribution, depending on the scattering length used. 
For the measurements at \SI{158}{a_0} the Thomas-Fermi radii $R_i(t)$ are transformed into their equivalent standard deviation $\sigma_i(t)$ using $\sigma_i(t) = R_i(t)/\sqrt{7}$~\cite{Corgier2018NJP, Corgier2020NJP}.
Individual data points are taken for a holding time spacing of \SI{1}{\milli\second} and a TOF spacing of \SI{5}{\milli\second}. 
Each measurement is repeated at least four times. 
At \SI{10}{a_0} we fit a Gaussian to the obtained density distribution.
For these measurements the TOF spacing is reduced to \SI{1}{\milli\second} at the expense of the holding time spacing which is increased to at least \SI{2}{\milli\second}, in order to obtain better statistics for the extracted ensemble expansion.
For each individual dataset, measurements are performed over the course of 12 hours of continuous operation.  
With this approach we ensure comparability within each dataset and avoid trap frequency drifts caused by thermal effects from power cycling the \SI{1960}{\nano\meter} laser in between measurement days.
To obtain the linear expansion rate for a given holding time, we only consider the data taken for more than \SI{10}{\milli\second} TOF, avoiding the resolution limitation of our detection system. Finally, the fitted expansion rates $v_i$ are transformed into 1D expansion energy using E$^{\text{1D}}_i = k_B T_i/2 =m v_i^2 / 2 $.

To simulate the behavior of the ensemble, we determine the trapping frequencies by fitting the oscillations of the ensemble width with respect to the holding time for a constant TOF in the trap frame and by projecting them into the rotated camera frame, afterwards.
Since the detection angle relative to the trap frame changes with respect to small deviations of the ODT beam alignment, the exact angle is evaluated for each measurement separately and fitted to the data, as well. 
We optimize the fit parameters on five different TOF in between \num{10} and \SI{25}{\milli\second} simultaneously with equal weighting, obtaining a single set of values, which provide the overall smallest error.
Based on the frequencies found, we perform simulations of the ensemble's behavior using the two approaches provided in the theoretical model section. 
The error bands stem from \num{1000} Monte-Carlo simulations within the obtained errors of trap frequencies, detection angle and scattering length (at \SI{10}{a_0}) as determined by fitting the ensemble width and the magnetic field characterization of the apparatus.  

For the advanced scenario we take experimental parameters and technical limitations of the setup, e.g. the rise time of the AOD, into account. 
We search for optimal 3D collimation by simulating a grid with a step size of \SI{36}{\micro\second} for the lensing and \SI{62}{\nano\second} for the DKC, using a simulated annealing algorithm \cite{Kirkpatrick1983Science} for the absolute minimum in each case. 
\subsection{Theoretical model}\label{sec:th_model}
For a scattering length of \SI{158}{a_0}, the interaction energy exceeds the kinetic energy of the ensemble, so the BEC dynamics is well described by the scaling equations as derived by Castin et al.~\cite{Castin1996PRL} and Kagan et al.~\cite{Kagan1997PRA}:
\begin{equation}
\ddot{\lambda}_i(t) + \omega_i^2(t) \lambda_i(t) = \frac{\omega_i^2(0)}{\lambda_i\lambda_x\lambda_y\lambda_z},
    \label{eqn:CastinDum}
\end{equation}
where the dimensionless variable $\lambda_i(t)=R_i(t)/R_i(0)$ characterizes the evolution of the size of the condensate in the direction $i \in \{x,y,z\}$. In this expression, $R_i(t)$ is the Thomas-Fermi radius in the direction $i$, and the initial radius is given by~\cite{Pethick2008}
\begin{equation}
R_i(0) = a_\text{osc}\frac{\bar{\omega}(0)}{\omega_i(0)}\left(\frac{15 N a}{a_\text{osc}}\right)^{1/5},
\end{equation}
with the average length of the quantum harmonic oscillator $a_\text{osc} = [\hbar/m\bar{\omega}(0)]^{1/2}$ and the geometric mean of the initial trapping frequencies $\bar{\omega}(0) = [\omega_x(0) \omega_y(0) \omega_z(0)]^{1/3}$. From the solution of Eq.\;(\ref{eqn:CastinDum}) we extract the standard deviations $\sigma_i(t) = R_i(t)/\sqrt{7}$ associated with the atomic density, that we compare with the experimental measurements obtained as described in the data acquisition and analysis section.

In the case of a scattering length of \SI{10}{a_0}, the Thomas-Fermi approximation is no longer suitable to accurately describe the dynamics of the BEC. 
Instead, we follow a variational approach and describe the BEC with a Gaussian ansatz~\cite{Perez1996PRL, Perez1997PRA}.
This leads in a harmonic trap to the following set of coupled differential equations
\begin{equation}
    \ddot{\sigma}_i(t)+\omega^2_i(t)\sigma_i(t) = \frac{\hbar^2}{4 m^2\sigma_i^3(t)} + \frac{\hbar^2aN}{4\sqrt{\pi}m^2}\frac{1}{\sigma_i\sigma_x\sigma_y\sigma_z},
    \label{eqn:Variationnal_approach}
\end{equation}
for the standard deviations $\sigma_i(t)$ of the atomic density. The time-independent version of Eq.~(\ref{eqn:Variationnal_approach}) is used to determine the initial size $\sigma_i(0)$ of the ensemble, which converges to the oscillator length for vanishing scattering length.
\section*{Data availability}
The data used in this manuscript is available from the corresponding author upon reasonable request.
\def\bibsection{\section*{\refname}} 
\bibliography{main.bib}

%apsrev4-2.bst 2019-01-14 (MD) hand-edited version of apsrev4-1.bst
%Control: key (0)
%Control: author (8) initials jnrlst
%Control: editor formatted (1) identically to author
%Control: production of article title (0) allowed
%Control: page (0) single
%Control: year (1) truncated
%Control: production of eprint (0) enabled
\begin{thebibliography}{83}%
\makeatletter
\providecommand \@ifxundefined [1]{%
 \@ifx{#1\undefined}
}%
\providecommand \@ifnum [1]{%
 \ifnum #1\expandafter \@firstoftwo
 \else \expandafter \@secondoftwo
 \fi
}%
\providecommand \@ifx [1]{%
 \ifx #1\expandafter \@firstoftwo
 \else \expandafter \@secondoftwo
 \fi
}%
\providecommand \natexlab [1]{#1}%
\providecommand \enquote  [1]{``#1''}%
\providecommand \bibnamefont  [1]{#1}%
\providecommand \bibfnamefont [1]{#1}%
\providecommand \citenamefont [1]{#1}%
\providecommand \href@noop [0]{\@secondoftwo}%
\providecommand \href [0]{\begingroup \@sanitize@url \@href}%
\providecommand \@href[1]{\@@startlink{#1}\@@href}%
\providecommand \@@href[1]{\endgroup#1\@@endlink}%
\providecommand \@sanitize@url [0]{\catcode `\\12\catcode `\$12\catcode
  `\&12\catcode `\#12\catcode `\^12\catcode `\_12\catcode `\%12\relax}%
\providecommand \@@startlink[1]{}%
\providecommand \@@endlink[0]{}%
\providecommand \url  [0]{\begingroup\@sanitize@url \@url }%
\providecommand \@url [1]{\endgroup\@href {#1}{\urlprefix }}%
\providecommand \urlprefix  [0]{URL }%
\providecommand \Eprint [0]{\href }%
\providecommand \doibase [0]{https://doi.org/}%
\providecommand \selectlanguage [0]{\@gobble}%
\providecommand \bibinfo  [0]{\@secondoftwo}%
\providecommand \bibfield  [0]{\@secondoftwo}%
\providecommand \translation [1]{[#1]}%
\providecommand \BibitemOpen [0]{}%
\providecommand \bibitemStop [0]{}%
\providecommand \bibitemNoStop [0]{.\EOS\space}%
\providecommand \EOS [0]{\spacefactor3000\relax}%
\providecommand \BibitemShut  [1]{\csname bibitem#1\endcsname}%
\let\auto@bib@innerbib\@empty
%</preamble>
\bibitem [{\citenamefont {Degen}\ \emph {et~al.}(2017)\citenamefont {Degen},
  \citenamefont {Reinhard},\ and\ \citenamefont {Cappellaro}}]{Degen2017RMP}%
  \BibitemOpen
  \bibfield  {author} {\bibinfo {author} {\bibfnamefont {C.~L.}\ \bibnamefont
  {Degen}}, \bibinfo {author} {\bibfnamefont {F.}~\bibnamefont {Reinhard}},\
  and\ \bibinfo {author} {\bibfnamefont {P.}~\bibnamefont {Cappellaro}},\
  }\bibfield  {title} {\bibinfo {title} {Quantum sensing},\ }\href
  {https://doi.org/10.1103/RevModPhys.89.035002} {\bibfield  {journal}
  {\bibinfo  {journal} {Rev. Mod. Phys.}\ }\textbf {\bibinfo {volume} {89}},\
  \bibinfo {pages} {035002} (\bibinfo {year} {2017})}\BibitemShut {NoStop}%
\bibitem [{\citenamefont {Mandel}\ \emph {et~al.}(2003)\citenamefont {Mandel},
  \citenamefont {Greiner}, \citenamefont {Widera}, \citenamefont {Rom},
  \citenamefont {H{\"a}nsch},\ and\ \citenamefont {Bloch}}]{Mandel2003Nature}%
  \BibitemOpen
  \bibfield  {author} {\bibinfo {author} {\bibfnamefont {O.}~\bibnamefont
  {Mandel}}, \bibinfo {author} {\bibfnamefont {M.}~\bibnamefont {Greiner}},
  \bibinfo {author} {\bibfnamefont {A.}~\bibnamefont {Widera}}, \bibinfo
  {author} {\bibfnamefont {T.}~\bibnamefont {Rom}}, \bibinfo {author}
  {\bibfnamefont {T.~W.}\ \bibnamefont {H{\"a}nsch}},\ and\ \bibinfo {author}
  {\bibfnamefont {I.}~\bibnamefont {Bloch}},\ }\bibfield  {title} {\bibinfo
  {title} {Controlled collisions for multi-particle entanglement of optically
  trapped atoms},\ }\href {https://doi.org/10.1038/nature02008} {\bibfield
  {journal} {\bibinfo  {journal} {Nature}\ }\textbf {\bibinfo {volume} {425}},\
  \bibinfo {pages} {937} (\bibinfo {year} {2003})}\BibitemShut {NoStop}%
\bibitem [{\citenamefont {Georgescu}\ \emph {et~al.}(2014)\citenamefont
  {Georgescu}, \citenamefont {Ashhab},\ and\ \citenamefont
  {Nori}}]{Georgescu2014RMP}%
  \BibitemOpen
  \bibfield  {author} {\bibinfo {author} {\bibfnamefont {I.~M.}\ \bibnamefont
  {Georgescu}}, \bibinfo {author} {\bibfnamefont {S.}~\bibnamefont {Ashhab}},\
  and\ \bibinfo {author} {\bibfnamefont {F.}~\bibnamefont {Nori}},\ }\bibfield
  {title} {\bibinfo {title} {Quantum simulation},\ }\href
  {https://doi.org/10.1103/RevModPhys.86.153} {\bibfield  {journal} {\bibinfo
  {journal} {Rev. Mod. Phys.}\ }\textbf {\bibinfo {volume} {86}},\ \bibinfo
  {pages} {153} (\bibinfo {year} {2014})}\BibitemShut {NoStop}%
\bibitem [{\citenamefont {Kasevich}\ and\ \citenamefont
  {Chu}(1991)}]{Kasevich1991PRL}%
  \BibitemOpen
  \bibfield  {author} {\bibinfo {author} {\bibfnamefont {M.}~\bibnamefont
  {Kasevich}}\ and\ \bibinfo {author} {\bibfnamefont {S.}~\bibnamefont {Chu}},\
  }\bibfield  {title} {\bibinfo {title} {Atomic interferometry using stimulated
  raman transitions},\ }\href {https://doi.org/10.1103/PhysRevLett.67.181}
  {\bibfield  {journal} {\bibinfo  {journal} {Phys. Rev. Lett.}\ }\textbf
  {\bibinfo {volume} {67}},\ \bibinfo {pages} {181} (\bibinfo {year}
  {1991})}\BibitemShut {NoStop}%
\bibitem [{\citenamefont {Kasevich}\ and\ \citenamefont
  {Chu}(1992)}]{Kasevich1992APB}%
  \BibitemOpen
  \bibfield  {author} {\bibinfo {author} {\bibfnamefont {M.}~\bibnamefont
  {Kasevich}}\ and\ \bibinfo {author} {\bibfnamefont {S.}~\bibnamefont {Chu}},\
  }\bibfield  {title} {\bibinfo {title} {Measurement of the gravitational
  acceleration of an atom with a light-pulse atom interferometer},\ }\href
  {https://doi.org/10.1007/BF00325375} {\bibfield  {journal} {\bibinfo
  {journal} {Applied Physics B}\ }\textbf {\bibinfo {volume} {54}},\ \bibinfo
  {pages} {321} (\bibinfo {year} {1992})}\BibitemShut {NoStop}%
\bibitem [{\citenamefont {Riehle}\ \emph {et~al.}(1991)\citenamefont {Riehle},
  \citenamefont {Kisters}, \citenamefont {Witte}, \citenamefont {Helmcke},\
  and\ \citenamefont {Bord{\'e}}}]{Riehle1991PRL}%
  \BibitemOpen
  \bibfield  {author} {\bibinfo {author} {\bibfnamefont {F.}~\bibnamefont
  {Riehle}}, \bibinfo {author} {\bibfnamefont {T.}~\bibnamefont {Kisters}},
  \bibinfo {author} {\bibfnamefont {A.}~\bibnamefont {Witte}}, \bibinfo
  {author} {\bibfnamefont {J.}~\bibnamefont {Helmcke}},\ and\ \bibinfo {author}
  {\bibfnamefont {C.~J.}\ \bibnamefont {Bord{\'e}}},\ }\bibfield  {title}
  {\bibinfo {title} {Optical ramsey spectroscopy in a rotating frame: Sagnac
  effect in a matter-wave interferometer},\ }\href
  {https://doi.org/10.1103/PhysRevLett.67.177} {\bibfield  {journal} {\bibinfo
  {journal} {Phys. Rev. Lett.}\ }\textbf {\bibinfo {volume} {67}},\ \bibinfo
  {pages} {177} (\bibinfo {year} {1991})}\BibitemShut {NoStop}%
\bibitem [{\citenamefont {Cronin}\ \emph {et~al.}(2009)\citenamefont {Cronin},
  \citenamefont {Schmiedmayer},\ and\ \citenamefont {Pritchard}}]{Cronin09RMP}%
  \BibitemOpen
  \bibfield  {author} {\bibinfo {author} {\bibfnamefont {A.~D.}\ \bibnamefont
  {Cronin}}, \bibinfo {author} {\bibfnamefont {J.}~\bibnamefont
  {Schmiedmayer}},\ and\ \bibinfo {author} {\bibfnamefont {D.~E.}\ \bibnamefont
  {Pritchard}},\ }\bibfield  {title} {\bibinfo {title} {Optics and
  interferometry with atoms and molecules},\ }\href
  {https://doi.org/10.1103/RevModPhys.81.1051} {\bibfield  {journal} {\bibinfo
  {journal} {Rev. Mod. Phys.}\ }\textbf {\bibinfo {volume} {81}},\ \bibinfo
  {pages} {1051} (\bibinfo {year} {2009})}\BibitemShut {NoStop}%
\bibitem [{\citenamefont {Schlippert}\ \emph {et~al.}(2014)\citenamefont
  {Schlippert}, \citenamefont {Hartwig}, \citenamefont {Albers}, \citenamefont
  {Richardson}, \citenamefont {Schubert}, \citenamefont {Roura}, \citenamefont
  {Schleich}, \citenamefont {Ertmer},\ and\ \citenamefont
  {Rasel}}]{Schlippert14PRL}%
  \BibitemOpen
  \bibfield  {author} {\bibinfo {author} {\bibfnamefont {D.}~\bibnamefont
  {Schlippert}}, \bibinfo {author} {\bibfnamefont {J.}~\bibnamefont {Hartwig}},
  \bibinfo {author} {\bibfnamefont {H.}~\bibnamefont {Albers}}, \bibinfo
  {author} {\bibfnamefont {L.~L.}\ \bibnamefont {Richardson}}, \bibinfo
  {author} {\bibfnamefont {C.}~\bibnamefont {Schubert}}, \bibinfo {author}
  {\bibfnamefont {A.}~\bibnamefont {Roura}}, \bibinfo {author} {\bibfnamefont
  {W.~P.}\ \bibnamefont {Schleich}}, \bibinfo {author} {\bibfnamefont
  {W.}~\bibnamefont {Ertmer}},\ and\ \bibinfo {author} {\bibfnamefont {E.~M.}\
  \bibnamefont {Rasel}},\ }\bibfield  {title} {\bibinfo {title} {Quantum test
  of the universality of free fall},\ }\href
  {https://doi.org/10.1103/PhysRevLett.112.203002} {\bibfield  {journal}
  {\bibinfo  {journal} {Phys. Rev. Lett.}\ }\textbf {\bibinfo {volume} {112}},\
  \bibinfo {pages} {203002} (\bibinfo {year} {2014})}\BibitemShut {NoStop}%
\bibitem [{\citenamefont {Tarallo}\ \emph {et~al.}(2014)\citenamefont
  {Tarallo}, \citenamefont {Mazzoni}, \citenamefont {Poli}, \citenamefont
  {Sutyrin}, \citenamefont {Zhang},\ and\ \citenamefont
  {Tino}}]{Tarallo2014PRL}%
  \BibitemOpen
  \bibfield  {author} {\bibinfo {author} {\bibfnamefont {M.~G.}\ \bibnamefont
  {Tarallo}}, \bibinfo {author} {\bibfnamefont {T.}~\bibnamefont {Mazzoni}},
  \bibinfo {author} {\bibfnamefont {N.}~\bibnamefont {Poli}}, \bibinfo {author}
  {\bibfnamefont {D.~V.}\ \bibnamefont {Sutyrin}}, \bibinfo {author}
  {\bibfnamefont {X.}~\bibnamefont {Zhang}},\ and\ \bibinfo {author}
  {\bibfnamefont {G.~M.}\ \bibnamefont {Tino}},\ }\bibfield  {title} {\bibinfo
  {title} {Test of einstein equivalence principle for 0-spin and
  half-integer-spin atoms: Search for spin-gravity coupling effects},\ }\href
  {https://doi.org/10.1103/PhysRevLett.113.023005} {\bibfield  {journal}
  {\bibinfo  {journal} {Phys. Rev. Lett.}\ }\textbf {\bibinfo {volume} {113}},\
  \bibinfo {pages} {023005} (\bibinfo {year} {2014})}\BibitemShut {NoStop}%
\bibitem [{\citenamefont {Albers}\ \emph {et~al.}(2020)\citenamefont {Albers},
  \citenamefont {Herbst}, \citenamefont {Richardson}, \citenamefont {Heine},
  \citenamefont {Nath}, \citenamefont {Hartwig}, \citenamefont {Schubert},
  \citenamefont {Vogt}, \citenamefont {Woltmann}, \citenamefont
  {L{\"a}mmerzahl}, \citenamefont {Herrmann}, \citenamefont {Ertmer},
  \citenamefont {Rasel},\ and\ \citenamefont {Schlippert}}]{Albers2020EPJD}%
  \BibitemOpen
  \bibfield  {author} {\bibinfo {author} {\bibfnamefont {H.}~\bibnamefont
  {Albers}}, \bibinfo {author} {\bibfnamefont {A.}~\bibnamefont {Herbst}},
  \bibinfo {author} {\bibfnamefont {L.~L.}\ \bibnamefont {Richardson}},
  \bibinfo {author} {\bibfnamefont {H.}~\bibnamefont {Heine}}, \bibinfo
  {author} {\bibfnamefont {D.}~\bibnamefont {Nath}}, \bibinfo {author}
  {\bibfnamefont {J.}~\bibnamefont {Hartwig}}, \bibinfo {author} {\bibfnamefont
  {C.}~\bibnamefont {Schubert}}, \bibinfo {author} {\bibfnamefont
  {C.}~\bibnamefont {Vogt}}, \bibinfo {author} {\bibfnamefont {M.}~\bibnamefont
  {Woltmann}}, \bibinfo {author} {\bibfnamefont {C.}~\bibnamefont
  {L{\"a}mmerzahl}}, \bibinfo {author} {\bibfnamefont {S.}~\bibnamefont
  {Herrmann}}, \bibinfo {author} {\bibfnamefont {W.}~\bibnamefont {Ertmer}},
  \bibinfo {author} {\bibfnamefont {E.~M.}\ \bibnamefont {Rasel}},\ and\
  \bibinfo {author} {\bibfnamefont {D.}~\bibnamefont {Schlippert}},\ }\bibfield
   {title} {\bibinfo {title} {Quantum test of the universality of free fall
  using rubidium and potassium},\ }\bibfield  {journal} {\bibinfo  {journal}
  {The European Physical Journal D}\ }\textbf {\bibinfo {volume} {74}},\ \href
  {https://doi.org/10.1140/epjd/e2020-10132-6} {10.1140/epjd/e2020-10132-6}
  (\bibinfo {year} {2020})\BibitemShut {NoStop}%
\bibitem [{\citenamefont {Asenbaum}\ \emph {et~al.}(2020)\citenamefont
  {Asenbaum}, \citenamefont {Overstreet}, \citenamefont {Kim}, \citenamefont
  {Curti},\ and\ \citenamefont {Kasevich}}]{Asenbaum2020PRL}%
  \BibitemOpen
  \bibfield  {author} {\bibinfo {author} {\bibfnamefont {P.}~\bibnamefont
  {Asenbaum}}, \bibinfo {author} {\bibfnamefont {C.}~\bibnamefont
  {Overstreet}}, \bibinfo {author} {\bibfnamefont {M.}~\bibnamefont {Kim}},
  \bibinfo {author} {\bibfnamefont {J.}~\bibnamefont {Curti}},\ and\ \bibinfo
  {author} {\bibfnamefont {M.~A.}\ \bibnamefont {Kasevich}},\ }\bibfield
  {title} {\bibinfo {title} {Atom-interferometric test of the equivalence
  principle at the ${10}^{\ensuremath{-}12}$ level},\ }\href
  {https://doi.org/10.1103/PhysRevLett.125.191101} {\bibfield  {journal}
  {\bibinfo  {journal} {Phys. Rev. Lett.}\ }\textbf {\bibinfo {volume} {125}},\
  \bibinfo {pages} {191101} (\bibinfo {year} {2020})}\BibitemShut {NoStop}%
\bibitem [{\citenamefont {Carlesso}\ \emph {et~al.}(2022)\citenamefont
  {Carlesso}, \citenamefont {Donadi}, \citenamefont {Ferialdi}, \citenamefont
  {Paternostro}, \citenamefont {Ulbricht},\ and\ \citenamefont
  {Bassi}}]{Carlesso2022NatPhys}%
  \BibitemOpen
  \bibfield  {author} {\bibinfo {author} {\bibfnamefont {M.}~\bibnamefont
  {Carlesso}}, \bibinfo {author} {\bibfnamefont {S.}~\bibnamefont {Donadi}},
  \bibinfo {author} {\bibfnamefont {L.}~\bibnamefont {Ferialdi}}, \bibinfo
  {author} {\bibfnamefont {M.}~\bibnamefont {Paternostro}}, \bibinfo {author}
  {\bibfnamefont {H.}~\bibnamefont {Ulbricht}},\ and\ \bibinfo {author}
  {\bibfnamefont {A.}~\bibnamefont {Bassi}},\ }\bibfield  {title} {\bibinfo
  {title} {Present status and future challenges of non-interferometric tests of
  collapse models},\ }\href {https://doi.org/10.1038/s41567-021-01489-5}
  {\bibfield  {journal} {\bibinfo  {journal} {Nature Physics}\ }\textbf
  {\bibinfo {volume} {18}},\ \bibinfo {pages} {243} (\bibinfo {year}
  {2022})}\BibitemShut {NoStop}%
\bibitem [{\citenamefont {Kovachy}\ \emph
  {et~al.}(2015{\natexlab{a}})\citenamefont {Kovachy}, \citenamefont
  {Asenbaum}, \citenamefont {Overstreet}, \citenamefont {Donnelly},
  \citenamefont {Dickerson}, \citenamefont {Sugarbaker}, \citenamefont
  {Hogan},\ and\ \citenamefont {Kasevich}}]{Kovachy15Nature}%
  \BibitemOpen
  \bibfield  {author} {\bibinfo {author} {\bibfnamefont {T.}~\bibnamefont
  {Kovachy}}, \bibinfo {author} {\bibfnamefont {P.}~\bibnamefont {Asenbaum}},
  \bibinfo {author} {\bibfnamefont {C.}~\bibnamefont {Overstreet}}, \bibinfo
  {author} {\bibfnamefont {C.~A.}\ \bibnamefont {Donnelly}}, \bibinfo {author}
  {\bibfnamefont {S.~M.}\ \bibnamefont {Dickerson}}, \bibinfo {author}
  {\bibfnamefont {A.}~\bibnamefont {Sugarbaker}}, \bibinfo {author}
  {\bibfnamefont {J.~M.}\ \bibnamefont {Hogan}},\ and\ \bibinfo {author}
  {\bibfnamefont {M.~A.}\ \bibnamefont {Kasevich}},\ }\bibfield  {title}
  {\bibinfo {title} {Quantum superposition at the half-metre scale},\ }\href
  {https://doi.org/10.1038/nature16155} {\bibfield  {journal} {\bibinfo
  {journal} {Nature}\ }\textbf {\bibinfo {volume} {528}},\ \bibinfo {pages}
  {530} (\bibinfo {year} {2015}{\natexlab{a}})}\BibitemShut {NoStop}%
\bibitem [{\citenamefont {Bassi}\ \emph {et~al.}(2013)\citenamefont {Bassi},
  \citenamefont {Lochan}, \citenamefont {Satin}, \citenamefont {Singh},\ and\
  \citenamefont {Ulbricht}}]{Bassi2013RMP}%
  \BibitemOpen
  \bibfield  {author} {\bibinfo {author} {\bibfnamefont {A.}~\bibnamefont
  {Bassi}}, \bibinfo {author} {\bibfnamefont {K.}~\bibnamefont {Lochan}},
  \bibinfo {author} {\bibfnamefont {S.}~\bibnamefont {Satin}}, \bibinfo
  {author} {\bibfnamefont {T.~P.}\ \bibnamefont {Singh}},\ and\ \bibinfo
  {author} {\bibfnamefont {H.}~\bibnamefont {Ulbricht}},\ }\bibfield  {title}
  {\bibinfo {title} {Models of wave-function collapse, underlying theories, and
  experimental tests},\ }\href {https://doi.org/10.1103/RevModPhys.85.471}
  {\bibfield  {journal} {\bibinfo  {journal} {Rev. Mod. Phys.}\ }\textbf
  {\bibinfo {volume} {85}},\ \bibinfo {pages} {471} (\bibinfo {year}
  {2013})}\BibitemShut {NoStop}%
\bibitem [{\citenamefont {Schrinski}\ \emph {et~al.}(2023)\citenamefont
  {Schrinski}, \citenamefont {Haslinger}, \citenamefont {Schmiedmayer},
  \citenamefont {Hornberger},\ and\ \citenamefont
  {Nimmrichter}}]{Schrinski23PRA}%
  \BibitemOpen
  \bibfield  {author} {\bibinfo {author} {\bibfnamefont {B.}~\bibnamefont
  {Schrinski}}, \bibinfo {author} {\bibfnamefont {P.}~\bibnamefont
  {Haslinger}}, \bibinfo {author} {\bibfnamefont {J.}~\bibnamefont
  {Schmiedmayer}}, \bibinfo {author} {\bibfnamefont {K.}~\bibnamefont
  {Hornberger}},\ and\ \bibinfo {author} {\bibfnamefont {S.}~\bibnamefont
  {Nimmrichter}},\ }\bibfield  {title} {\bibinfo {title} {Testing collapse
  models with bose-einstein-condensate interferometry},\ }\href
  {https://doi.org/10.1103/PhysRevA.107.043320} {\bibfield  {journal} {\bibinfo
   {journal} {Phys. Rev. A}\ }\textbf {\bibinfo {volume} {107}},\ \bibinfo
  {pages} {043320} (\bibinfo {year} {2023})}\BibitemShut {NoStop}%
\bibitem [{\citenamefont {Rosi}\ \emph {et~al.}(2014)\citenamefont {Rosi},
  \citenamefont {Sorrentino}, \citenamefont {Cacciapuoti}, \citenamefont
  {Prevedelli},\ and\ \citenamefont {Tino}}]{Rosi14Nature}%
  \BibitemOpen
  \bibfield  {author} {\bibinfo {author} {\bibfnamefont {G.}~\bibnamefont
  {Rosi}}, \bibinfo {author} {\bibfnamefont {F.}~\bibnamefont {Sorrentino}},
  \bibinfo {author} {\bibfnamefont {L.}~\bibnamefont {Cacciapuoti}}, \bibinfo
  {author} {\bibfnamefont {M.}~\bibnamefont {Prevedelli}},\ and\ \bibinfo
  {author} {\bibfnamefont {G.~M.}\ \bibnamefont {Tino}},\ }\bibfield  {title}
  {\bibinfo {title} {Precision measurement of the newtonian gravitational
  constant using cold atoms},\ }\href {https://doi.org/10.1038/nature13433}
  {\bibfield  {journal} {\bibinfo  {journal} {Nature}\ }\textbf {\bibinfo
  {volume} {510}},\ \bibinfo {pages} {518} (\bibinfo {year}
  {2014})}\BibitemShut {NoStop}%
\bibitem [{\citenamefont {Parker}\ \emph {et~al.}(2018)\citenamefont {Parker},
  \citenamefont {Yu}, \citenamefont {Zhong}, \citenamefont {Estey},\ and\
  \citenamefont {M\"uller}}]{Parker2018Science}%
  \BibitemOpen
  \bibfield  {author} {\bibinfo {author} {\bibfnamefont {R.~H.}\ \bibnamefont
  {Parker}}, \bibinfo {author} {\bibfnamefont {C.}~\bibnamefont {Yu}}, \bibinfo
  {author} {\bibfnamefont {W.}~\bibnamefont {Zhong}}, \bibinfo {author}
  {\bibfnamefont {B.}~\bibnamefont {Estey}},\ and\ \bibinfo {author}
  {\bibfnamefont {H.}~\bibnamefont {M\"uller}},\ }\bibfield  {title} {\bibinfo
  {title} {Measurement of the fine-structure constant as a test of the standard
  model},\ }\href {https://doi.org/10.1126/science.aap7706} {\bibfield
  {journal} {\bibinfo  {journal} {Science}\ }\textbf {\bibinfo {volume}
  {360}},\ \bibinfo {pages} {191} (\bibinfo {year} {2018})}\BibitemShut
  {NoStop}%
\bibitem [{\citenamefont {Morel}\ \emph {et~al.}(2020)\citenamefont {Morel},
  \citenamefont {Yao}, \citenamefont {Clad{\'e}},\ and\ \citenamefont
  {Guellati-Kh{\'e}lifa}}]{Morel2020Nature}%
  \BibitemOpen
  \bibfield  {author} {\bibinfo {author} {\bibfnamefont {L.}~\bibnamefont
  {Morel}}, \bibinfo {author} {\bibfnamefont {Z.}~\bibnamefont {Yao}}, \bibinfo
  {author} {\bibfnamefont {P.}~\bibnamefont {Clad{\'e}}},\ and\ \bibinfo
  {author} {\bibfnamefont {S.}~\bibnamefont {Guellati-Kh{\'e}lifa}},\
  }\bibfield  {title} {\bibinfo {title} {Determination of the fine-structure
  constant with an accuracy of 81 parts per trillion},\ }\href
  {https://doi.org/10.1038/s41586-020-2964-7} {\bibfield  {journal} {\bibinfo
  {journal} {Nature}\ }\textbf {\bibinfo {volume} {588}},\ \bibinfo {pages}
  {61} (\bibinfo {year} {2020})}\BibitemShut {NoStop}%
\bibitem [{\citenamefont {Gustavson}\ \emph {et~al.}(1997)\citenamefont
  {Gustavson}, \citenamefont {Bouyer},\ and\ \citenamefont
  {Kasevich}}]{Gustavson97PRL}%
  \BibitemOpen
  \bibfield  {author} {\bibinfo {author} {\bibfnamefont {T.~L.}\ \bibnamefont
  {Gustavson}}, \bibinfo {author} {\bibfnamefont {P.}~\bibnamefont {Bouyer}},\
  and\ \bibinfo {author} {\bibfnamefont {M.~A.}\ \bibnamefont {Kasevich}},\
  }\bibfield  {title} {\bibinfo {title} {Precision rotation measurements with
  an atom interferometer gyroscope},\ }\href
  {https://doi.org/10.1103/physrevlett.78.2046} {\bibfield  {journal} {\bibinfo
   {journal} {Physical Review Letters}\ }\textbf {\bibinfo {volume} {78}},\
  \bibinfo {pages} {2046} (\bibinfo {year} {1997})}\BibitemShut {NoStop}%
\bibitem [{\citenamefont {Canuel}\ \emph {et~al.}(2006)\citenamefont {Canuel},
  \citenamefont {Leduc}, \citenamefont {Holleville}, \citenamefont {Gauguet},
  \citenamefont {Fils}, \citenamefont {Virdis}, \citenamefont {Clairon},
  \citenamefont {Dimarcq}, \citenamefont {Bord\'e}, \citenamefont {Landragin},\
  and\ \citenamefont {Bouyer}}]{Canuel06PRL}%
  \BibitemOpen
  \bibfield  {author} {\bibinfo {author} {\bibfnamefont {B.}~\bibnamefont
  {Canuel}}, \bibinfo {author} {\bibfnamefont {F.}~\bibnamefont {Leduc}},
  \bibinfo {author} {\bibfnamefont {D.}~\bibnamefont {Holleville}}, \bibinfo
  {author} {\bibfnamefont {A.}~\bibnamefont {Gauguet}}, \bibinfo {author}
  {\bibfnamefont {J.}~\bibnamefont {Fils}}, \bibinfo {author} {\bibfnamefont
  {A.}~\bibnamefont {Virdis}}, \bibinfo {author} {\bibfnamefont
  {A.}~\bibnamefont {Clairon}}, \bibinfo {author} {\bibfnamefont
  {N.}~\bibnamefont {Dimarcq}}, \bibinfo {author} {\bibfnamefont {C.~J.}\
  \bibnamefont {Bord\'e}}, \bibinfo {author} {\bibfnamefont {A.}~\bibnamefont
  {Landragin}},\ and\ \bibinfo {author} {\bibfnamefont {P.}~\bibnamefont
  {Bouyer}},\ }\bibfield  {title} {\bibinfo {title} {Six-axis inertial sensor
  using cold-atom interferometry},\ }\href
  {https://doi.org/10.1103/PhysRevLett.97.010402} {\bibfield  {journal}
  {\bibinfo  {journal} {Phys. Rev. Lett.}\ }\textbf {\bibinfo {volume} {97}},\
  \bibinfo {pages} {010402} (\bibinfo {year} {2006})}\BibitemShut {NoStop}%
\bibitem [{\citenamefont {Dickerson}\ \emph {et~al.}(2013)\citenamefont
  {Dickerson}, \citenamefont {Hogan}, \citenamefont {Sugarbaker}, \citenamefont
  {Johnson},\ and\ \citenamefont {Kasevich}}]{Dickerson13PRL}%
  \BibitemOpen
  \bibfield  {author} {\bibinfo {author} {\bibfnamefont {S.~M.}\ \bibnamefont
  {Dickerson}}, \bibinfo {author} {\bibfnamefont {J.~M.}\ \bibnamefont
  {Hogan}}, \bibinfo {author} {\bibfnamefont {A.}~\bibnamefont {Sugarbaker}},
  \bibinfo {author} {\bibfnamefont {D.~M.~S.}\ \bibnamefont {Johnson}},\ and\
  \bibinfo {author} {\bibfnamefont {M.~A.}\ \bibnamefont {Kasevich}},\
  }\bibfield  {title} {\bibinfo {title} {Multiaxis inertial sensing with
  long-time point source atom interferometry},\ }\href
  {https://doi.org/10.1103/PhysRevLett.111.083001} {\bibfield  {journal}
  {\bibinfo  {journal} {Phys. Rev. Lett.}\ }\textbf {\bibinfo {volume} {111}},\
  \bibinfo {pages} {083001} (\bibinfo {year} {2013})}\BibitemShut {NoStop}%
\bibitem [{\citenamefont {Dutta}\ \emph {et~al.}(2016)\citenamefont {Dutta},
  \citenamefont {Savoie}, \citenamefont {Fang}, \citenamefont {Venon},
  \citenamefont {Garrido~Alzar}, \citenamefont {Geiger},\ and\ \citenamefont
  {Landragin}}]{Dutta16PRL}%
  \BibitemOpen
  \bibfield  {author} {\bibinfo {author} {\bibfnamefont {I.}~\bibnamefont
  {Dutta}}, \bibinfo {author} {\bibfnamefont {D.}~\bibnamefont {Savoie}},
  \bibinfo {author} {\bibfnamefont {B.}~\bibnamefont {Fang}}, \bibinfo {author}
  {\bibfnamefont {B.}~\bibnamefont {Venon}}, \bibinfo {author} {\bibfnamefont
  {C.}~\bibnamefont {Garrido~Alzar}}, \bibinfo {author} {\bibfnamefont
  {R.}~\bibnamefont {Geiger}},\ and\ \bibinfo {author} {\bibfnamefont
  {A.}~\bibnamefont {Landragin}},\ }\bibfield  {title} {\bibinfo {title}
  {Continuous cold-atom inertial sensor with $1\text{ }\text{
  }\mathrm{nrad}/\mathrm{sec}$ rotation stability},\ }\href
  {https://doi.org/10.1103/PhysRevLett.116.183003} {\bibfield  {journal}
  {\bibinfo  {journal} {Phys. Rev. Lett.}\ }\textbf {\bibinfo {volume} {116}},\
  \bibinfo {pages} {183003} (\bibinfo {year} {2016})}\BibitemShut {NoStop}%
\bibitem [{\citenamefont {Savoie}\ \emph {et~al.}(2018)\citenamefont {Savoie},
  \citenamefont {Altorio}, \citenamefont {Fang}, \citenamefont {Sidorenkov},
  \citenamefont {Geiger},\ and\ \citenamefont {Landragin}}]{Savoie2018SciAdv}%
  \BibitemOpen
  \bibfield  {author} {\bibinfo {author} {\bibfnamefont {D.}~\bibnamefont
  {Savoie}}, \bibinfo {author} {\bibfnamefont {M.}~\bibnamefont {Altorio}},
  \bibinfo {author} {\bibfnamefont {B.}~\bibnamefont {Fang}}, \bibinfo {author}
  {\bibfnamefont {L.~A.}\ \bibnamefont {Sidorenkov}}, \bibinfo {author}
  {\bibfnamefont {R.}~\bibnamefont {Geiger}},\ and\ \bibinfo {author}
  {\bibfnamefont {A.}~\bibnamefont {Landragin}},\ }\bibfield  {title} {\bibinfo
  {title} {Interleaved atom interferometry for high-sensitivity inertial
  measurements},\ }\href {https://doi.org/10.1126/sciadv.aau7948} {\bibfield
  {journal} {\bibinfo  {journal} {Science advances}\ }\textbf {\bibinfo
  {volume} {4}},\ \bibinfo {pages} {eaau7948} (\bibinfo {year}
  {2018})}\BibitemShut {NoStop}%
\bibitem [{\citenamefont {{Le Gou{\"e}t}}\ \emph {et~al.}(2008)\citenamefont
  {{Le Gou{\"e}t}}, \citenamefont {Mehlst{\"a}ubler}, \citenamefont {Kim},
  \citenamefont {Merlet}, \citenamefont {Clairon}, \citenamefont {Landragin},\
  and\ \citenamefont {{Pereira dos Santos}}}]{LeGouet2008APB}%
  \BibitemOpen
  \bibfield  {author} {\bibinfo {author} {\bibfnamefont {J.}~\bibnamefont {{Le
  Gou{\"e}t}}}, \bibinfo {author} {\bibfnamefont {T.~E.}\ \bibnamefont
  {Mehlst{\"a}ubler}}, \bibinfo {author} {\bibfnamefont {J.}~\bibnamefont
  {Kim}}, \bibinfo {author} {\bibfnamefont {S.}~\bibnamefont {Merlet}},
  \bibinfo {author} {\bibfnamefont {A.}~\bibnamefont {Clairon}}, \bibinfo
  {author} {\bibfnamefont {A.}~\bibnamefont {Landragin}},\ and\ \bibinfo
  {author} {\bibfnamefont {F.}~\bibnamefont {{Pereira dos Santos}}},\
  }\bibfield  {title} {\bibinfo {title} {Limits to the sensitivity of a low
  noise compact atomic gravimeter},\ }\href
  {https://doi.org/10.1007/s00340-008-3088-1} {\bibfield  {journal} {\bibinfo
  {journal} {Applied Physics B}\ }\textbf {\bibinfo {volume} {92}},\ \bibinfo
  {pages} {133} (\bibinfo {year} {2008})}\BibitemShut {NoStop}%
\bibitem [{\citenamefont {Hu}\ \emph {et~al.}(2013)\citenamefont {Hu},
  \citenamefont {Sun}, \citenamefont {Duan}, \citenamefont {Zhou},
  \citenamefont {Chen}, \citenamefont {Zhan}, \citenamefont {Zhang},\ and\
  \citenamefont {Luo}}]{Hu2013PRA}%
  \BibitemOpen
  \bibfield  {author} {\bibinfo {author} {\bibfnamefont {Z.-K.}\ \bibnamefont
  {Hu}}, \bibinfo {author} {\bibfnamefont {B.-L.}\ \bibnamefont {Sun}},
  \bibinfo {author} {\bibfnamefont {X.-C.}\ \bibnamefont {Duan}}, \bibinfo
  {author} {\bibfnamefont {M.-K.}\ \bibnamefont {Zhou}}, \bibinfo {author}
  {\bibfnamefont {L.-L.}\ \bibnamefont {Chen}}, \bibinfo {author}
  {\bibfnamefont {S.}~\bibnamefont {Zhan}}, \bibinfo {author} {\bibfnamefont
  {Q.-Z.}\ \bibnamefont {Zhang}},\ and\ \bibinfo {author} {\bibfnamefont
  {J.}~\bibnamefont {Luo}},\ }\bibfield  {title} {\bibinfo {title}
  {Demonstration of an ultrahigh-sensitivity atom-interferometry absolute
  gravimeter},\ }\href {https://doi.org/10.1103/PhysRevA.88.043610} {\bibfield
  {journal} {\bibinfo  {journal} {Phys. Rev. A}\ }\textbf {\bibinfo {volume}
  {88}},\ \bibinfo {pages} {043610} (\bibinfo {year} {2013})}\BibitemShut
  {NoStop}%
\bibitem [{\citenamefont {M{\'e}noret}\ \emph {et~al.}(2018)\citenamefont
  {M{\'e}noret}, \citenamefont {Vermeulen}, \citenamefont {{Le Moigne}},
  \citenamefont {Bonvalot}, \citenamefont {Bouyer}, \citenamefont {Landragin},\
  and\ \citenamefont {Desruelle}}]{Menoret2018SR}%
  \BibitemOpen
  \bibfield  {author} {\bibinfo {author} {\bibfnamefont {V.}~\bibnamefont
  {M{\'e}noret}}, \bibinfo {author} {\bibfnamefont {P.}~\bibnamefont
  {Vermeulen}}, \bibinfo {author} {\bibfnamefont {N.}~\bibnamefont {{Le
  Moigne}}}, \bibinfo {author} {\bibfnamefont {S.}~\bibnamefont {Bonvalot}},
  \bibinfo {author} {\bibfnamefont {P.}~\bibnamefont {Bouyer}}, \bibinfo
  {author} {\bibfnamefont {A.}~\bibnamefont {Landragin}},\ and\ \bibinfo
  {author} {\bibfnamefont {B.}~\bibnamefont {Desruelle}},\ }\bibfield  {title}
  {\bibinfo {title} {Gravity measurements below 10-9 g with a transportable
  absolute quantum gravimeter},\ }\href
  {https://doi.org/10.1038/s41598-018-30608-1} {\bibfield  {journal} {\bibinfo
  {journal} {Scientific reports}\ }\textbf {\bibinfo {volume} {8}},\ \bibinfo
  {pages} {12300} (\bibinfo {year} {2018})}\BibitemShut {NoStop}%
\bibitem [{\citenamefont {Louchet-Chauvet}\ \emph {et~al.}(2011)\citenamefont
  {Louchet-Chauvet}, \citenamefont {Farah}, \citenamefont {Bodart},
  \citenamefont {Clairon}, \citenamefont {Landragin}, \citenamefont {Merlet},\
  and\ \citenamefont {Santos}}]{Louchet2011NJP}%
  \BibitemOpen
  \bibfield  {author} {\bibinfo {author} {\bibfnamefont {A.}~\bibnamefont
  {Louchet-Chauvet}}, \bibinfo {author} {\bibfnamefont {T.}~\bibnamefont
  {Farah}}, \bibinfo {author} {\bibfnamefont {Q.}~\bibnamefont {Bodart}},
  \bibinfo {author} {\bibfnamefont {A.}~\bibnamefont {Clairon}}, \bibinfo
  {author} {\bibfnamefont {A.}~\bibnamefont {Landragin}}, \bibinfo {author}
  {\bibfnamefont {S.}~\bibnamefont {Merlet}},\ and\ \bibinfo {author}
  {\bibfnamefont {F.~P.~D.}\ \bibnamefont {Santos}},\ }\bibfield  {title}
  {\bibinfo {title} {The influence of transverse motion within an atomic
  gravimeter},\ }\href {https://doi.org/10.1088/1367-2630/13/6/065025}
  {\bibfield  {journal} {\bibinfo  {journal} {New Journal of Physics}\ }\textbf
  {\bibinfo {volume} {13}},\ \bibinfo {pages} {065025} (\bibinfo {year}
  {2011})}\BibitemShut {NoStop}%
\bibitem [{\citenamefont {Schkolnik}\ \emph {et~al.}(2015)\citenamefont
  {Schkolnik}, \citenamefont {Leykauf}, \citenamefont {Hauth}, \citenamefont
  {Freier},\ and\ \citenamefont {Peters}}]{Schkolnik2015APB}%
  \BibitemOpen
  \bibfield  {author} {\bibinfo {author} {\bibfnamefont {V.}~\bibnamefont
  {Schkolnik}}, \bibinfo {author} {\bibfnamefont {B.}~\bibnamefont {Leykauf}},
  \bibinfo {author} {\bibfnamefont {M.}~\bibnamefont {Hauth}}, \bibinfo
  {author} {\bibfnamefont {C.}~\bibnamefont {Freier}},\ and\ \bibinfo {author}
  {\bibfnamefont {A.}~\bibnamefont {Peters}},\ }\bibfield  {title} {\bibinfo
  {title} {The effect of wavefront aberrations in atom interferometry},\ }\href
  {https://doi.org/10.1007/s00340-015-6138-5} {\bibfield  {journal} {\bibinfo
  {journal} {Applied Physics B}\ }\textbf {\bibinfo {volume} {120}},\ \bibinfo
  {pages} {311} (\bibinfo {year} {2015})}\BibitemShut {NoStop}%
\bibitem [{\citenamefont {Anderson}\ \emph {et~al.}(1995)\citenamefont
  {Anderson}, \citenamefont {Ensher}, \citenamefont {Matthews}, \citenamefont
  {Wieman},\ and\ \citenamefont {Cornell}}]{Anderson95Science}%
  \BibitemOpen
  \bibfield  {author} {\bibinfo {author} {\bibfnamefont {M.}~\bibnamefont
  {Anderson}}, \bibinfo {author} {\bibfnamefont {J.}~\bibnamefont {Ensher}},
  \bibinfo {author} {\bibfnamefont {M.}~\bibnamefont {Matthews}}, \bibinfo
  {author} {\bibfnamefont {C.}~\bibnamefont {Wieman}},\ and\ \bibinfo {author}
  {\bibfnamefont {E.}~\bibnamefont {Cornell}},\ }\bibfield  {title} {\bibinfo
  {title} {{Observation of Bose-Einstein Condensation in a Dilute Atomic
  Vapor}},\ }\href {https://doi.org/10.1126/science.269.5221.198} {\bibfield
  {journal} {\bibinfo  {journal} {Science}\ }\textbf {\bibinfo {volume}
  {269}},\ \bibinfo {pages} {198} (\bibinfo {year} {1995})}\BibitemShut
  {NoStop}%
\bibitem [{\citenamefont {Davis}\ \emph {et~al.}(1995)\citenamefont {Davis},
  \citenamefont {Mewes}, \citenamefont {Joffe}, \citenamefont {Andrews},\ and\
  \citenamefont {Ketterle}}]{Davis95PRL}%
  \BibitemOpen
  \bibfield  {author} {\bibinfo {author} {\bibfnamefont {K.~B.}\ \bibnamefont
  {Davis}}, \bibinfo {author} {\bibfnamefont {M.-O.}\ \bibnamefont {Mewes}},
  \bibinfo {author} {\bibfnamefont {M.~A.}\ \bibnamefont {Joffe}}, \bibinfo
  {author} {\bibfnamefont {M.~R.}\ \bibnamefont {Andrews}},\ and\ \bibinfo
  {author} {\bibfnamefont {W.}~\bibnamefont {Ketterle}},\ }\bibfield  {title}
  {\bibinfo {title} {Evaporative cooling of sodium atoms},\ }\href
  {https://doi.org/10.1103/physrevlett.74.5202} {\bibfield  {journal} {\bibinfo
   {journal} {Physical Review Letters}\ }\textbf {\bibinfo {volume} {74}},\
  \bibinfo {pages} {5202} (\bibinfo {year} {1995})}\BibitemShut {NoStop}%
\bibitem [{\citenamefont {Schlippert}\ \emph {et~al.}(2021)\citenamefont
  {Schlippert}, \citenamefont {Meiners}, \citenamefont {Rengelink},
  \citenamefont {Schubert}, \citenamefont {Tell}, \citenamefont {Wodey},
  \citenamefont {Zipfel}, \citenamefont {Ertmer},\ and\ \citenamefont
  {Rasel}}]{Schlippert2020}%
  \BibitemOpen
  \bibfield  {author} {\bibinfo {author} {\bibfnamefont {D.}~\bibnamefont
  {Schlippert}}, \bibinfo {author} {\bibfnamefont {C.}~\bibnamefont {Meiners}},
  \bibinfo {author} {\bibfnamefont {R.}~\bibnamefont {Rengelink}}, \bibinfo
  {author} {\bibfnamefont {C.}~\bibnamefont {Schubert}}, \bibinfo {author}
  {\bibfnamefont {D.}~\bibnamefont {Tell}}, \bibinfo {author} {\bibfnamefont
  {{\'{E}}.}~\bibnamefont {Wodey}}, \bibinfo {author} {\bibfnamefont
  {K.}~\bibnamefont {Zipfel}}, \bibinfo {author} {\bibfnamefont
  {W.}~\bibnamefont {Ertmer}},\ and\ \bibinfo {author} {\bibfnamefont
  {E.}~\bibnamefont {Rasel}},\ }\bibfield  {title} {\bibinfo {title}
  {Matter-wave interferometry for inertial sensing and tests of fundamental
  physics},\ }in\ \href {https://doi.org/10.1142/9789811213984_0010} {\emph
  {\bibinfo {booktitle} {{CPT} and Lorentz Symmetry}}}\ (\bibinfo  {publisher}
  {{WORLD} {SCIENTIFIC}},\ \bibinfo {year} {2021})\BibitemShut {NoStop}%
\bibitem [{\citenamefont {Hensel}\ \emph {et~al.}(2021)\citenamefont {Hensel},
  \citenamefont {Loriani}, \citenamefont {Schubert}, \citenamefont {Fitzek},
  \citenamefont {Abend}, \citenamefont {Ahlers}, \citenamefont {Siem{\ss}},
  \citenamefont {Hammerer}, \citenamefont {Rasel},\ and\ \citenamefont
  {Gaaloul}}]{Hensel2021}%
  \BibitemOpen
  \bibfield  {author} {\bibinfo {author} {\bibfnamefont {T.}~\bibnamefont
  {Hensel}}, \bibinfo {author} {\bibfnamefont {S.}~\bibnamefont {Loriani}},
  \bibinfo {author} {\bibfnamefont {C.}~\bibnamefont {Schubert}}, \bibinfo
  {author} {\bibfnamefont {F.}~\bibnamefont {Fitzek}}, \bibinfo {author}
  {\bibfnamefont {S.}~\bibnamefont {Abend}}, \bibinfo {author} {\bibfnamefont
  {H.}~\bibnamefont {Ahlers}}, \bibinfo {author} {\bibfnamefont {J.-N.}\
  \bibnamefont {Siem{\ss}}}, \bibinfo {author} {\bibfnamefont {K.}~\bibnamefont
  {Hammerer}}, \bibinfo {author} {\bibfnamefont {E.~M.}\ \bibnamefont
  {Rasel}},\ and\ \bibinfo {author} {\bibfnamefont {N.}~\bibnamefont
  {Gaaloul}},\ }\bibfield  {title} {\bibinfo {title} {Inertial sensing with
  quantum gases: a comparative performance study of condensed versus thermal
  sources for atom interferometry},\ }\bibfield  {journal} {\bibinfo  {journal}
  {The European Physical Journal D}\ }\textbf {\bibinfo {volume} {75}},\ \href
  {https://doi.org/10.1140/epjd/s10053-021-00069-9}
  {10.1140/epjd/s10053-021-00069-9} (\bibinfo {year} {2021})\BibitemShut
  {NoStop}%
\bibitem [{\citenamefont {Weber}\ \emph {et~al.}(2003)\citenamefont {Weber},
  \citenamefont {Herbig}, \citenamefont {Mark}, \citenamefont {N{\"a}gerl},\
  and\ \citenamefont {Grimm}}]{Weber2003Science}%
  \BibitemOpen
  \bibfield  {author} {\bibinfo {author} {\bibfnamefont {T.}~\bibnamefont
  {Weber}}, \bibinfo {author} {\bibfnamefont {J.}~\bibnamefont {Herbig}},
  \bibinfo {author} {\bibfnamefont {M.}~\bibnamefont {Mark}}, \bibinfo {author}
  {\bibfnamefont {H.-C.}\ \bibnamefont {N{\"a}gerl}},\ and\ \bibinfo {author}
  {\bibfnamefont {R.}~\bibnamefont {Grimm}},\ }\bibfield  {title} {\bibinfo
  {title} {Bose-einstein condensation of cesium},\ }\href
  {https://doi.org/10.1126/science.1079699} {\bibfield  {journal} {\bibinfo
  {journal} {Science (New York, N.Y.)}\ }\textbf {\bibinfo {volume} {299}},\
  \bibinfo {pages} {232} (\bibinfo {year} {2003})}\BibitemShut {NoStop}%
\bibitem [{\citenamefont {Hardman}\ \emph {et~al.}(2016)\citenamefont
  {Hardman}, \citenamefont {Everitt}, \citenamefont {McDonald}, \citenamefont
  {Manju}, \citenamefont {Wigley}, \citenamefont {Sooriyabandara},
  \citenamefont {Kuhn}, \citenamefont {Debs}, \citenamefont {Close},\ and\
  \citenamefont {Robins}}]{Hardman2016PRL}%
  \BibitemOpen
  \bibfield  {author} {\bibinfo {author} {\bibfnamefont {K.~S.}\ \bibnamefont
  {Hardman}}, \bibinfo {author} {\bibfnamefont {P.~J.}\ \bibnamefont
  {Everitt}}, \bibinfo {author} {\bibfnamefont {G.~D.}\ \bibnamefont
  {McDonald}}, \bibinfo {author} {\bibfnamefont {P.}~\bibnamefont {Manju}},
  \bibinfo {author} {\bibfnamefont {P.~B.}\ \bibnamefont {Wigley}}, \bibinfo
  {author} {\bibfnamefont {M.~A.}\ \bibnamefont {Sooriyabandara}}, \bibinfo
  {author} {\bibfnamefont {C.~C.~N.}\ \bibnamefont {Kuhn}}, \bibinfo {author}
  {\bibfnamefont {J.~E.}\ \bibnamefont {Debs}}, \bibinfo {author}
  {\bibfnamefont {J.~D.}\ \bibnamefont {Close}},\ and\ \bibinfo {author}
  {\bibfnamefont {N.~P.}\ \bibnamefont {Robins}},\ }\bibfield  {title}
  {\bibinfo {title} {Simultaneous precision gravimetry and magnetic gradiometry
  with a bose-einstein condensate: A high precision, quantum sensor},\ }\href
  {https://doi.org/10.1103/PhysRevLett.117.138501} {\bibfield  {journal}
  {\bibinfo  {journal} {Phys. Rev. Lett.}\ }\textbf {\bibinfo {volume} {117}},\
  \bibinfo {pages} {138501} (\bibinfo {year} {2016})}\BibitemShut {NoStop}%
\bibitem [{\citenamefont {Gochnauer}\ \emph {et~al.}(2021)\citenamefont
  {Gochnauer}, \citenamefont {Rahman}, \citenamefont {Wirth-Singh},\ and\
  \citenamefont {Gupta}}]{Gochnauer2021Atoms}%
  \BibitemOpen
  \bibfield  {author} {\bibinfo {author} {\bibfnamefont {D.}~\bibnamefont
  {Gochnauer}}, \bibinfo {author} {\bibfnamefont {T.}~\bibnamefont {Rahman}},
  \bibinfo {author} {\bibfnamefont {A.}~\bibnamefont {Wirth-Singh}},\ and\
  \bibinfo {author} {\bibfnamefont {S.}~\bibnamefont {Gupta}},\ }\bibfield
  {title} {\bibinfo {title} {Interferometry in an atomic fountain with
  ytterbium bose--einstein condensates},\ }\href
  {https://doi.org/10.3390/atoms9030058} {\bibfield  {journal} {\bibinfo
  {journal} {Atoms}\ }\textbf {\bibinfo {volume} {9}},\ \bibinfo {pages} {58}
  (\bibinfo {year} {2021})}\BibitemShut {NoStop}%
\bibitem [{\citenamefont {Aguilera}\ \emph {et~al.}(2014)\citenamefont
  {Aguilera}, \citenamefont {Ahlers}, \citenamefont {Battelier}, \citenamefont
  {Bawamia}, \citenamefont {Bertoldi}, \citenamefont {Bondarescu},
  \citenamefont {Bongs}, \citenamefont {Bouyer}, \citenamefont {Braxmaier},
  \citenamefont {Cacciapuoti}, \citenamefont {Chaloner}, \citenamefont
  {Chwalla}, \citenamefont {Ertmer}, \citenamefont {Franz}, \citenamefont
  {Gaaloul}, \citenamefont {Gehler}, \citenamefont {Gerardi}, \citenamefont
  {Gesa}, \citenamefont {G{\"u}rlebeck}, \citenamefont {Hartwig}, \citenamefont
  {Hauth}, \citenamefont {Hellmig}, \citenamefont {Herr}, \citenamefont
  {Herrmann}, \citenamefont {Heske}, \citenamefont {Hinton}, \citenamefont
  {Ireland}, \citenamefont {Jetzer}, \citenamefont {Johann}, \citenamefont
  {Krutzik}, \citenamefont {Kubelka}, \citenamefont {L{\"a}mmerzahl},
  \citenamefont {Landragin}, \citenamefont {Lloro}, \citenamefont {Massonnet},
  \citenamefont {Mateos}, \citenamefont {Milke}, \citenamefont {Nofrarias},
  \citenamefont {Oswald}, \citenamefont {Peters}, \citenamefont
  {Posso-Trujillo}, \citenamefont {Rasel}, \citenamefont {Rocco}, \citenamefont
  {Roura}, \citenamefont {Rudolph}, \citenamefont {Schleich}, \citenamefont
  {Schubert}, \citenamefont {Schuldt}, \citenamefont {Seidel}, \citenamefont
  {Sengstock}, \citenamefont {Sopuerta}, \citenamefont {Sorrentino},
  \citenamefont {Summers}, \citenamefont {Tino}, \citenamefont {Trenkel},
  \citenamefont {Uzunoglu}, \citenamefont {von Klitzing}, \citenamefont
  {Walser}, \citenamefont {Wendrich}, \citenamefont {Wenzlawski}, \citenamefont
  {We{\ss}els}, \citenamefont {Wicht}, \citenamefont {Wille}, \citenamefont
  {Williams}, \citenamefont {Windpassinger},\ and\ \citenamefont
  {Zahzam}}]{Aguilera2014CQG}%
  \BibitemOpen
  \bibfield  {author} {\bibinfo {author} {\bibfnamefont {D.~N.}\ \bibnamefont
  {Aguilera}}, \bibinfo {author} {\bibfnamefont {H.}~\bibnamefont {Ahlers}},
  \bibinfo {author} {\bibfnamefont {B.}~\bibnamefont {Battelier}}, \bibinfo
  {author} {\bibfnamefont {A.}~\bibnamefont {Bawamia}}, \bibinfo {author}
  {\bibfnamefont {A.}~\bibnamefont {Bertoldi}}, \bibinfo {author}
  {\bibfnamefont {R.}~\bibnamefont {Bondarescu}}, \bibinfo {author}
  {\bibfnamefont {K.}~\bibnamefont {Bongs}}, \bibinfo {author} {\bibfnamefont
  {P.}~\bibnamefont {Bouyer}}, \bibinfo {author} {\bibfnamefont
  {C.}~\bibnamefont {Braxmaier}}, \bibinfo {author} {\bibfnamefont
  {L.}~\bibnamefont {Cacciapuoti}}, \bibinfo {author} {\bibfnamefont
  {C.}~\bibnamefont {Chaloner}}, \bibinfo {author} {\bibfnamefont
  {M.}~\bibnamefont {Chwalla}}, \bibinfo {author} {\bibfnamefont
  {W.}~\bibnamefont {Ertmer}}, \bibinfo {author} {\bibfnamefont
  {M.}~\bibnamefont {Franz}}, \bibinfo {author} {\bibfnamefont
  {N.}~\bibnamefont {Gaaloul}}, \bibinfo {author} {\bibfnamefont
  {M.}~\bibnamefont {Gehler}}, \bibinfo {author} {\bibfnamefont
  {D.}~\bibnamefont {Gerardi}}, \bibinfo {author} {\bibfnamefont
  {L.}~\bibnamefont {Gesa}}, \bibinfo {author} {\bibfnamefont {N.}~\bibnamefont
  {G{\"u}rlebeck}}, \bibinfo {author} {\bibfnamefont {J.}~\bibnamefont
  {Hartwig}}, \bibinfo {author} {\bibfnamefont {M.}~\bibnamefont {Hauth}},
  \bibinfo {author} {\bibfnamefont {O.}~\bibnamefont {Hellmig}}, \bibinfo
  {author} {\bibfnamefont {W.}~\bibnamefont {Herr}}, \bibinfo {author}
  {\bibfnamefont {S.}~\bibnamefont {Herrmann}}, \bibinfo {author}
  {\bibfnamefont {A.}~\bibnamefont {Heske}}, \bibinfo {author} {\bibfnamefont
  {A.}~\bibnamefont {Hinton}}, \bibinfo {author} {\bibfnamefont
  {P.}~\bibnamefont {Ireland}}, \bibinfo {author} {\bibfnamefont
  {P.}~\bibnamefont {Jetzer}}, \bibinfo {author} {\bibfnamefont
  {U.}~\bibnamefont {Johann}}, \bibinfo {author} {\bibfnamefont
  {M.}~\bibnamefont {Krutzik}}, \bibinfo {author} {\bibfnamefont
  {A.}~\bibnamefont {Kubelka}}, \bibinfo {author} {\bibfnamefont
  {C.}~\bibnamefont {L{\"a}mmerzahl}}, \bibinfo {author} {\bibfnamefont
  {A.}~\bibnamefont {Landragin}}, \bibinfo {author} {\bibfnamefont
  {I.}~\bibnamefont {Lloro}}, \bibinfo {author} {\bibfnamefont
  {D.}~\bibnamefont {Massonnet}}, \bibinfo {author} {\bibfnamefont
  {I.}~\bibnamefont {Mateos}}, \bibinfo {author} {\bibfnamefont
  {A.}~\bibnamefont {Milke}}, \bibinfo {author} {\bibfnamefont
  {M.}~\bibnamefont {Nofrarias}}, \bibinfo {author} {\bibfnamefont
  {M.}~\bibnamefont {Oswald}}, \bibinfo {author} {\bibfnamefont
  {A.}~\bibnamefont {Peters}}, \bibinfo {author} {\bibfnamefont
  {K.}~\bibnamefont {Posso-Trujillo}}, \bibinfo {author} {\bibfnamefont
  {E.}~\bibnamefont {Rasel}}, \bibinfo {author} {\bibfnamefont
  {E.}~\bibnamefont {Rocco}}, \bibinfo {author} {\bibfnamefont
  {A.}~\bibnamefont {Roura}}, \bibinfo {author} {\bibfnamefont
  {J.}~\bibnamefont {Rudolph}}, \bibinfo {author} {\bibfnamefont
  {W.}~\bibnamefont {Schleich}}, \bibinfo {author} {\bibfnamefont
  {C.}~\bibnamefont {Schubert}}, \bibinfo {author} {\bibfnamefont
  {T.}~\bibnamefont {Schuldt}}, \bibinfo {author} {\bibfnamefont
  {S.}~\bibnamefont {Seidel}}, \bibinfo {author} {\bibfnamefont
  {K.}~\bibnamefont {Sengstock}}, \bibinfo {author} {\bibfnamefont {C.~F.}\
  \bibnamefont {Sopuerta}}, \bibinfo {author} {\bibfnamefont {F.}~\bibnamefont
  {Sorrentino}}, \bibinfo {author} {\bibfnamefont {D.}~\bibnamefont {Summers}},
  \bibinfo {author} {\bibfnamefont {G.~M.}\ \bibnamefont {Tino}}, \bibinfo
  {author} {\bibfnamefont {C.}~\bibnamefont {Trenkel}}, \bibinfo {author}
  {\bibfnamefont {N.}~\bibnamefont {Uzunoglu}}, \bibinfo {author}
  {\bibfnamefont {W.}~\bibnamefont {von Klitzing}}, \bibinfo {author}
  {\bibfnamefont {R.}~\bibnamefont {Walser}}, \bibinfo {author} {\bibfnamefont
  {T.}~\bibnamefont {Wendrich}}, \bibinfo {author} {\bibfnamefont
  {A.}~\bibnamefont {Wenzlawski}}, \bibinfo {author} {\bibfnamefont
  {P.}~\bibnamefont {We{\ss}els}}, \bibinfo {author} {\bibfnamefont
  {A.}~\bibnamefont {Wicht}}, \bibinfo {author} {\bibfnamefont
  {E.}~\bibnamefont {Wille}}, \bibinfo {author} {\bibfnamefont
  {M.}~\bibnamefont {Williams}}, \bibinfo {author} {\bibfnamefont
  {P.}~\bibnamefont {Windpassinger}},\ and\ \bibinfo {author} {\bibfnamefont
  {N.}~\bibnamefont {Zahzam}},\ }\bibfield  {title} {\bibinfo {title}
  {Ste-quest---test of the universality of free fall using cold atom
  interferometry},\ }\href {https://doi.org/10.1088/0264-9381/31/11/115010}
  {\bibfield  {journal} {\bibinfo  {journal} {Classical and Quantum Gravity}\
  }\textbf {\bibinfo {volume} {31}},\ \bibinfo {pages} {115010} (\bibinfo
  {year} {2014})}\BibitemShut {NoStop}%
\bibitem [{\citenamefont {Trimeche}\ \emph {et~al.}(2019)\citenamefont
  {Trimeche}, \citenamefont {Battelier}, \citenamefont {Becker}, \citenamefont
  {Bertoldi}, \citenamefont {Bouyer}, \citenamefont {Braxmaier}, \citenamefont
  {Charron}, \citenamefont {Corgier}, \citenamefont {Cornelius}, \citenamefont
  {Douch}, \citenamefont {Gaaloul}, \citenamefont {Herrmann}, \citenamefont
  {M{\"u}ller}, \citenamefont {Rasel}, \citenamefont {Schubert}, \citenamefont
  {Wu},\ and\ \citenamefont {{Pereira dos Santos}}}]{Trimeche2019CQG}%
  \BibitemOpen
  \bibfield  {author} {\bibinfo {author} {\bibfnamefont {A.}~\bibnamefont
  {Trimeche}}, \bibinfo {author} {\bibfnamefont {B.}~\bibnamefont {Battelier}},
  \bibinfo {author} {\bibfnamefont {D.}~\bibnamefont {Becker}}, \bibinfo
  {author} {\bibfnamefont {A.}~\bibnamefont {Bertoldi}}, \bibinfo {author}
  {\bibfnamefont {P.}~\bibnamefont {Bouyer}}, \bibinfo {author} {\bibfnamefont
  {C.}~\bibnamefont {Braxmaier}}, \bibinfo {author} {\bibfnamefont
  {E.}~\bibnamefont {Charron}}, \bibinfo {author} {\bibfnamefont
  {R.}~\bibnamefont {Corgier}}, \bibinfo {author} {\bibfnamefont
  {M.}~\bibnamefont {Cornelius}}, \bibinfo {author} {\bibfnamefont
  {K.}~\bibnamefont {Douch}}, \bibinfo {author} {\bibfnamefont
  {N.}~\bibnamefont {Gaaloul}}, \bibinfo {author} {\bibfnamefont
  {S.}~\bibnamefont {Herrmann}}, \bibinfo {author} {\bibfnamefont
  {J.}~\bibnamefont {M{\"u}ller}}, \bibinfo {author} {\bibfnamefont
  {E.}~\bibnamefont {Rasel}}, \bibinfo {author} {\bibfnamefont
  {C.}~\bibnamefont {Schubert}}, \bibinfo {author} {\bibfnamefont
  {H.}~\bibnamefont {Wu}},\ and\ \bibinfo {author} {\bibfnamefont
  {F.}~\bibnamefont {{Pereira dos Santos}}},\ }\bibfield  {title} {\bibinfo
  {title} {Concept study and preliminary design of a cold atom interferometer
  for space gravity gradiometry},\ }\href
  {https://doi.org/10.1088/1361-6382/ab4548} {\bibfield  {journal} {\bibinfo
  {journal} {Classical and Quantum Gravity}\ }\textbf {\bibinfo {volume}
  {36}},\ \bibinfo {pages} {215004} (\bibinfo {year} {2019})}\BibitemShut
  {NoStop}%
\bibitem [{\citenamefont {Loriani}\ \emph {et~al.}(2019)\citenamefont
  {Loriani}, \citenamefont {Schlippert}, \citenamefont {Schubert},
  \citenamefont {Abend}, \citenamefont {Ahlers}, \citenamefont {Ertmer},
  \citenamefont {Rudolph}, \citenamefont {Hogan}, \citenamefont {Kasevich},
  \citenamefont {Rasel},\ and\ \citenamefont {Gaaloul}}]{Loriani2019NJP}%
  \BibitemOpen
  \bibfield  {author} {\bibinfo {author} {\bibfnamefont {S.}~\bibnamefont
  {Loriani}}, \bibinfo {author} {\bibfnamefont {D.}~\bibnamefont {Schlippert}},
  \bibinfo {author} {\bibfnamefont {C.}~\bibnamefont {Schubert}}, \bibinfo
  {author} {\bibfnamefont {S.}~\bibnamefont {Abend}}, \bibinfo {author}
  {\bibfnamefont {H.}~\bibnamefont {Ahlers}}, \bibinfo {author} {\bibfnamefont
  {W.}~\bibnamefont {Ertmer}}, \bibinfo {author} {\bibfnamefont
  {J.}~\bibnamefont {Rudolph}}, \bibinfo {author} {\bibfnamefont {J.~M.}\
  \bibnamefont {Hogan}}, \bibinfo {author} {\bibfnamefont {M.~A.}\ \bibnamefont
  {Kasevich}}, \bibinfo {author} {\bibfnamefont {E.~M.}\ \bibnamefont
  {Rasel}},\ and\ \bibinfo {author} {\bibfnamefont {N.}~\bibnamefont
  {Gaaloul}},\ }\bibfield  {title} {\bibinfo {title} {Atomic source selection
  in space-borne gravitational wave detection},\ }\href
  {https://doi.org/10.1088/1367-2630/ab22d0} {\bibfield  {journal} {\bibinfo
  {journal} {New Journal of Physics}\ }\textbf {\bibinfo {volume} {21}},\
  \bibinfo {pages} {063030} (\bibinfo {year} {2019})}\BibitemShut {NoStop}%
\bibitem [{\citenamefont {Corgier}\ \emph {et~al.}(2020)\citenamefont
  {Corgier}, \citenamefont {Loriani}, \citenamefont {Ahlers}, \citenamefont
  {Posso-Trujillo}, \citenamefont {Schubert}, \citenamefont {Rasel},
  \citenamefont {Charron},\ and\ \citenamefont {Gaaloul}}]{Corgier2020NJP}%
  \BibitemOpen
  \bibfield  {author} {\bibinfo {author} {\bibfnamefont {R.}~\bibnamefont
  {Corgier}}, \bibinfo {author} {\bibfnamefont {S.}~\bibnamefont {Loriani}},
  \bibinfo {author} {\bibfnamefont {H.}~\bibnamefont {Ahlers}}, \bibinfo
  {author} {\bibfnamefont {K.}~\bibnamefont {Posso-Trujillo}}, \bibinfo
  {author} {\bibfnamefont {C.}~\bibnamefont {Schubert}}, \bibinfo {author}
  {\bibfnamefont {E.~M.}\ \bibnamefont {Rasel}}, \bibinfo {author}
  {\bibfnamefont {E.}~\bibnamefont {Charron}},\ and\ \bibinfo {author}
  {\bibfnamefont {N.}~\bibnamefont {Gaaloul}},\ }\bibfield  {title} {\bibinfo
  {title} {Interacting quantum mixtures for precision atom interferometry},\
  }\href {https://doi.org/10.1088/1367-2630/abcbc8} {\bibfield  {journal}
  {\bibinfo  {journal} {New Journal of Physics}\ }\textbf {\bibinfo {volume}
  {22}},\ \bibinfo {pages} {123008} (\bibinfo {year} {2020})}\BibitemShut
  {NoStop}%
\bibitem [{\citenamefont {Struckmann}\ \emph {et~al.}(2024)\citenamefont
  {Struckmann}, \citenamefont {Corgier}, \citenamefont {Loriani}, \citenamefont
  {Kleinsteinberg}, \citenamefont {Gox}, \citenamefont {Giese}, \citenamefont
  {M\'etris}, \citenamefont {Gaaloul},\ and\ \citenamefont
  {Wolf}}]{Struckmann2024PRD}%
  \BibitemOpen
  \bibfield  {author} {\bibinfo {author} {\bibfnamefont {C.}~\bibnamefont
  {Struckmann}}, \bibinfo {author} {\bibfnamefont {R.}~\bibnamefont {Corgier}},
  \bibinfo {author} {\bibfnamefont {S.}~\bibnamefont {Loriani}}, \bibinfo
  {author} {\bibfnamefont {G.}~\bibnamefont {Kleinsteinberg}}, \bibinfo
  {author} {\bibfnamefont {N.}~\bibnamefont {Gox}}, \bibinfo {author}
  {\bibfnamefont {E.}~\bibnamefont {Giese}}, \bibinfo {author} {\bibfnamefont
  {G.}~\bibnamefont {M\'etris}}, \bibinfo {author} {\bibfnamefont
  {N.}~\bibnamefont {Gaaloul}},\ and\ \bibinfo {author} {\bibfnamefont
  {P.}~\bibnamefont {Wolf}},\ }\bibfield  {title} {\bibinfo {title} {Platform
  and environment requirements of a satellite quantum test of the weak
  equivalence principle at the ${10}^{\ensuremath{-}17}$ level},\ }\href
  {https://doi.org/10.1103/PhysRevD.109.064010} {\bibfield  {journal} {\bibinfo
   {journal} {Phys. Rev. D}\ }\textbf {\bibinfo {volume} {109}},\ \bibinfo
  {pages} {064010} (\bibinfo {year} {2024})}\BibitemShut {NoStop}%
\bibitem [{\citenamefont {Leanhardt}\ \emph {et~al.}(2003)\citenamefont
  {Leanhardt}, \citenamefont {Pasquini}, \citenamefont {Saba}, \citenamefont
  {Schirotzek}, \citenamefont {Shin}, \citenamefont {Kielpinski}, \citenamefont
  {Pritchard},\ and\ \citenamefont {Ketterle}}]{Leanhardt2003Science}%
  \BibitemOpen
  \bibfield  {author} {\bibinfo {author} {\bibfnamefont {A.~E.}\ \bibnamefont
  {Leanhardt}}, \bibinfo {author} {\bibfnamefont {T.~A.}\ \bibnamefont
  {Pasquini}}, \bibinfo {author} {\bibfnamefont {M.}~\bibnamefont {Saba}},
  \bibinfo {author} {\bibfnamefont {A.}~\bibnamefont {Schirotzek}}, \bibinfo
  {author} {\bibfnamefont {Y.}~\bibnamefont {Shin}}, \bibinfo {author}
  {\bibfnamefont {D.}~\bibnamefont {Kielpinski}}, \bibinfo {author}
  {\bibfnamefont {D.~E.}\ \bibnamefont {Pritchard}},\ and\ \bibinfo {author}
  {\bibfnamefont {W.}~\bibnamefont {Ketterle}},\ }\bibfield  {title} {\bibinfo
  {title} {Cooling bose-einstein condensates below 500 picokelvin},\ }\href
  {https://doi.org/10.1126/science.1088827} {\bibfield  {journal} {\bibinfo
  {journal} {Science (New York, N.Y.)}\ }\textbf {\bibinfo {volume} {301}},\
  \bibinfo {pages} {1513} (\bibinfo {year} {2003})}\BibitemShut {NoStop}%
\bibitem [{\citenamefont {Medley}\ \emph {et~al.}(2011)\citenamefont {Medley},
  \citenamefont {Weld}, \citenamefont {Miyake}, \citenamefont {Pritchard},\
  and\ \citenamefont {Ketterle}}]{Medley2011PRL}%
  \BibitemOpen
  \bibfield  {author} {\bibinfo {author} {\bibfnamefont {P.}~\bibnamefont
  {Medley}}, \bibinfo {author} {\bibfnamefont {D.~M.}\ \bibnamefont {Weld}},
  \bibinfo {author} {\bibfnamefont {H.}~\bibnamefont {Miyake}}, \bibinfo
  {author} {\bibfnamefont {D.~E.}\ \bibnamefont {Pritchard}},\ and\ \bibinfo
  {author} {\bibfnamefont {W.}~\bibnamefont {Ketterle}},\ }\bibfield  {title}
  {\bibinfo {title} {Spin gradient demagnetization cooling of ultracold
  atoms},\ }\href {https://doi.org/10.1103/PhysRevLett.106.195301} {\bibfield
  {journal} {\bibinfo  {journal} {Phys. Rev. Lett.}\ }\textbf {\bibinfo
  {volume} {106}},\ \bibinfo {pages} {195301} (\bibinfo {year}
  {2011})}\BibitemShut {NoStop}%
\bibitem [{\citenamefont {Ammann}\ and\ \citenamefont
  {Christensen}(1997)}]{Ammann1997PRL}%
  \BibitemOpen
  \bibfield  {author} {\bibinfo {author} {\bibfnamefont {H.}~\bibnamefont
  {Ammann}}\ and\ \bibinfo {author} {\bibfnamefont {N.}~\bibnamefont
  {Christensen}},\ }\bibfield  {title} {\bibinfo {title} {Delta kick cooling: A
  new method for cooling atoms},\ }\href
  {https://doi.org/10.1103/PhysRevLett.78.2088} {\bibfield  {journal} {\bibinfo
   {journal} {Phys. Rev. Lett.}\ }\textbf {\bibinfo {volume} {78}},\ \bibinfo
  {pages} {2088} (\bibinfo {year} {1997})}\BibitemShut {NoStop}%
\bibitem [{\citenamefont {Kalnins}\ \emph {et~al.}(2005)\citenamefont
  {Kalnins}, \citenamefont {Amini},\ and\ \citenamefont
  {Gould}}]{Kalnins2005PRA}%
  \BibitemOpen
  \bibfield  {author} {\bibinfo {author} {\bibfnamefont {J.~G.}\ \bibnamefont
  {Kalnins}}, \bibinfo {author} {\bibfnamefont {J.~M.}\ \bibnamefont {Amini}},\
  and\ \bibinfo {author} {\bibfnamefont {H.}~\bibnamefont {Gould}},\ }\bibfield
   {title} {\bibinfo {title} {Focusing a fountain of neutral cesium atoms with
  an electrostatic lens triplet},\ }\href
  {https://doi.org/10.1103/PhysRevA.72.043406} {\bibfield  {journal} {\bibinfo
  {journal} {Phys. Rev. A}\ }\textbf {\bibinfo {volume} {72}},\ \bibinfo
  {pages} {043406} (\bibinfo {year} {2005})}\BibitemShut {NoStop}%
\bibitem [{\citenamefont {M\"untinga}\ \emph {et~al.}(2013)\citenamefont
  {M\"untinga}, \citenamefont {Ahlers}, \citenamefont {Krutzik}, \citenamefont
  {Wenzlawski}, \citenamefont {Arnold}, \citenamefont {Becker}, \citenamefont
  {Bongs}, \citenamefont {Dittus}, \citenamefont {Duncker}, \citenamefont
  {Gaaloul}, \citenamefont {Gherasim}, \citenamefont {Giese}, \citenamefont
  {Grzeschik}, \citenamefont {H\"ansch}, \citenamefont {Hellmig}, \citenamefont
  {Herr}, \citenamefont {Herrmann}, \citenamefont {Kajari}, \citenamefont
  {Kleinert}, \citenamefont {L\"ammerzahl}, \citenamefont {Lewoczko-Adamczyk},
  \citenamefont {Malcolm}, \citenamefont {Meyer}, \citenamefont {Nolte},
  \citenamefont {Peters}, \citenamefont {Popp}, \citenamefont {Reichel},
  \citenamefont {Roura}, \citenamefont {Rudolph}, \citenamefont {Schiemangk},
  \citenamefont {Schneider}, \citenamefont {Seidel}, \citenamefont {Sengstock},
  \citenamefont {Tamma}, \citenamefont {Valenzuela}, \citenamefont {Vogel},
  \citenamefont {Walser}, \citenamefont {Wendrich}, \citenamefont
  {Windpassinger}, \citenamefont {Zeller}, \citenamefont {van Zoest},
  \citenamefont {Ertmer}, \citenamefont {Schleich},\ and\ \citenamefont
  {Rasel}}]{Muntinga2013PRL}%
  \BibitemOpen
  \bibfield  {author} {\bibinfo {author} {\bibfnamefont {H.}~\bibnamefont
  {M\"untinga}}, \bibinfo {author} {\bibfnamefont {H.}~\bibnamefont {Ahlers}},
  \bibinfo {author} {\bibfnamefont {M.}~\bibnamefont {Krutzik}}, \bibinfo
  {author} {\bibfnamefont {A.}~\bibnamefont {Wenzlawski}}, \bibinfo {author}
  {\bibfnamefont {S.}~\bibnamefont {Arnold}}, \bibinfo {author} {\bibfnamefont
  {D.}~\bibnamefont {Becker}}, \bibinfo {author} {\bibfnamefont
  {K.}~\bibnamefont {Bongs}}, \bibinfo {author} {\bibfnamefont
  {H.}~\bibnamefont {Dittus}}, \bibinfo {author} {\bibfnamefont
  {H.}~\bibnamefont {Duncker}}, \bibinfo {author} {\bibfnamefont
  {N.}~\bibnamefont {Gaaloul}}, \bibinfo {author} {\bibfnamefont
  {C.}~\bibnamefont {Gherasim}}, \bibinfo {author} {\bibfnamefont
  {E.}~\bibnamefont {Giese}}, \bibinfo {author} {\bibfnamefont
  {C.}~\bibnamefont {Grzeschik}}, \bibinfo {author} {\bibfnamefont {T.~W.}\
  \bibnamefont {H\"ansch}}, \bibinfo {author} {\bibfnamefont {O.}~\bibnamefont
  {Hellmig}}, \bibinfo {author} {\bibfnamefont {W.}~\bibnamefont {Herr}},
  \bibinfo {author} {\bibfnamefont {S.}~\bibnamefont {Herrmann}}, \bibinfo
  {author} {\bibfnamefont {E.}~\bibnamefont {Kajari}}, \bibinfo {author}
  {\bibfnamefont {S.}~\bibnamefont {Kleinert}}, \bibinfo {author}
  {\bibfnamefont {C.}~\bibnamefont {L\"ammerzahl}}, \bibinfo {author}
  {\bibfnamefont {W.}~\bibnamefont {Lewoczko-Adamczyk}}, \bibinfo {author}
  {\bibfnamefont {J.}~\bibnamefont {Malcolm}}, \bibinfo {author} {\bibfnamefont
  {N.}~\bibnamefont {Meyer}}, \bibinfo {author} {\bibfnamefont
  {R.}~\bibnamefont {Nolte}}, \bibinfo {author} {\bibfnamefont
  {A.}~\bibnamefont {Peters}}, \bibinfo {author} {\bibfnamefont
  {M.}~\bibnamefont {Popp}}, \bibinfo {author} {\bibfnamefont {J.}~\bibnamefont
  {Reichel}}, \bibinfo {author} {\bibfnamefont {A.}~\bibnamefont {Roura}},
  \bibinfo {author} {\bibfnamefont {J.}~\bibnamefont {Rudolph}}, \bibinfo
  {author} {\bibfnamefont {M.}~\bibnamefont {Schiemangk}}, \bibinfo {author}
  {\bibfnamefont {M.}~\bibnamefont {Schneider}}, \bibinfo {author}
  {\bibfnamefont {S.~T.}\ \bibnamefont {Seidel}}, \bibinfo {author}
  {\bibfnamefont {K.}~\bibnamefont {Sengstock}}, \bibinfo {author}
  {\bibfnamefont {V.}~\bibnamefont {Tamma}}, \bibinfo {author} {\bibfnamefont
  {T.}~\bibnamefont {Valenzuela}}, \bibinfo {author} {\bibfnamefont
  {A.}~\bibnamefont {Vogel}}, \bibinfo {author} {\bibfnamefont
  {R.}~\bibnamefont {Walser}}, \bibinfo {author} {\bibfnamefont
  {T.}~\bibnamefont {Wendrich}}, \bibinfo {author} {\bibfnamefont
  {P.}~\bibnamefont {Windpassinger}}, \bibinfo {author} {\bibfnamefont
  {W.}~\bibnamefont {Zeller}}, \bibinfo {author} {\bibfnamefont
  {T.}~\bibnamefont {van Zoest}}, \bibinfo {author} {\bibfnamefont
  {W.}~\bibnamefont {Ertmer}}, \bibinfo {author} {\bibfnamefont {W.~P.}\
  \bibnamefont {Schleich}},\ and\ \bibinfo {author} {\bibfnamefont {E.~M.}\
  \bibnamefont {Rasel}},\ }\bibfield  {title} {\bibinfo {title} {Interferometry
  with bose-einstein condensates in microgravity},\ }\href
  {https://doi.org/10.1103/PhysRevLett.110.093602} {\bibfield  {journal}
  {\bibinfo  {journal} {Phys. Rev. Lett.}\ }\textbf {\bibinfo {volume} {110}},\
  \bibinfo {pages} {093602} (\bibinfo {year} {2013})}\BibitemShut {NoStop}%
\bibitem [{\citenamefont {Ketterle}\ and\ \citenamefont
  {Druten}(1996)}]{Ketterle96AAMOP}%
  \BibitemOpen
  \bibfield  {author} {\bibinfo {author} {\bibfnamefont {W.}~\bibnamefont
  {Ketterle}}\ and\ \bibinfo {author} {\bibfnamefont {N.~V.}\ \bibnamefont
  {Druten}},\ }\bibfield  {title} {\bibinfo {title} {Evaporative cooling of
  trapped atoms},\ }in\ \href {https://doi.org/10.1016/s1049-250x(08)60101-9}
  {\emph {\bibinfo {booktitle} {Advances In Atomic, Molecular, and Optical
  Physics}}}\ (\bibinfo  {publisher} {Elsevier},\ \bibinfo {year} {1996})\ pp.\
  \bibinfo {pages} {181--236}\BibitemShut {NoStop}%
\bibitem [{\citenamefont {Kovachy}\ \emph
  {et~al.}(2015{\natexlab{b}})\citenamefont {Kovachy}, \citenamefont {Hogan},
  \citenamefont {Sugarbaker}, \citenamefont {Dickerson}, \citenamefont
  {Donnelly}, \citenamefont {Overstreet},\ and\ \citenamefont
  {Kasevich}}]{Kovachy2015PRL}%
  \BibitemOpen
  \bibfield  {author} {\bibinfo {author} {\bibfnamefont {T.}~\bibnamefont
  {Kovachy}}, \bibinfo {author} {\bibfnamefont {J.~M.}\ \bibnamefont {Hogan}},
  \bibinfo {author} {\bibfnamefont {A.}~\bibnamefont {Sugarbaker}}, \bibinfo
  {author} {\bibfnamefont {S.~M.}\ \bibnamefont {Dickerson}}, \bibinfo {author}
  {\bibfnamefont {C.~A.}\ \bibnamefont {Donnelly}}, \bibinfo {author}
  {\bibfnamefont {C.}~\bibnamefont {Overstreet}},\ and\ \bibinfo {author}
  {\bibfnamefont {M.~A.}\ \bibnamefont {Kasevich}},\ }\bibfield  {title}
  {\bibinfo {title} {Matter wave lensing to picokelvin temperatures},\ }\href
  {https://doi.org/10.1103/PhysRevLett.114.143004} {\bibfield  {journal}
  {\bibinfo  {journal} {Phys. Rev. Lett.}\ }\textbf {\bibinfo {volume} {114}},\
  \bibinfo {pages} {143004} (\bibinfo {year} {2015}{\natexlab{b}})}\BibitemShut
  {NoStop}%
\bibitem [{\citenamefont {Deppner}\ \emph {et~al.}(2021)\citenamefont
  {Deppner}, \citenamefont {Herr}, \citenamefont {Cornelius}, \citenamefont
  {Stromberger}, \citenamefont {Sternke}, \citenamefont {Grzeschik},
  \citenamefont {Grote}, \citenamefont {Rudolph}, \citenamefont {Herrmann},
  \citenamefont {Krutzik}, \citenamefont {Wenzlawski}, \citenamefont {Corgier},
  \citenamefont {Charron}, \citenamefont {Gu\'ery-Odelin}, \citenamefont
  {Gaaloul}, \citenamefont {L\"ammerzahl}, \citenamefont {Peters},
  \citenamefont {Windpassinger},\ and\ \citenamefont {Rasel}}]{Deppner2021PRL}%
  \BibitemOpen
  \bibfield  {author} {\bibinfo {author} {\bibfnamefont {C.}~\bibnamefont
  {Deppner}}, \bibinfo {author} {\bibfnamefont {W.}~\bibnamefont {Herr}},
  \bibinfo {author} {\bibfnamefont {M.}~\bibnamefont {Cornelius}}, \bibinfo
  {author} {\bibfnamefont {P.}~\bibnamefont {Stromberger}}, \bibinfo {author}
  {\bibfnamefont {T.}~\bibnamefont {Sternke}}, \bibinfo {author} {\bibfnamefont
  {C.}~\bibnamefont {Grzeschik}}, \bibinfo {author} {\bibfnamefont
  {A.}~\bibnamefont {Grote}}, \bibinfo {author} {\bibfnamefont
  {J.}~\bibnamefont {Rudolph}}, \bibinfo {author} {\bibfnamefont
  {S.}~\bibnamefont {Herrmann}}, \bibinfo {author} {\bibfnamefont
  {M.}~\bibnamefont {Krutzik}}, \bibinfo {author} {\bibfnamefont
  {A.}~\bibnamefont {Wenzlawski}}, \bibinfo {author} {\bibfnamefont
  {R.}~\bibnamefont {Corgier}}, \bibinfo {author} {\bibfnamefont
  {E.}~\bibnamefont {Charron}}, \bibinfo {author} {\bibfnamefont
  {D.}~\bibnamefont {Gu\'ery-Odelin}}, \bibinfo {author} {\bibfnamefont
  {N.}~\bibnamefont {Gaaloul}}, \bibinfo {author} {\bibfnamefont
  {C.}~\bibnamefont {L\"ammerzahl}}, \bibinfo {author} {\bibfnamefont
  {A.}~\bibnamefont {Peters}}, \bibinfo {author} {\bibfnamefont
  {P.}~\bibnamefont {Windpassinger}},\ and\ \bibinfo {author} {\bibfnamefont
  {E.~M.}\ \bibnamefont {Rasel}},\ }\bibfield  {title} {\bibinfo {title}
  {Collective-mode enhanced matter-wave optics},\ }\href
  {https://doi.org/10.1103/PhysRevLett.127.100401} {\bibfield  {journal}
  {\bibinfo  {journal} {Phys. Rev. Lett.}\ }\textbf {\bibinfo {volume} {127}},\
  \bibinfo {pages} {100401} (\bibinfo {year} {2021})}\BibitemShut {NoStop}%
\bibitem [{\citenamefont {Gaaloul}\ \emph {et~al.}(2022)\citenamefont
  {Gaaloul}, \citenamefont {Meister}, \citenamefont {Corgier}, \citenamefont
  {Pichery}, \citenamefont {Boegel}, \citenamefont {Herr}, \citenamefont
  {Ahlers}, \citenamefont {Charron}, \citenamefont {Williams}, \citenamefont
  {Thompson}, \citenamefont {Schleich}, \citenamefont {Rasel},\ and\
  \citenamefont {Bigelow}}]{Gaaloul2022NatComm}%
  \BibitemOpen
  \bibfield  {author} {\bibinfo {author} {\bibfnamefont {N.}~\bibnamefont
  {Gaaloul}}, \bibinfo {author} {\bibfnamefont {M.}~\bibnamefont {Meister}},
  \bibinfo {author} {\bibfnamefont {R.}~\bibnamefont {Corgier}}, \bibinfo
  {author} {\bibfnamefont {A.}~\bibnamefont {Pichery}}, \bibinfo {author}
  {\bibfnamefont {P.}~\bibnamefont {Boegel}}, \bibinfo {author} {\bibfnamefont
  {W.}~\bibnamefont {Herr}}, \bibinfo {author} {\bibfnamefont {H.}~\bibnamefont
  {Ahlers}}, \bibinfo {author} {\bibfnamefont {E.}~\bibnamefont {Charron}},
  \bibinfo {author} {\bibfnamefont {J.~R.}\ \bibnamefont {Williams}}, \bibinfo
  {author} {\bibfnamefont {R.~J.}\ \bibnamefont {Thompson}}, \bibinfo {author}
  {\bibfnamefont {W.~P.}\ \bibnamefont {Schleich}}, \bibinfo {author}
  {\bibfnamefont {E.~M.}\ \bibnamefont {Rasel}},\ and\ \bibinfo {author}
  {\bibfnamefont {N.~P.}\ \bibnamefont {Bigelow}},\ }\bibfield  {title}
  {\bibinfo {title} {A space-based quantum gas laboratory at picokelvin energy
  scales},\ }\href {https://doi.org/10.1038/s41467-022-35274-6} {\bibfield
  {journal} {\bibinfo  {journal} {Nature communications}\ }\textbf {\bibinfo
  {volume} {13}},\ \bibinfo {pages} {7889} (\bibinfo {year}
  {2022})}\BibitemShut {NoStop}%
\bibitem [{\citenamefont {Inouye}\ \emph {et~al.}(1998)\citenamefont {Inouye},
  \citenamefont {Andrews}, \citenamefont {Stenger}, \citenamefont {Miesner},
  \citenamefont {Stamper-Kurn},\ and\ \citenamefont
  {Ketterle}}]{Inouye1998Nature}%
  \BibitemOpen
  \bibfield  {author} {\bibinfo {author} {\bibfnamefont {S.}~\bibnamefont
  {Inouye}}, \bibinfo {author} {\bibfnamefont {M.~R.}\ \bibnamefont {Andrews}},
  \bibinfo {author} {\bibfnamefont {J.}~\bibnamefont {Stenger}}, \bibinfo
  {author} {\bibfnamefont {H.-J.}\ \bibnamefont {Miesner}}, \bibinfo {author}
  {\bibfnamefont {D.~M.}\ \bibnamefont {Stamper-Kurn}},\ and\ \bibinfo {author}
  {\bibfnamefont {W.}~\bibnamefont {Ketterle}},\ }\bibfield  {title} {\bibinfo
  {title} {Observation of feshbach resonances in a bose-einstein condensate},\
  }\href {https://doi.org/10.1038/32354} {\bibfield  {journal} {\bibinfo
  {journal} {Nature}\ }\textbf {\bibinfo {volume} {392}},\ \bibinfo {pages}
  {151} (\bibinfo {year} {1998})}\BibitemShut {NoStop}%
\bibitem [{\citenamefont {Masi}\ \emph {et~al.}(2021)\citenamefont {Masi},
  \citenamefont {Petrucciani}, \citenamefont {Burchianti}, \citenamefont
  {Fort}, \citenamefont {Inguscio}, \citenamefont {Marconi}, \citenamefont
  {Modugno}, \citenamefont {Preti}, \citenamefont {Trypogeorgos}, \citenamefont
  {Fattori},\ and\ \citenamefont {Minardi}}]{Masi2021PRR}%
  \BibitemOpen
  \bibfield  {author} {\bibinfo {author} {\bibfnamefont {L.}~\bibnamefont
  {Masi}}, \bibinfo {author} {\bibfnamefont {T.}~\bibnamefont {Petrucciani}},
  \bibinfo {author} {\bibfnamefont {A.}~\bibnamefont {Burchianti}}, \bibinfo
  {author} {\bibfnamefont {C.}~\bibnamefont {Fort}}, \bibinfo {author}
  {\bibfnamefont {M.}~\bibnamefont {Inguscio}}, \bibinfo {author}
  {\bibfnamefont {L.}~\bibnamefont {Marconi}}, \bibinfo {author} {\bibfnamefont
  {G.}~\bibnamefont {Modugno}}, \bibinfo {author} {\bibfnamefont
  {N.}~\bibnamefont {Preti}}, \bibinfo {author} {\bibfnamefont
  {D.}~\bibnamefont {Trypogeorgos}}, \bibinfo {author} {\bibfnamefont
  {M.}~\bibnamefont {Fattori}},\ and\ \bibinfo {author} {\bibfnamefont
  {F.}~\bibnamefont {Minardi}},\ }\bibfield  {title} {\bibinfo {title}
  {Multimode trapped interferometer with noninteracting bose-einstein
  condensates},\ }\href {https://doi.org/10.1103/PhysRevResearch.3.043188}
  {\bibfield  {journal} {\bibinfo  {journal} {Phys. Rev. Research}\ }\textbf
  {\bibinfo {volume} {3}},\ \bibinfo {pages} {043188} (\bibinfo {year}
  {2021})}\BibitemShut {NoStop}%
\bibitem [{\citenamefont {Albers}\ \emph {et~al.}(2022)\citenamefont {Albers},
  \citenamefont {Corgier}, \citenamefont {Herbst}, \citenamefont {Rajagopalan},
  \citenamefont {Schubert}, \citenamefont {Vogt}, \citenamefont {Woltmann},
  \citenamefont {Lämmerzahl}, \citenamefont {Herrmann}, \citenamefont
  {Charron}, \citenamefont {Ertmer}, \citenamefont {Rasel}, \citenamefont
  {Gaaloul},\ and\ \citenamefont {Schlippert}}]{Albers2022Commun}%
  \BibitemOpen
  \bibfield  {author} {\bibinfo {author} {\bibfnamefont {H.}~\bibnamefont
  {Albers}}, \bibinfo {author} {\bibfnamefont {R.}~\bibnamefont {Corgier}},
  \bibinfo {author} {\bibfnamefont {A.}~\bibnamefont {Herbst}}, \bibinfo
  {author} {\bibfnamefont {A.}~\bibnamefont {Rajagopalan}}, \bibinfo {author}
  {\bibfnamefont {C.}~\bibnamefont {Schubert}}, \bibinfo {author}
  {\bibfnamefont {C.}~\bibnamefont {Vogt}}, \bibinfo {author} {\bibfnamefont
  {M.}~\bibnamefont {Woltmann}}, \bibinfo {author} {\bibfnamefont
  {C.}~\bibnamefont {Lämmerzahl}}, \bibinfo {author} {\bibfnamefont
  {S.}~\bibnamefont {Herrmann}}, \bibinfo {author} {\bibfnamefont
  {E.}~\bibnamefont {Charron}}, \bibinfo {author} {\bibfnamefont
  {W.}~\bibnamefont {Ertmer}}, \bibinfo {author} {\bibfnamefont {E.~M.}\
  \bibnamefont {Rasel}}, \bibinfo {author} {\bibfnamefont {N.}~\bibnamefont
  {Gaaloul}},\ and\ \bibinfo {author} {\bibfnamefont {D.}~\bibnamefont
  {Schlippert}},\ }\bibfield  {title} {\bibinfo {title} {All-optical
  matter-wave lens using time-averaged potentials},\ }\href
  {https://doi.org/10.1038/s42005-022-00825-2} {\bibfield  {journal} {\bibinfo
  {journal} {Communications Physics}\ }\textbf {\bibinfo {volume} {5}},\
  \bibinfo {pages} {60} (\bibinfo {year} {2022})}\BibitemShut {NoStop}%
\bibitem [{\citenamefont {Roy}\ \emph {et~al.}(2016)\citenamefont {Roy},
  \citenamefont {Green}, \citenamefont {Bowler},\ and\ \citenamefont
  {Gupta}}]{Roy2016PRA}%
  \BibitemOpen
  \bibfield  {author} {\bibinfo {author} {\bibfnamefont {R.}~\bibnamefont
  {Roy}}, \bibinfo {author} {\bibfnamefont {A.}~\bibnamefont {Green}}, \bibinfo
  {author} {\bibfnamefont {R.}~\bibnamefont {Bowler}},\ and\ \bibinfo {author}
  {\bibfnamefont {S.}~\bibnamefont {Gupta}},\ }\bibfield  {title} {\bibinfo
  {title} {Rapid cooling to quantum degeneracy in dynamically shaped atom
  traps},\ }\href {https://doi.org/10.1103/PhysRevA.93.043403} {\bibfield
  {journal} {\bibinfo  {journal} {Phys. Rev. A}\ }\textbf {\bibinfo {volume}
  {93}},\ \bibinfo {pages} {043403} (\bibinfo {year} {2016})}\BibitemShut
  {NoStop}%
\bibitem [{\citenamefont {Jin}\ \emph {et~al.}(1996)\citenamefont {Jin},
  \citenamefont {Ensher}, \citenamefont {Matthews}, \citenamefont {Wieman},\
  and\ \citenamefont {Cornell}}]{Jin1996PRL}%
  \BibitemOpen
  \bibfield  {author} {\bibinfo {author} {\bibfnamefont {D.~S.}\ \bibnamefont
  {Jin}}, \bibinfo {author} {\bibfnamefont {J.~R.}\ \bibnamefont {Ensher}},
  \bibinfo {author} {\bibfnamefont {M.~R.}\ \bibnamefont {Matthews}}, \bibinfo
  {author} {\bibfnamefont {C.~E.}\ \bibnamefont {Wieman}},\ and\ \bibinfo
  {author} {\bibfnamefont {E.~A.}\ \bibnamefont {Cornell}},\ }\bibfield
  {title} {\bibinfo {title} {Collective excitations of a bose-einstein
  condensate in a dilute gas},\ }\href
  {https://doi.org/10.1103/PhysRevLett.77.420} {\bibfield  {journal} {\bibinfo
  {journal} {Phys. Rev. Lett.}\ }\textbf {\bibinfo {volume} {77}},\ \bibinfo
  {pages} {420} (\bibinfo {year} {1996})}\BibitemShut {NoStop}%
\bibitem [{\citenamefont {Mewes}\ \emph {et~al.}(1996)\citenamefont {Mewes},
  \citenamefont {Andrews}, \citenamefont {van Druten}, \citenamefont {Kurn},
  \citenamefont {Durfee}, \citenamefont {Townsend},\ and\ \citenamefont
  {Ketterle}}]{Mewes1996PRL}%
  \BibitemOpen
  \bibfield  {author} {\bibinfo {author} {\bibfnamefont {M.-O.}\ \bibnamefont
  {Mewes}}, \bibinfo {author} {\bibfnamefont {M.~R.}\ \bibnamefont {Andrews}},
  \bibinfo {author} {\bibfnamefont {N.~J.}\ \bibnamefont {van Druten}},
  \bibinfo {author} {\bibfnamefont {D.~M.}\ \bibnamefont {Kurn}}, \bibinfo
  {author} {\bibfnamefont {D.~S.}\ \bibnamefont {Durfee}}, \bibinfo {author}
  {\bibfnamefont {C.~G.}\ \bibnamefont {Townsend}},\ and\ \bibinfo {author}
  {\bibfnamefont {W.}~\bibnamefont {Ketterle}},\ }\bibfield  {title} {\bibinfo
  {title} {Collective excitations of a bose-einstein condensate in a magnetic
  trap},\ }\href {https://doi.org/10.1103/PhysRevLett.77.988} {\bibfield
  {journal} {\bibinfo  {journal} {Phys. Rev. Lett.}\ }\textbf {\bibinfo
  {volume} {77}},\ \bibinfo {pages} {988} (\bibinfo {year} {1996})}\BibitemShut
  {NoStop}%
\bibitem [{\citenamefont {Egorov}\ \emph {et~al.}(2013)\citenamefont {Egorov},
  \citenamefont {Opanchuk}, \citenamefont {Drummond}, \citenamefont {Hall},
  \citenamefont {Hannaford},\ and\ \citenamefont {Sidorov}}]{EgorovPRA2013}%
  \BibitemOpen
  \bibfield  {author} {\bibinfo {author} {\bibfnamefont {M.}~\bibnamefont
  {Egorov}}, \bibinfo {author} {\bibfnamefont {B.}~\bibnamefont {Opanchuk}},
  \bibinfo {author} {\bibfnamefont {P.}~\bibnamefont {Drummond}}, \bibinfo
  {author} {\bibfnamefont {B.~V.}\ \bibnamefont {Hall}}, \bibinfo {author}
  {\bibfnamefont {P.}~\bibnamefont {Hannaford}},\ and\ \bibinfo {author}
  {\bibfnamefont {A.~I.}\ \bibnamefont {Sidorov}},\ }\bibfield  {title}
  {\bibinfo {title} {Measurement of $s$-wave scattering lengths in a
  two-component bose-einstein condensate},\ }\href
  {https://doi.org/10.1103/PhysRevA.87.053614} {\bibfield  {journal} {\bibinfo
  {journal} {Phys. Rev. A}\ }\textbf {\bibinfo {volume} {87}},\ \bibinfo
  {pages} {053614} (\bibinfo {year} {2013})}\BibitemShut {NoStop}%
\bibitem [{\citenamefont {Kraemer}\ \emph {et~al.}(2004)\citenamefont
  {Kraemer}, \citenamefont {Herbig}, \citenamefont {Mark}, \citenamefont
  {Weber}, \citenamefont {Chin}, \citenamefont {N{\"a}gerl},\ and\
  \citenamefont {Grimm}}]{Kraemer2004APB}%
  \BibitemOpen
  \bibfield  {author} {\bibinfo {author} {\bibfnamefont {T.}~\bibnamefont
  {Kraemer}}, \bibinfo {author} {\bibfnamefont {J.}~\bibnamefont {Herbig}},
  \bibinfo {author} {\bibfnamefont {M.}~\bibnamefont {Mark}}, \bibinfo {author}
  {\bibfnamefont {T.}~\bibnamefont {Weber}}, \bibinfo {author} {\bibfnamefont
  {C.}~\bibnamefont {Chin}}, \bibinfo {author} {\bibfnamefont {H.-C.}\
  \bibnamefont {N{\"a}gerl}},\ and\ \bibinfo {author} {\bibfnamefont
  {R.}~\bibnamefont {Grimm}},\ }\bibfield  {title} {\bibinfo {title} {Optimized
  production of a cesium bose--einstein condensate},\ }\href
  {https://doi.org/10.1007/s00340-004-1657-5} {\bibfield  {journal} {\bibinfo
  {journal} {Applied Physics B}\ }\textbf {\bibinfo {volume} {79}},\ \bibinfo
  {pages} {1013} (\bibinfo {year} {2004})}\BibitemShut {NoStop}%
\bibitem [{\citenamefont {Roati}\ \emph {et~al.}(2007)\citenamefont {Roati},
  \citenamefont {Zaccanti}, \citenamefont {D'Errico}, \citenamefont {Catani},
  \citenamefont {Modugno}, \citenamefont {Simoni}, \citenamefont {Inguscio},\
  and\ \citenamefont {Modugno}}]{Roati2007PRL}%
  \BibitemOpen
  \bibfield  {author} {\bibinfo {author} {\bibfnamefont {G.}~\bibnamefont
  {Roati}}, \bibinfo {author} {\bibfnamefont {M.}~\bibnamefont {Zaccanti}},
  \bibinfo {author} {\bibfnamefont {C.}~\bibnamefont {D'Errico}}, \bibinfo
  {author} {\bibfnamefont {J.}~\bibnamefont {Catani}}, \bibinfo {author}
  {\bibfnamefont {M.}~\bibnamefont {Modugno}}, \bibinfo {author} {\bibfnamefont
  {A.}~\bibnamefont {Simoni}}, \bibinfo {author} {\bibfnamefont
  {M.}~\bibnamefont {Inguscio}},\ and\ \bibinfo {author} {\bibfnamefont
  {G.}~\bibnamefont {Modugno}},\ }\bibfield  {title} {\bibinfo {title}
  {$^{39}\mathrm{K}$ bose-einstein condensate with tunable interactions},\
  }\href {https://doi.org/10.1103/PhysRevLett.99.010403} {\bibfield  {journal}
  {\bibinfo  {journal} {Phys. Rev. Lett.}\ }\textbf {\bibinfo {volume} {99}},\
  \bibinfo {pages} {010403} (\bibinfo {year} {2007})}\BibitemShut {NoStop}%
\bibitem [{\citenamefont {Hogan}\ \emph {et~al.}(2011)\citenamefont {Hogan},
  \citenamefont {Johnson}, \citenamefont {Dickerson}, \citenamefont {Kovachy},
  \citenamefont {Sugarbaker}, \citenamefont {Chiow}, \citenamefont {Graham},
  \citenamefont {Kasevich}, \citenamefont {Saif}, \citenamefont {Rajendran},
  \citenamefont {Bouyer}, \citenamefont {Seery}, \citenamefont {Feinberg},\
  and\ \citenamefont {Keski-Kuha}}]{Hogan2011GRG}%
  \BibitemOpen
  \bibfield  {author} {\bibinfo {author} {\bibfnamefont {J.~M.}\ \bibnamefont
  {Hogan}}, \bibinfo {author} {\bibfnamefont {D.~M.~S.}\ \bibnamefont
  {Johnson}}, \bibinfo {author} {\bibfnamefont {S.}~\bibnamefont {Dickerson}},
  \bibinfo {author} {\bibfnamefont {T.}~\bibnamefont {Kovachy}}, \bibinfo
  {author} {\bibfnamefont {A.}~\bibnamefont {Sugarbaker}}, \bibinfo {author}
  {\bibfnamefont {S.-w.}\ \bibnamefont {Chiow}}, \bibinfo {author}
  {\bibfnamefont {P.~W.}\ \bibnamefont {Graham}}, \bibinfo {author}
  {\bibfnamefont {M.~A.}\ \bibnamefont {Kasevich}}, \bibinfo {author}
  {\bibfnamefont {B.}~\bibnamefont {Saif}}, \bibinfo {author} {\bibfnamefont
  {S.}~\bibnamefont {Rajendran}}, \bibinfo {author} {\bibfnamefont
  {P.}~\bibnamefont {Bouyer}}, \bibinfo {author} {\bibfnamefont {B.~D.}\
  \bibnamefont {Seery}}, \bibinfo {author} {\bibfnamefont {L.}~\bibnamefont
  {Feinberg}},\ and\ \bibinfo {author} {\bibfnamefont {R.}~\bibnamefont
  {Keski-Kuha}},\ }\bibfield  {title} {\bibinfo {title} {An atomic
  gravitational wave interferometric sensor in low earth orbit (agis-leo)},\
  }\href {https://doi.org/10.1007/s10714-011-1182-x} {\bibfield  {journal}
  {\bibinfo  {journal} {General Relativity and Gravitation}\ }\textbf {\bibinfo
  {volume} {43}},\ \bibinfo {pages} {1953} (\bibinfo {year}
  {2011})}\BibitemShut {NoStop}%
\bibitem [{\citenamefont {Canuel}\ \emph {et~al.}(2018)\citenamefont {Canuel},
  \citenamefont {Bertoldi}, \citenamefont {Amand}, \citenamefont {{Di Pozzo
  Borgo}}, \citenamefont {Chantrait}, \citenamefont {Danquigny}, \citenamefont
  {{Dovale {\'A}lvarez}}, \citenamefont {Fang}, \citenamefont {Freise},
  \citenamefont {Geiger}, \citenamefont {Gillot}, \citenamefont {Henry},
  \citenamefont {Hinderer}, \citenamefont {Holleville}, \citenamefont {Junca},
  \citenamefont {Lef{\`e}vre}, \citenamefont {Merzougui}, \citenamefont
  {Mielec}, \citenamefont {Monfret}, \citenamefont {Pelisson}, \citenamefont
  {Prevedelli}, \citenamefont {Reynaud}, \citenamefont {Riou}, \citenamefont
  {Rogister}, \citenamefont {Rosat}, \citenamefont {Cormier}, \citenamefont
  {Landragin}, \citenamefont {Chaibi}, \citenamefont {Gaffet},\ and\
  \citenamefont {Bouyer}}]{Canuel2018SciRep}%
  \BibitemOpen
  \bibfield  {author} {\bibinfo {author} {\bibfnamefont {B.}~\bibnamefont
  {Canuel}}, \bibinfo {author} {\bibfnamefont {A.}~\bibnamefont {Bertoldi}},
  \bibinfo {author} {\bibfnamefont {L.}~\bibnamefont {Amand}}, \bibinfo
  {author} {\bibfnamefont {E.}~\bibnamefont {{Di Pozzo Borgo}}}, \bibinfo
  {author} {\bibfnamefont {T.}~\bibnamefont {Chantrait}}, \bibinfo {author}
  {\bibfnamefont {C.}~\bibnamefont {Danquigny}}, \bibinfo {author}
  {\bibfnamefont {M.}~\bibnamefont {{Dovale {\'A}lvarez}}}, \bibinfo {author}
  {\bibfnamefont {B.}~\bibnamefont {Fang}}, \bibinfo {author} {\bibfnamefont
  {A.}~\bibnamefont {Freise}}, \bibinfo {author} {\bibfnamefont
  {R.}~\bibnamefont {Geiger}}, \bibinfo {author} {\bibfnamefont
  {J.}~\bibnamefont {Gillot}}, \bibinfo {author} {\bibfnamefont
  {S.}~\bibnamefont {Henry}}, \bibinfo {author} {\bibfnamefont
  {J.}~\bibnamefont {Hinderer}}, \bibinfo {author} {\bibfnamefont
  {D.}~\bibnamefont {Holleville}}, \bibinfo {author} {\bibfnamefont
  {J.}~\bibnamefont {Junca}}, \bibinfo {author} {\bibfnamefont
  {G.}~\bibnamefont {Lef{\`e}vre}}, \bibinfo {author} {\bibfnamefont
  {M.}~\bibnamefont {Merzougui}}, \bibinfo {author} {\bibfnamefont
  {N.}~\bibnamefont {Mielec}}, \bibinfo {author} {\bibfnamefont
  {T.}~\bibnamefont {Monfret}}, \bibinfo {author} {\bibfnamefont
  {S.}~\bibnamefont {Pelisson}}, \bibinfo {author} {\bibfnamefont
  {M.}~\bibnamefont {Prevedelli}}, \bibinfo {author} {\bibfnamefont
  {S.}~\bibnamefont {Reynaud}}, \bibinfo {author} {\bibfnamefont
  {I.}~\bibnamefont {Riou}}, \bibinfo {author} {\bibfnamefont {Y.}~\bibnamefont
  {Rogister}}, \bibinfo {author} {\bibfnamefont {S.}~\bibnamefont {Rosat}},
  \bibinfo {author} {\bibfnamefont {E.}~\bibnamefont {Cormier}}, \bibinfo
  {author} {\bibfnamefont {A.}~\bibnamefont {Landragin}}, \bibinfo {author}
  {\bibfnamefont {W.}~\bibnamefont {Chaibi}}, \bibinfo {author} {\bibfnamefont
  {S.}~\bibnamefont {Gaffet}},\ and\ \bibinfo {author} {\bibfnamefont
  {P.}~\bibnamefont {Bouyer}},\ }\bibfield  {title} {\bibinfo {title}
  {Exploring gravity with the miga large scale atom interferometer},\ }\href
  {https://doi.org/10.1038/s41598-018-32165-z} {\bibfield  {journal} {\bibinfo
  {journal} {Scientific reports}\ }\textbf {\bibinfo {volume} {8}},\ \bibinfo
  {pages} {14064} (\bibinfo {year} {2018})}\BibitemShut {NoStop}%
\bibitem [{\citenamefont {Zhan}\ \emph {et~al.}(2019)\citenamefont {Zhan} \emph
  {et~al.}}]{Zhan2019quq}%
  \BibitemOpen
  \bibfield  {author} {\bibinfo {author} {\bibfnamefont {M.-S.}\ \bibnamefont
  {Zhan}} \emph {et~al.},\ }\bibfield  {title} {\bibinfo {title} {{ZAIGA:
  Zhaoshan Long-baseline Atom Interferometer Gravitation Antenna}},\ }\href
  {https://doi.org/10.1142/S0218271819400054} {\bibfield  {journal} {\bibinfo
  {journal} {Int. J. Mod. Phys. D}\ }\textbf {\bibinfo {volume} {29}},\
  \bibinfo {pages} {1940005} (\bibinfo {year} {2019})},\ \Eprint
  {https://arxiv.org/abs/1903.09288} {arXiv:1903.09288 [physics.atom-ph]}
  \BibitemShut {NoStop}%
\bibitem [{\citenamefont {Schubert}\ \emph {et~al.}(2019)\citenamefont
  {Schubert}, \citenamefont {Schlippert}, \citenamefont {Abend}, \citenamefont
  {Giese}, \citenamefont {Roura}, \citenamefont {Schleich}, \citenamefont
  {Ertmer},\ and\ \citenamefont {Rasel}}]{schubert_scalable_2019}%
  \BibitemOpen
  \bibfield  {author} {\bibinfo {author} {\bibfnamefont {C.}~\bibnamefont
  {Schubert}}, \bibinfo {author} {\bibfnamefont {D.}~\bibnamefont
  {Schlippert}}, \bibinfo {author} {\bibfnamefont {S.}~\bibnamefont {Abend}},
  \bibinfo {author} {\bibfnamefont {E.}~\bibnamefont {Giese}}, \bibinfo
  {author} {\bibfnamefont {A.}~\bibnamefont {Roura}}, \bibinfo {author}
  {\bibfnamefont {W.~P.}\ \bibnamefont {Schleich}}, \bibinfo {author}
  {\bibfnamefont {W.}~\bibnamefont {Ertmer}},\ and\ \bibinfo {author}
  {\bibfnamefont {E.~M.}\ \bibnamefont {Rasel}},\ }\href
  {https://doi.org/10.48550/arXiv.1909.01951} {\bibinfo {title} {Scalable,
  symmetric atom interferometer for infrasound gravitational wave detection}}
  (\bibinfo {year} {2019}),\ \bibinfo {note} {arXiv:1909.01951
  [quant-ph]}\BibitemShut {NoStop}%
\bibitem [{\citenamefont {Canuel}\ \emph {et~al.}(2020)\citenamefont {Canuel},
  \citenamefont {Abend}, \citenamefont {Amaro-Seoane}, \citenamefont
  {Badaracco}, \citenamefont {Beaufils}, \citenamefont {Bertoldi},
  \citenamefont {Bongs}, \citenamefont {Bouyer}, \citenamefont {Braxmaier},
  \citenamefont {Chaibi}, \citenamefont {Christensen}, \citenamefont {Fitzek},
  \citenamefont {Flouris}, \citenamefont {Gaaloul}, \citenamefont {Gaffet},
  \citenamefont {Alzar}, \citenamefont {Geiger}, \citenamefont
  {Guellati-Khelifa}, \citenamefont {Hammerer}, \citenamefont {Harms},
  \citenamefont {Hinderer}, \citenamefont {Holynski}, \citenamefont {Junca},
  \citenamefont {Katsanevas}, \citenamefont {Klempt}, \citenamefont
  {Kozanitis}, \citenamefont {Krutzik}, \citenamefont {Landragin},
  \citenamefont {Roche}, \citenamefont {Leykauf}, \citenamefont {Lien},
  \citenamefont {Loriani}, \citenamefont {Merlet}, \citenamefont {Merzougui},
  \citenamefont {Nofrarias}, \citenamefont {Papadakos}, \citenamefont {dos
  Santos}, \citenamefont {Peters}, \citenamefont {Plexousakis}, \citenamefont
  {Prevedelli}, \citenamefont {Rasel}, \citenamefont {Rogister}, \citenamefont
  {Rosat}, \citenamefont {Roura}, \citenamefont {Sabulsky}, \citenamefont
  {Schkolnik}, \citenamefont {Schlippert}, \citenamefont {Schubert},
  \citenamefont {Sidorenkov}, \citenamefont {Siemß}, \citenamefont {Sopuerta},
  \citenamefont {Sorrentino}, \citenamefont {Struckmann}, \citenamefont {Tino},
  \citenamefont {Tsagkatakis}, \citenamefont {Viceré}, \citenamefont {von
  Klitzing}, \citenamefont {Woerner},\ and\ \citenamefont
  {Zou}}]{Canuel2020CQG}%
  \BibitemOpen
  \bibfield  {author} {\bibinfo {author} {\bibfnamefont {B.}~\bibnamefont
  {Canuel}}, \bibinfo {author} {\bibfnamefont {S.}~\bibnamefont {Abend}},
  \bibinfo {author} {\bibfnamefont {P.}~\bibnamefont {Amaro-Seoane}}, \bibinfo
  {author} {\bibfnamefont {F.}~\bibnamefont {Badaracco}}, \bibinfo {author}
  {\bibfnamefont {Q.}~\bibnamefont {Beaufils}}, \bibinfo {author}
  {\bibfnamefont {A.}~\bibnamefont {Bertoldi}}, \bibinfo {author}
  {\bibfnamefont {K.}~\bibnamefont {Bongs}}, \bibinfo {author} {\bibfnamefont
  {P.}~\bibnamefont {Bouyer}}, \bibinfo {author} {\bibfnamefont
  {C.}~\bibnamefont {Braxmaier}}, \bibinfo {author} {\bibfnamefont
  {W.}~\bibnamefont {Chaibi}}, \bibinfo {author} {\bibfnamefont
  {N.}~\bibnamefont {Christensen}}, \bibinfo {author} {\bibfnamefont
  {F.}~\bibnamefont {Fitzek}}, \bibinfo {author} {\bibfnamefont
  {G.}~\bibnamefont {Flouris}}, \bibinfo {author} {\bibfnamefont
  {N.}~\bibnamefont {Gaaloul}}, \bibinfo {author} {\bibfnamefont
  {S.}~\bibnamefont {Gaffet}}, \bibinfo {author} {\bibfnamefont {C.~L.~G.}\
  \bibnamefont {Alzar}}, \bibinfo {author} {\bibfnamefont {R.}~\bibnamefont
  {Geiger}}, \bibinfo {author} {\bibfnamefont {S.}~\bibnamefont
  {Guellati-Khelifa}}, \bibinfo {author} {\bibfnamefont {K.}~\bibnamefont
  {Hammerer}}, \bibinfo {author} {\bibfnamefont {J.}~\bibnamefont {Harms}},
  \bibinfo {author} {\bibfnamefont {J.}~\bibnamefont {Hinderer}}, \bibinfo
  {author} {\bibfnamefont {M.}~\bibnamefont {Holynski}}, \bibinfo {author}
  {\bibfnamefont {J.}~\bibnamefont {Junca}}, \bibinfo {author} {\bibfnamefont
  {S.}~\bibnamefont {Katsanevas}}, \bibinfo {author} {\bibfnamefont
  {C.}~\bibnamefont {Klempt}}, \bibinfo {author} {\bibfnamefont
  {C.}~\bibnamefont {Kozanitis}}, \bibinfo {author} {\bibfnamefont
  {M.}~\bibnamefont {Krutzik}}, \bibinfo {author} {\bibfnamefont
  {A.}~\bibnamefont {Landragin}}, \bibinfo {author} {\bibfnamefont {I.~L.}\
  \bibnamefont {Roche}}, \bibinfo {author} {\bibfnamefont {B.}~\bibnamefont
  {Leykauf}}, \bibinfo {author} {\bibfnamefont {Y.-H.}\ \bibnamefont {Lien}},
  \bibinfo {author} {\bibfnamefont {S.}~\bibnamefont {Loriani}}, \bibinfo
  {author} {\bibfnamefont {S.}~\bibnamefont {Merlet}}, \bibinfo {author}
  {\bibfnamefont {M.}~\bibnamefont {Merzougui}}, \bibinfo {author}
  {\bibfnamefont {M.}~\bibnamefont {Nofrarias}}, \bibinfo {author}
  {\bibfnamefont {P.}~\bibnamefont {Papadakos}}, \bibinfo {author}
  {\bibfnamefont {F.~P.}\ \bibnamefont {dos Santos}}, \bibinfo {author}
  {\bibfnamefont {A.}~\bibnamefont {Peters}}, \bibinfo {author} {\bibfnamefont
  {D.}~\bibnamefont {Plexousakis}}, \bibinfo {author} {\bibfnamefont
  {M.}~\bibnamefont {Prevedelli}}, \bibinfo {author} {\bibfnamefont {E.~M.}\
  \bibnamefont {Rasel}}, \bibinfo {author} {\bibfnamefont {Y.}~\bibnamefont
  {Rogister}}, \bibinfo {author} {\bibfnamefont {S.}~\bibnamefont {Rosat}},
  \bibinfo {author} {\bibfnamefont {A.}~\bibnamefont {Roura}}, \bibinfo
  {author} {\bibfnamefont {D.~O.}\ \bibnamefont {Sabulsky}}, \bibinfo {author}
  {\bibfnamefont {V.}~\bibnamefont {Schkolnik}}, \bibinfo {author}
  {\bibfnamefont {D.}~\bibnamefont {Schlippert}}, \bibinfo {author}
  {\bibfnamefont {C.}~\bibnamefont {Schubert}}, \bibinfo {author}
  {\bibfnamefont {L.}~\bibnamefont {Sidorenkov}}, \bibinfo {author}
  {\bibfnamefont {J.-N.}\ \bibnamefont {Siemß}}, \bibinfo {author}
  {\bibfnamefont {C.~F.}\ \bibnamefont {Sopuerta}}, \bibinfo {author}
  {\bibfnamefont {F.}~\bibnamefont {Sorrentino}}, \bibinfo {author}
  {\bibfnamefont {C.}~\bibnamefont {Struckmann}}, \bibinfo {author}
  {\bibfnamefont {G.~M.}\ \bibnamefont {Tino}}, \bibinfo {author}
  {\bibfnamefont {G.}~\bibnamefont {Tsagkatakis}}, \bibinfo {author}
  {\bibfnamefont {A.}~\bibnamefont {Viceré}}, \bibinfo {author} {\bibfnamefont
  {W.}~\bibnamefont {von Klitzing}}, \bibinfo {author} {\bibfnamefont
  {L.}~\bibnamefont {Woerner}},\ and\ \bibinfo {author} {\bibfnamefont
  {X.}~\bibnamefont {Zou}},\ }\bibfield  {title} {\bibinfo {title} {Elgar—a
  european laboratory for gravitation and atom-interferometric research},\
  }\href {https://doi.org/10.1088/1361-6382/aba80e} {\bibfield  {journal}
  {\bibinfo  {journal} {Classical and Quantum Gravity}\ }\textbf {\bibinfo
  {volume} {37}},\ \bibinfo {pages} {225017} (\bibinfo {year}
  {2020})}\BibitemShut {NoStop}%
\bibitem [{\citenamefont {Badurina}\ \emph {et~al.}(2020)\citenamefont
  {Badurina}, \citenamefont {Bentine}, \citenamefont {Blas}, \citenamefont
  {Bongs}, \citenamefont {Bortoletto}, \citenamefont {Bowcock}, \citenamefont
  {Bridges}, \citenamefont {Bowden}, \citenamefont {Buchmueller}, \citenamefont
  {Burrage}, \citenamefont {Coleman}, \citenamefont {Elertas}, \citenamefont
  {Ellis}, \citenamefont {Foot}, \citenamefont {Gibson}, \citenamefont
  {Haehnelt}, \citenamefont {Harte}, \citenamefont {Hedges}, \citenamefont
  {Hobson}, \citenamefont {Holynski}, \citenamefont {Jones}, \citenamefont
  {Langlois}, \citenamefont {Lellouch}, \citenamefont {Lewicki}, \citenamefont
  {Maiolino}, \citenamefont {Majewski}, \citenamefont {Malik}, \citenamefont
  {March-Russell}, \citenamefont {McCabe}, \citenamefont {Newbold},
  \citenamefont {Sauer}, \citenamefont {Schneider}, \citenamefont {Shipsey},
  \citenamefont {Singh}, \citenamefont {Uchida}, \citenamefont {Valenzuela},
  \citenamefont {van~der Grinten}, \citenamefont {Vaskonen}, \citenamefont
  {Vossebeld}, \citenamefont {Weatherill},\ and\ \citenamefont
  {Wilmut}}]{Badurina2020JCA}%
  \BibitemOpen
  \bibfield  {author} {\bibinfo {author} {\bibfnamefont {L.}~\bibnamefont
  {Badurina}}, \bibinfo {author} {\bibfnamefont {E.}~\bibnamefont {Bentine}},
  \bibinfo {author} {\bibfnamefont {D.}~\bibnamefont {Blas}}, \bibinfo {author}
  {\bibfnamefont {K.}~\bibnamefont {Bongs}}, \bibinfo {author} {\bibfnamefont
  {D.}~\bibnamefont {Bortoletto}}, \bibinfo {author} {\bibfnamefont
  {T.}~\bibnamefont {Bowcock}}, \bibinfo {author} {\bibfnamefont
  {K.}~\bibnamefont {Bridges}}, \bibinfo {author} {\bibfnamefont
  {W.}~\bibnamefont {Bowden}}, \bibinfo {author} {\bibfnamefont
  {O.}~\bibnamefont {Buchmueller}}, \bibinfo {author} {\bibfnamefont
  {C.}~\bibnamefont {Burrage}}, \bibinfo {author} {\bibfnamefont
  {J.}~\bibnamefont {Coleman}}, \bibinfo {author} {\bibfnamefont
  {G.}~\bibnamefont {Elertas}}, \bibinfo {author} {\bibfnamefont
  {J.}~\bibnamefont {Ellis}}, \bibinfo {author} {\bibfnamefont
  {C.}~\bibnamefont {Foot}}, \bibinfo {author} {\bibfnamefont {V.}~\bibnamefont
  {Gibson}}, \bibinfo {author} {\bibfnamefont {M.}~\bibnamefont {Haehnelt}},
  \bibinfo {author} {\bibfnamefont {T.}~\bibnamefont {Harte}}, \bibinfo
  {author} {\bibfnamefont {S.}~\bibnamefont {Hedges}}, \bibinfo {author}
  {\bibfnamefont {R.}~\bibnamefont {Hobson}}, \bibinfo {author} {\bibfnamefont
  {M.}~\bibnamefont {Holynski}}, \bibinfo {author} {\bibfnamefont
  {T.}~\bibnamefont {Jones}}, \bibinfo {author} {\bibfnamefont
  {M.}~\bibnamefont {Langlois}}, \bibinfo {author} {\bibfnamefont
  {S.}~\bibnamefont {Lellouch}}, \bibinfo {author} {\bibfnamefont
  {M.}~\bibnamefont {Lewicki}}, \bibinfo {author} {\bibfnamefont
  {R.}~\bibnamefont {Maiolino}}, \bibinfo {author} {\bibfnamefont
  {P.}~\bibnamefont {Majewski}}, \bibinfo {author} {\bibfnamefont
  {S.}~\bibnamefont {Malik}}, \bibinfo {author} {\bibfnamefont
  {J.}~\bibnamefont {March-Russell}}, \bibinfo {author} {\bibfnamefont
  {C.}~\bibnamefont {McCabe}}, \bibinfo {author} {\bibfnamefont
  {D.}~\bibnamefont {Newbold}}, \bibinfo {author} {\bibfnamefont
  {B.}~\bibnamefont {Sauer}}, \bibinfo {author} {\bibfnamefont
  {U.}~\bibnamefont {Schneider}}, \bibinfo {author} {\bibfnamefont
  {I.}~\bibnamefont {Shipsey}}, \bibinfo {author} {\bibfnamefont
  {Y.}~\bibnamefont {Singh}}, \bibinfo {author} {\bibfnamefont
  {M.}~\bibnamefont {Uchida}}, \bibinfo {author} {\bibfnamefont
  {T.}~\bibnamefont {Valenzuela}}, \bibinfo {author} {\bibfnamefont
  {M.}~\bibnamefont {van~der Grinten}}, \bibinfo {author} {\bibfnamefont
  {V.}~\bibnamefont {Vaskonen}}, \bibinfo {author} {\bibfnamefont
  {J.}~\bibnamefont {Vossebeld}}, \bibinfo {author} {\bibfnamefont
  {D.}~\bibnamefont {Weatherill}},\ and\ \bibinfo {author} {\bibfnamefont
  {I.}~\bibnamefont {Wilmut}},\ }\bibfield  {title} {\bibinfo {title} {Aion: an
  atom interferometer observatory and network},\ }\href
  {https://doi.org/10.1088/1475-7516/2020/05/011} {\bibfield  {journal}
  {\bibinfo  {journal} {Journal of Cosmology and Astroparticle Physics}\
  }\textbf {\bibinfo {volume} {2020}}\bibinfo  {number} { (05)},\ \bibinfo
  {pages} {011}}\BibitemShut {NoStop}%
\bibitem [{\citenamefont {Ahlers}\ \emph {et~al.}(2022)\citenamefont {Ahlers},
  \citenamefont {Badurina}, \citenamefont {Bassi}, \citenamefont {Battelier},
  \citenamefont {Beaufils}, \citenamefont {Bongs}, \citenamefont {Bouyer},
  \citenamefont {Braxmaier}, \citenamefont {Buchmueller}, \citenamefont
  {Carlesso}, \citenamefont {Charron}, \citenamefont {Chiofalo}, \citenamefont
  {Corgier}, \citenamefont {Donadi}, \citenamefont {Droz}, \citenamefont
  {Ecoffet}, \citenamefont {Ellis}, \citenamefont {Estève}, \citenamefont
  {Gaaloul}, \citenamefont {Gerardi}, \citenamefont {Giese}, \citenamefont
  {Grosse}, \citenamefont {Hees}, \citenamefont {Hensel}, \citenamefont {Herr},
  \citenamefont {Jetzer}, \citenamefont {Kleinsteinberg}, \citenamefont
  {Klempt}, \citenamefont {Lecomte}, \citenamefont {Lopes}, \citenamefont
  {Loriani}, \citenamefont {Métris}, \citenamefont {Martin}, \citenamefont
  {Martín}, \citenamefont {Müller}, \citenamefont {Nofrarias}, \citenamefont
  {Santos}, \citenamefont {Rasel}, \citenamefont {Robert}, \citenamefont
  {Saks}, \citenamefont {Salter}, \citenamefont {Schlippert}, \citenamefont
  {Schubert}, \citenamefont {Schuldt}, \citenamefont {Sopuerta}, \citenamefont
  {Struckmann}, \citenamefont {Tino}, \citenamefont {Valenzuela}, \citenamefont
  {von Klitzing}, \citenamefont {Wörner}, \citenamefont {Wolf}, \citenamefont
  {Yu},\ and\ \citenamefont {Zelan}}]{Ahlers2022}%
  \BibitemOpen
\bibfield  {number} {  }\bibfield  {author} {\bibinfo {author} {\bibfnamefont
  {H.}~\bibnamefont {Ahlers}}, \bibinfo {author} {\bibfnamefont
  {L.}~\bibnamefont {Badurina}}, \bibinfo {author} {\bibfnamefont
  {A.}~\bibnamefont {Bassi}}, \bibinfo {author} {\bibfnamefont
  {B.}~\bibnamefont {Battelier}}, \bibinfo {author} {\bibfnamefont
  {Q.}~\bibnamefont {Beaufils}}, \bibinfo {author} {\bibfnamefont
  {K.}~\bibnamefont {Bongs}}, \bibinfo {author} {\bibfnamefont
  {P.}~\bibnamefont {Bouyer}}, \bibinfo {author} {\bibfnamefont
  {C.}~\bibnamefont {Braxmaier}}, \bibinfo {author} {\bibfnamefont
  {O.}~\bibnamefont {Buchmueller}}, \bibinfo {author} {\bibfnamefont
  {M.}~\bibnamefont {Carlesso}}, \bibinfo {author} {\bibfnamefont
  {E.}~\bibnamefont {Charron}}, \bibinfo {author} {\bibfnamefont {M.~L.}\
  \bibnamefont {Chiofalo}}, \bibinfo {author} {\bibfnamefont {R.}~\bibnamefont
  {Corgier}}, \bibinfo {author} {\bibfnamefont {S.}~\bibnamefont {Donadi}},
  \bibinfo {author} {\bibfnamefont {F.}~\bibnamefont {Droz}}, \bibinfo {author}
  {\bibfnamefont {R.}~\bibnamefont {Ecoffet}}, \bibinfo {author} {\bibfnamefont
  {J.}~\bibnamefont {Ellis}}, \bibinfo {author} {\bibfnamefont
  {F.}~\bibnamefont {Estève}}, \bibinfo {author} {\bibfnamefont
  {N.}~\bibnamefont {Gaaloul}}, \bibinfo {author} {\bibfnamefont
  {D.}~\bibnamefont {Gerardi}}, \bibinfo {author} {\bibfnamefont
  {E.}~\bibnamefont {Giese}}, \bibinfo {author} {\bibfnamefont
  {J.}~\bibnamefont {Grosse}}, \bibinfo {author} {\bibfnamefont
  {A.}~\bibnamefont {Hees}}, \bibinfo {author} {\bibfnamefont {T.}~\bibnamefont
  {Hensel}}, \bibinfo {author} {\bibfnamefont {W.}~\bibnamefont {Herr}},
  \bibinfo {author} {\bibfnamefont {P.}~\bibnamefont {Jetzer}}, \bibinfo
  {author} {\bibfnamefont {G.}~\bibnamefont {Kleinsteinberg}}, \bibinfo
  {author} {\bibfnamefont {C.}~\bibnamefont {Klempt}}, \bibinfo {author}
  {\bibfnamefont {S.}~\bibnamefont {Lecomte}}, \bibinfo {author} {\bibfnamefont
  {L.}~\bibnamefont {Lopes}}, \bibinfo {author} {\bibfnamefont
  {S.}~\bibnamefont {Loriani}}, \bibinfo {author} {\bibfnamefont
  {G.}~\bibnamefont {Métris}}, \bibinfo {author} {\bibfnamefont
  {T.}~\bibnamefont {Martin}}, \bibinfo {author} {\bibfnamefont
  {V.}~\bibnamefont {Martín}}, \bibinfo {author} {\bibfnamefont
  {G.}~\bibnamefont {Müller}}, \bibinfo {author} {\bibfnamefont
  {M.}~\bibnamefont {Nofrarias}}, \bibinfo {author} {\bibfnamefont {F.~P.~D.}\
  \bibnamefont {Santos}}, \bibinfo {author} {\bibfnamefont {E.~M.}\
  \bibnamefont {Rasel}}, \bibinfo {author} {\bibfnamefont {A.}~\bibnamefont
  {Robert}}, \bibinfo {author} {\bibfnamefont {N.}~\bibnamefont {Saks}},
  \bibinfo {author} {\bibfnamefont {M.}~\bibnamefont {Salter}}, \bibinfo
  {author} {\bibfnamefont {D.}~\bibnamefont {Schlippert}}, \bibinfo {author}
  {\bibfnamefont {C.}~\bibnamefont {Schubert}}, \bibinfo {author}
  {\bibfnamefont {T.}~\bibnamefont {Schuldt}}, \bibinfo {author} {\bibfnamefont
  {C.~F.}\ \bibnamefont {Sopuerta}}, \bibinfo {author} {\bibfnamefont
  {C.}~\bibnamefont {Struckmann}}, \bibinfo {author} {\bibfnamefont {G.~M.}\
  \bibnamefont {Tino}}, \bibinfo {author} {\bibfnamefont {T.}~\bibnamefont
  {Valenzuela}}, \bibinfo {author} {\bibfnamefont {W.}~\bibnamefont {von
  Klitzing}}, \bibinfo {author} {\bibfnamefont {L.}~\bibnamefont {Wörner}},
  \bibinfo {author} {\bibfnamefont {P.}~\bibnamefont {Wolf}}, \bibinfo {author}
  {\bibfnamefont {N.}~\bibnamefont {Yu}},\ and\ \bibinfo {author}
  {\bibfnamefont {M.}~\bibnamefont {Zelan}},\ }\href
  {https://doi.org/10.48550/ARXIV.2211.15412} {\bibinfo {title} {Ste-quest:
  Space time explorer and quantum equivalence principle space test}} (\bibinfo
  {year} {2022}),\ \Eprint {https://arxiv.org/abs/2211.15412} {arXiv:2211.15412
  [physics.space-ph]} \BibitemShut {NoStop}%
\bibitem [{\citenamefont {El-Neaj}\ \emph {et~al.}(2020)\citenamefont
  {El-Neaj}, \citenamefont {Alpigiani}, \citenamefont {Amairi-Pyka},
  \citenamefont {Ara{\'u}jo}, \citenamefont {Bala{\v{z}}}, \citenamefont
  {Bassi}, \citenamefont {Bathe-Peters}, \citenamefont {Battelier},
  \citenamefont {Beli{\'c}}, \citenamefont {Bentine}, \citenamefont {Bernabeu},
  \citenamefont {Bertoldi}, \citenamefont {Bingham}, \citenamefont {Blas},
  \citenamefont {Bolpasi}, \citenamefont {Bongs}, \citenamefont {Bose},
  \citenamefont {Bouyer}, \citenamefont {Bowcock}, \citenamefont {Bowden},
  \citenamefont {Buchmueller}, \citenamefont {Burrage}, \citenamefont {Calmet},
  \citenamefont {Canuel}, \citenamefont {Caramete}, \citenamefont {Carroll},
  \citenamefont {Cella}, \citenamefont {Charmandaris}, \citenamefont
  {Chattopadhyay}, \citenamefont {Chen}, \citenamefont {Chiofalo},
  \citenamefont {Coleman}, \citenamefont {Cotter}, \citenamefont {Cui},
  \citenamefont {Derevianko}, \citenamefont {{De Roeck}}, \citenamefont
  {Djordjevic}, \citenamefont {Dornan}, \citenamefont {Doser}, \citenamefont
  {Drougkakis}, \citenamefont {Dunningham}, \citenamefont {Dutan},
  \citenamefont {Easo}, \citenamefont {Elertas}, \citenamefont {Ellis},
  \citenamefont {{El Sawy}}, \citenamefont {Fassi}, \citenamefont {Felea},
  \citenamefont {Feng}, \citenamefont {Flack}, \citenamefont {Foot},
  \citenamefont {Fuentes}, \citenamefont {Gaaloul}, \citenamefont {Gauguet},
  \citenamefont {Geiger}, \citenamefont {Gibson}, \citenamefont {Giudice},
  \citenamefont {Goldwin}, \citenamefont {Grachov}, \citenamefont {Graham},
  \citenamefont {Grasso}, \citenamefont {{van der Grinten}}, \citenamefont
  {G{\"u}ndogan}, \citenamefont {Haehnelt}, \citenamefont {Harte},
  \citenamefont {Hees}, \citenamefont {Hobson}, \citenamefont {Hogan},
  \citenamefont {Holst}, \citenamefont {Holynski}, \citenamefont {Kasevich},
  \citenamefont {Kavanagh}, \citenamefont {{von Klitzing}}, \citenamefont
  {Kovachy}, \citenamefont {Krikler}, \citenamefont {Krutzik}, \citenamefont
  {Lewicki}, \citenamefont {Lien}, \citenamefont {Liu}, \citenamefont
  {Luciano}, \citenamefont {Magnon}, \citenamefont {Mahmoud}, \citenamefont
  {Malik}, \citenamefont {McCabe}, \citenamefont {Mitchell}, \citenamefont
  {Pahl}, \citenamefont {Pal}, \citenamefont {Pandey}, \citenamefont
  {Papazoglou}, \citenamefont {Paternostro}, \citenamefont {Penning},
  \citenamefont {Peters}, \citenamefont {Prevedelli}, \citenamefont
  {Puthiya-Veettil}, \citenamefont {Quenby}, \citenamefont {Rasel},
  \citenamefont {Ravenhall}, \citenamefont {Ringwood}, \citenamefont {Roura},
  \citenamefont {Sabulsky}, \citenamefont {Sameed}, \citenamefont {Sauer},
  \citenamefont {Sch{\"a}ffer}, \citenamefont {Schiller}, \citenamefont
  {Schkolnik}, \citenamefont {Schlippert}, \citenamefont {Schubert},
  \citenamefont {Sfar}, \citenamefont {Shayeghi}, \citenamefont {Shipsey},
  \citenamefont {Signorini}, \citenamefont {Singh}, \citenamefont
  {Soares-Santos}, \citenamefont {Sorrentino}, \citenamefont {Sumner},
  \citenamefont {Tassis}, \citenamefont {Tentindo}, \citenamefont {Tino},
  \citenamefont {Tinsley}, \citenamefont {Unwin}, \citenamefont {Valenzuela},
  \citenamefont {Vasilakis}, \citenamefont {Vaskonen}, \citenamefont {Vogt},
  \citenamefont {Webber-Date}, \citenamefont {Wenzlawski}, \citenamefont
  {Windpassinger}, \citenamefont {Woltmann}, \citenamefont {Yazgan},
  \citenamefont {Zhan}, \citenamefont {Zou},\ and\ \citenamefont
  {Zupan}}]{ElNeaj2020EPJQ}%
  \BibitemOpen
  \bibfield  {author} {\bibinfo {author} {\bibfnamefont {Y.~A.}\ \bibnamefont
  {El-Neaj}}, \bibinfo {author} {\bibfnamefont {C.}~\bibnamefont {Alpigiani}},
  \bibinfo {author} {\bibfnamefont {S.}~\bibnamefont {Amairi-Pyka}}, \bibinfo
  {author} {\bibfnamefont {H.}~\bibnamefont {Ara{\'u}jo}}, \bibinfo {author}
  {\bibfnamefont {A.}~\bibnamefont {Bala{\v{z}}}}, \bibinfo {author}
  {\bibfnamefont {A.}~\bibnamefont {Bassi}}, \bibinfo {author} {\bibfnamefont
  {L.}~\bibnamefont {Bathe-Peters}}, \bibinfo {author} {\bibfnamefont
  {B.}~\bibnamefont {Battelier}}, \bibinfo {author} {\bibfnamefont
  {A.}~\bibnamefont {Beli{\'c}}}, \bibinfo {author} {\bibfnamefont
  {E.}~\bibnamefont {Bentine}}, \bibinfo {author} {\bibfnamefont
  {J.}~\bibnamefont {Bernabeu}}, \bibinfo {author} {\bibfnamefont
  {A.}~\bibnamefont {Bertoldi}}, \bibinfo {author} {\bibfnamefont
  {R.}~\bibnamefont {Bingham}}, \bibinfo {author} {\bibfnamefont
  {D.}~\bibnamefont {Blas}}, \bibinfo {author} {\bibfnamefont {V.}~\bibnamefont
  {Bolpasi}}, \bibinfo {author} {\bibfnamefont {K.}~\bibnamefont {Bongs}},
  \bibinfo {author} {\bibfnamefont {S.}~\bibnamefont {Bose}}, \bibinfo {author}
  {\bibfnamefont {P.}~\bibnamefont {Bouyer}}, \bibinfo {author} {\bibfnamefont
  {T.}~\bibnamefont {Bowcock}}, \bibinfo {author} {\bibfnamefont
  {W.}~\bibnamefont {Bowden}}, \bibinfo {author} {\bibfnamefont
  {O.}~\bibnamefont {Buchmueller}}, \bibinfo {author} {\bibfnamefont
  {C.}~\bibnamefont {Burrage}}, \bibinfo {author} {\bibfnamefont
  {X.}~\bibnamefont {Calmet}}, \bibinfo {author} {\bibfnamefont
  {B.}~\bibnamefont {Canuel}}, \bibinfo {author} {\bibfnamefont {L.-I.}\
  \bibnamefont {Caramete}}, \bibinfo {author} {\bibfnamefont {A.}~\bibnamefont
  {Carroll}}, \bibinfo {author} {\bibfnamefont {G.}~\bibnamefont {Cella}},
  \bibinfo {author} {\bibfnamefont {V.}~\bibnamefont {Charmandaris}}, \bibinfo
  {author} {\bibfnamefont {S.}~\bibnamefont {Chattopadhyay}}, \bibinfo {author}
  {\bibfnamefont {X.}~\bibnamefont {Chen}}, \bibinfo {author} {\bibfnamefont
  {M.~L.}\ \bibnamefont {Chiofalo}}, \bibinfo {author} {\bibfnamefont
  {J.}~\bibnamefont {Coleman}}, \bibinfo {author} {\bibfnamefont
  {J.}~\bibnamefont {Cotter}}, \bibinfo {author} {\bibfnamefont
  {Y.}~\bibnamefont {Cui}}, \bibinfo {author} {\bibfnamefont {A.}~\bibnamefont
  {Derevianko}}, \bibinfo {author} {\bibfnamefont {A.}~\bibnamefont {{De
  Roeck}}}, \bibinfo {author} {\bibfnamefont {G.~S.}\ \bibnamefont
  {Djordjevic}}, \bibinfo {author} {\bibfnamefont {P.}~\bibnamefont {Dornan}},
  \bibinfo {author} {\bibfnamefont {M.}~\bibnamefont {Doser}}, \bibinfo
  {author} {\bibfnamefont {I.}~\bibnamefont {Drougkakis}}, \bibinfo {author}
  {\bibfnamefont {J.}~\bibnamefont {Dunningham}}, \bibinfo {author}
  {\bibfnamefont {I.}~\bibnamefont {Dutan}}, \bibinfo {author} {\bibfnamefont
  {S.}~\bibnamefont {Easo}}, \bibinfo {author} {\bibfnamefont {G.}~\bibnamefont
  {Elertas}}, \bibinfo {author} {\bibfnamefont {J.}~\bibnamefont {Ellis}},
  \bibinfo {author} {\bibfnamefont {M.}~\bibnamefont {{El Sawy}}}, \bibinfo
  {author} {\bibfnamefont {F.}~\bibnamefont {Fassi}}, \bibinfo {author}
  {\bibfnamefont {D.}~\bibnamefont {Felea}}, \bibinfo {author} {\bibfnamefont
  {C.-H.}\ \bibnamefont {Feng}}, \bibinfo {author} {\bibfnamefont
  {R.}~\bibnamefont {Flack}}, \bibinfo {author} {\bibfnamefont
  {C.}~\bibnamefont {Foot}}, \bibinfo {author} {\bibfnamefont {I.}~\bibnamefont
  {Fuentes}}, \bibinfo {author} {\bibfnamefont {N.}~\bibnamefont {Gaaloul}},
  \bibinfo {author} {\bibfnamefont {A.}~\bibnamefont {Gauguet}}, \bibinfo
  {author} {\bibfnamefont {R.}~\bibnamefont {Geiger}}, \bibinfo {author}
  {\bibfnamefont {V.}~\bibnamefont {Gibson}}, \bibinfo {author} {\bibfnamefont
  {G.}~\bibnamefont {Giudice}}, \bibinfo {author} {\bibfnamefont
  {J.}~\bibnamefont {Goldwin}}, \bibinfo {author} {\bibfnamefont
  {O.}~\bibnamefont {Grachov}}, \bibinfo {author} {\bibfnamefont {P.~W.}\
  \bibnamefont {Graham}}, \bibinfo {author} {\bibfnamefont {D.}~\bibnamefont
  {Grasso}}, \bibinfo {author} {\bibfnamefont {M.}~\bibnamefont {{van der
  Grinten}}}, \bibinfo {author} {\bibfnamefont {M.}~\bibnamefont
  {G{\"u}ndogan}}, \bibinfo {author} {\bibfnamefont {M.~G.}\ \bibnamefont
  {Haehnelt}}, \bibinfo {author} {\bibfnamefont {T.}~\bibnamefont {Harte}},
  \bibinfo {author} {\bibfnamefont {A.}~\bibnamefont {Hees}}, \bibinfo {author}
  {\bibfnamefont {R.}~\bibnamefont {Hobson}}, \bibinfo {author} {\bibfnamefont
  {J.}~\bibnamefont {Hogan}}, \bibinfo {author} {\bibfnamefont
  {B.}~\bibnamefont {Holst}}, \bibinfo {author} {\bibfnamefont
  {M.}~\bibnamefont {Holynski}}, \bibinfo {author} {\bibfnamefont
  {M.}~\bibnamefont {Kasevich}}, \bibinfo {author} {\bibfnamefont {B.~J.}\
  \bibnamefont {Kavanagh}}, \bibinfo {author} {\bibfnamefont {W.}~\bibnamefont
  {{von Klitzing}}}, \bibinfo {author} {\bibfnamefont {T.}~\bibnamefont
  {Kovachy}}, \bibinfo {author} {\bibfnamefont {B.}~\bibnamefont {Krikler}},
  \bibinfo {author} {\bibfnamefont {M.}~\bibnamefont {Krutzik}}, \bibinfo
  {author} {\bibfnamefont {M.}~\bibnamefont {Lewicki}}, \bibinfo {author}
  {\bibfnamefont {Y.-H.}\ \bibnamefont {Lien}}, \bibinfo {author}
  {\bibfnamefont {M.}~\bibnamefont {Liu}}, \bibinfo {author} {\bibfnamefont
  {G.~G.}\ \bibnamefont {Luciano}}, \bibinfo {author} {\bibfnamefont
  {A.}~\bibnamefont {Magnon}}, \bibinfo {author} {\bibfnamefont {M.~A.}\
  \bibnamefont {Mahmoud}}, \bibinfo {author} {\bibfnamefont {S.}~\bibnamefont
  {Malik}}, \bibinfo {author} {\bibfnamefont {C.}~\bibnamefont {McCabe}},
  \bibinfo {author} {\bibfnamefont {J.}~\bibnamefont {Mitchell}}, \bibinfo
  {author} {\bibfnamefont {J.}~\bibnamefont {Pahl}}, \bibinfo {author}
  {\bibfnamefont {D.}~\bibnamefont {Pal}}, \bibinfo {author} {\bibfnamefont
  {S.}~\bibnamefont {Pandey}}, \bibinfo {author} {\bibfnamefont
  {D.}~\bibnamefont {Papazoglou}}, \bibinfo {author} {\bibfnamefont
  {M.}~\bibnamefont {Paternostro}}, \bibinfo {author} {\bibfnamefont
  {B.}~\bibnamefont {Penning}}, \bibinfo {author} {\bibfnamefont
  {A.}~\bibnamefont {Peters}}, \bibinfo {author} {\bibfnamefont
  {M.}~\bibnamefont {Prevedelli}}, \bibinfo {author} {\bibfnamefont
  {V.}~\bibnamefont {Puthiya-Veettil}}, \bibinfo {author} {\bibfnamefont
  {J.}~\bibnamefont {Quenby}}, \bibinfo {author} {\bibfnamefont
  {E.}~\bibnamefont {Rasel}}, \bibinfo {author} {\bibfnamefont
  {S.}~\bibnamefont {Ravenhall}}, \bibinfo {author} {\bibfnamefont
  {J.}~\bibnamefont {Ringwood}}, \bibinfo {author} {\bibfnamefont
  {A.}~\bibnamefont {Roura}}, \bibinfo {author} {\bibfnamefont
  {D.}~\bibnamefont {Sabulsky}}, \bibinfo {author} {\bibfnamefont
  {M.}~\bibnamefont {Sameed}}, \bibinfo {author} {\bibfnamefont
  {B.}~\bibnamefont {Sauer}}, \bibinfo {author} {\bibfnamefont {S.~A.}\
  \bibnamefont {Sch{\"a}ffer}}, \bibinfo {author} {\bibfnamefont
  {S.}~\bibnamefont {Schiller}}, \bibinfo {author} {\bibfnamefont
  {V.}~\bibnamefont {Schkolnik}}, \bibinfo {author} {\bibfnamefont
  {D.}~\bibnamefont {Schlippert}}, \bibinfo {author} {\bibfnamefont
  {C.}~\bibnamefont {Schubert}}, \bibinfo {author} {\bibfnamefont {H.~R.}\
  \bibnamefont {Sfar}}, \bibinfo {author} {\bibfnamefont {A.}~\bibnamefont
  {Shayeghi}}, \bibinfo {author} {\bibfnamefont {I.}~\bibnamefont {Shipsey}},
  \bibinfo {author} {\bibfnamefont {C.}~\bibnamefont {Signorini}}, \bibinfo
  {author} {\bibfnamefont {Y.}~\bibnamefont {Singh}}, \bibinfo {author}
  {\bibfnamefont {M.}~\bibnamefont {Soares-Santos}}, \bibinfo {author}
  {\bibfnamefont {F.}~\bibnamefont {Sorrentino}}, \bibinfo {author}
  {\bibfnamefont {T.}~\bibnamefont {Sumner}}, \bibinfo {author} {\bibfnamefont
  {K.}~\bibnamefont {Tassis}}, \bibinfo {author} {\bibfnamefont
  {S.}~\bibnamefont {Tentindo}}, \bibinfo {author} {\bibfnamefont {G.~M.}\
  \bibnamefont {Tino}}, \bibinfo {author} {\bibfnamefont {J.~N.}\ \bibnamefont
  {Tinsley}}, \bibinfo {author} {\bibfnamefont {J.}~\bibnamefont {Unwin}},
  \bibinfo {author} {\bibfnamefont {T.}~\bibnamefont {Valenzuela}}, \bibinfo
  {author} {\bibfnamefont {G.}~\bibnamefont {Vasilakis}}, \bibinfo {author}
  {\bibfnamefont {V.}~\bibnamefont {Vaskonen}}, \bibinfo {author}
  {\bibfnamefont {C.}~\bibnamefont {Vogt}}, \bibinfo {author} {\bibfnamefont
  {A.}~\bibnamefont {Webber-Date}}, \bibinfo {author} {\bibfnamefont
  {A.}~\bibnamefont {Wenzlawski}}, \bibinfo {author} {\bibfnamefont
  {P.}~\bibnamefont {Windpassinger}}, \bibinfo {author} {\bibfnamefont
  {M.}~\bibnamefont {Woltmann}}, \bibinfo {author} {\bibfnamefont
  {E.}~\bibnamefont {Yazgan}}, \bibinfo {author} {\bibfnamefont {M.-S.}\
  \bibnamefont {Zhan}}, \bibinfo {author} {\bibfnamefont {X.}~\bibnamefont
  {Zou}},\ and\ \bibinfo {author} {\bibfnamefont {J.}~\bibnamefont {Zupan}},\
  }\bibfield  {title} {\bibinfo {title} {Aedge: Atomic experiment for dark
  matter and gravity exploration in space},\ }\bibfield  {journal} {\bibinfo
  {journal} {EPJ Quantum Technology}\ }\textbf {\bibinfo {volume} {7}},\ \href
  {https://doi.org/10.1140/epjqt/s40507-020-0080-0}
  {10.1140/epjqt/s40507-020-0080-0} (\bibinfo {year} {2020})\BibitemShut
  {NoStop}%
\bibitem [{\citenamefont {Du}\ \emph {et~al.}(2022)\citenamefont {Du},
  \citenamefont {Murgui}, \citenamefont {Pardo}, \citenamefont {Wang},\ and\
  \citenamefont {Zurek}}]{Du2022PRD}%
  \BibitemOpen
  \bibfield  {author} {\bibinfo {author} {\bibfnamefont {Y.}~\bibnamefont
  {Du}}, \bibinfo {author} {\bibfnamefont {C.}~\bibnamefont {Murgui}}, \bibinfo
  {author} {\bibfnamefont {K.}~\bibnamefont {Pardo}}, \bibinfo {author}
  {\bibfnamefont {Y.}~\bibnamefont {Wang}},\ and\ \bibinfo {author}
  {\bibfnamefont {K.~M.}\ \bibnamefont {Zurek}},\ }\bibfield  {title} {\bibinfo
  {title} {Atom interferometer tests of dark matter},\ }\href
  {https://doi.org/10.1103/PhysRevD.106.095041} {\bibfield  {journal} {\bibinfo
   {journal} {Phys. Rev. D}\ }\textbf {\bibinfo {volume} {106}},\ \bibinfo
  {pages} {095041} (\bibinfo {year} {2022})}\BibitemShut {NoStop}%
\bibitem [{\citenamefont {Badurina}\ \emph {et~al.}(2023)\citenamefont
  {Badurina}, \citenamefont {Gibson}, \citenamefont {McCabe},\ and\
  \citenamefont {Mitchell}}]{Badurina2023PRD}%
  \BibitemOpen
  \bibfield  {author} {\bibinfo {author} {\bibfnamefont {L.}~\bibnamefont
  {Badurina}}, \bibinfo {author} {\bibfnamefont {V.}~\bibnamefont {Gibson}},
  \bibinfo {author} {\bibfnamefont {C.}~\bibnamefont {McCabe}},\ and\ \bibinfo
  {author} {\bibfnamefont {J.}~\bibnamefont {Mitchell}},\ }\bibfield  {title}
  {\bibinfo {title} {Ultralight dark matter searches at the sub-hz frontier
  with atom multigradiometry},\ }\href
  {https://doi.org/10.1103/PhysRevD.107.055002} {\bibfield  {journal} {\bibinfo
   {journal} {Phys. Rev. D}\ }\textbf {\bibinfo {volume} {107}},\ \bibinfo
  {pages} {055002} (\bibinfo {year} {2023})}\BibitemShut {NoStop}%
\bibitem [{\citenamefont {Chu}\ \emph {et~al.}(1986)\citenamefont {Chu},
  \citenamefont {Bjorkholm}, \citenamefont {Ashkin}, \citenamefont {Gordon},\
  and\ \citenamefont {Hollberg}}]{Chu86OL}%
  \BibitemOpen
  \bibfield  {author} {\bibinfo {author} {\bibfnamefont {S.}~\bibnamefont
  {Chu}}, \bibinfo {author} {\bibfnamefont {J.~E.}\ \bibnamefont {Bjorkholm}},
  \bibinfo {author} {\bibfnamefont {A.}~\bibnamefont {Ashkin}}, \bibinfo
  {author} {\bibfnamefont {J.~P.}\ \bibnamefont {Gordon}},\ and\ \bibinfo
  {author} {\bibfnamefont {L.~W.}\ \bibnamefont {Hollberg}},\ }\bibfield
  {title} {\bibinfo {title} {Proposal for optically cooling atoms to
  temperatures of the order of 10{\textsuperscript{-6}} k},\ }\href
  {https://doi.org/10.1364/ol.11.000073} {\bibfield  {journal} {\bibinfo
  {journal} {Optics Letters}\ }\textbf {\bibinfo {volume} {11}},\ \bibinfo
  {pages} {73} (\bibinfo {year} {1986})}\BibitemShut {NoStop}%
\bibitem [{\citenamefont {Corgier}\ \emph {et~al.}(2018)\citenamefont
  {Corgier}, \citenamefont {Amri}, \citenamefont {Herr}, \citenamefont
  {Ahlers}, \citenamefont {Rudolph}, \citenamefont {Gu{\'{e}}ry-Odelin},
  \citenamefont {Rasel}, \citenamefont {Charron},\ and\ \citenamefont
  {Gaaloul}}]{Corgier2018NJP}%
  \BibitemOpen
  \bibfield  {author} {\bibinfo {author} {\bibfnamefont {R.}~\bibnamefont
  {Corgier}}, \bibinfo {author} {\bibfnamefont {S.}~\bibnamefont {Amri}},
  \bibinfo {author} {\bibfnamefont {W.}~\bibnamefont {Herr}}, \bibinfo {author}
  {\bibfnamefont {H.}~\bibnamefont {Ahlers}}, \bibinfo {author} {\bibfnamefont
  {J.}~\bibnamefont {Rudolph}}, \bibinfo {author} {\bibfnamefont
  {D.}~\bibnamefont {Gu{\'{e}}ry-Odelin}}, \bibinfo {author} {\bibfnamefont
  {E.~M.}\ \bibnamefont {Rasel}}, \bibinfo {author} {\bibfnamefont
  {E.}~\bibnamefont {Charron}},\ and\ \bibinfo {author} {\bibfnamefont
  {N.}~\bibnamefont {Gaaloul}},\ }\bibfield  {title} {\bibinfo {title} {Fast
  manipulation of bose{\textendash}einstein condensates with an atom chip},\
  }\href {https://doi.org/10.1088/1367-2630/aabdfc} {\bibfield  {journal}
  {\bibinfo  {journal} {New J. Phys.}\ }\textbf {\bibinfo {volume} {20}},\
  \bibinfo {pages} {055002} (\bibinfo {year} {2018})}\BibitemShut {NoStop}%
\bibitem [{\citenamefont {Herbst}\ \emph {et~al.}(2024)\citenamefont {Herbst},
  \citenamefont {Estrampes}, \citenamefont {Albers}, \citenamefont
  {Vollenkemper}, \citenamefont {Stolzenberg}, \citenamefont {Bode},
  \citenamefont {Charron}, \citenamefont {Rasel}, \citenamefont {Gaaloul},\
  and\ \citenamefont {Schlippert}}]{Herbst2024PRR}%
  \BibitemOpen
  \bibfield  {author} {\bibinfo {author} {\bibfnamefont {A.}~\bibnamefont
  {Herbst}}, \bibinfo {author} {\bibfnamefont {T.}~\bibnamefont {Estrampes}},
  \bibinfo {author} {\bibfnamefont {H.}~\bibnamefont {Albers}}, \bibinfo
  {author} {\bibfnamefont {V.}~\bibnamefont {Vollenkemper}}, \bibinfo {author}
  {\bibfnamefont {K.}~\bibnamefont {Stolzenberg}}, \bibinfo {author}
  {\bibfnamefont {S.}~\bibnamefont {Bode}}, \bibinfo {author} {\bibfnamefont
  {E.}~\bibnamefont {Charron}}, \bibinfo {author} {\bibfnamefont {E.~M.}\
  \bibnamefont {Rasel}}, \bibinfo {author} {\bibfnamefont {N.}~\bibnamefont
  {Gaaloul}},\ and\ \bibinfo {author} {\bibfnamefont {D.}~\bibnamefont
  {Schlippert}},\ }\bibfield  {title} {\bibinfo {title} {High-flux source
  system for matter-wave interferometry exploiting tunable interactions},\
  }\href {https://doi.org/10.1103/PhysRevResearch.6.013139} {\bibfield
  {journal} {\bibinfo  {journal} {Phys. Rev. Res.}\ }\textbf {\bibinfo {volume}
  {6}},\ \bibinfo {pages} {013139} (\bibinfo {year} {2024})}\BibitemShut
  {NoStop}%
\bibitem [{\citenamefont {Kirkpatrick}\ \emph {et~al.}(1983)\citenamefont
  {Kirkpatrick}, \citenamefont {Gelatt},\ and\ \citenamefont
  {Vecchi}}]{Kirkpatrick1983Science}%
  \BibitemOpen
  \bibfield  {author} {\bibinfo {author} {\bibfnamefont {S.}~\bibnamefont
  {Kirkpatrick}}, \bibinfo {author} {\bibfnamefont {C.~D.}\ \bibnamefont
  {Gelatt}},\ and\ \bibinfo {author} {\bibfnamefont {M.~P.}\ \bibnamefont
  {Vecchi}},\ }\bibfield  {title} {\bibinfo {title} {Optimization by simulated
  annealing},\ }\href {https://doi.org/10.1126/science.220.4598.671} {\bibfield
   {journal} {\bibinfo  {journal} {Science}\ }\textbf {\bibinfo {volume}
  {220}},\ \bibinfo {pages} {671} (\bibinfo {year} {1983})}\BibitemShut
  {NoStop}%
\bibitem [{\citenamefont {Catani}\ \emph {et~al.}(2006)\citenamefont {Catani},
  \citenamefont {Maioli}, \citenamefont {De~Sarlo}, \citenamefont {Minardi},\
  and\ \citenamefont {Inguscio}}]{Catani2006PRA}%
  \BibitemOpen
  \bibfield  {author} {\bibinfo {author} {\bibfnamefont {J.}~\bibnamefont
  {Catani}}, \bibinfo {author} {\bibfnamefont {P.}~\bibnamefont {Maioli}},
  \bibinfo {author} {\bibfnamefont {L.}~\bibnamefont {De~Sarlo}}, \bibinfo
  {author} {\bibfnamefont {F.}~\bibnamefont {Minardi}},\ and\ \bibinfo {author}
  {\bibfnamefont {M.}~\bibnamefont {Inguscio}},\ }\bibfield  {title} {\bibinfo
  {title} {Intense slow beams of bosonic potassium isotopes},\ }\href
  {https://doi.org/10.1103/PhysRevA.73.033415} {\bibfield  {journal} {\bibinfo
  {journal} {Phys. Rev. A}\ }\textbf {\bibinfo {volume} {73}},\ \bibinfo
  {pages} {033415} (\bibinfo {year} {2006})}\BibitemShut {NoStop}%
\bibitem [{\citenamefont {Herbst}\ \emph {et~al.}(2022)\citenamefont {Herbst},
  \citenamefont {Albers}, \citenamefont {Stolzenberg}, \citenamefont {Bode},\
  and\ \citenamefont {Schlippert}}]{Herbst2022PRA}%
  \BibitemOpen
  \bibfield  {author} {\bibinfo {author} {\bibfnamefont {A.}~\bibnamefont
  {Herbst}}, \bibinfo {author} {\bibfnamefont {H.}~\bibnamefont {Albers}},
  \bibinfo {author} {\bibfnamefont {K.}~\bibnamefont {Stolzenberg}}, \bibinfo
  {author} {\bibfnamefont {S.}~\bibnamefont {Bode}},\ and\ \bibinfo {author}
  {\bibfnamefont {D.}~\bibnamefont {Schlippert}},\ }\bibfield  {title}
  {\bibinfo {title} {Rapid generation of all-optical $^{39}\mathrm{K}$
  bose-einstein condensates using a low-field feshbach resonance},\ }\href
  {https://doi.org/10.1103/PhysRevA.106.043320} {\bibfield  {journal} {\bibinfo
   {journal} {Phys. Rev. A}\ }\textbf {\bibinfo {volume} {106}},\ \bibinfo
  {pages} {043320} (\bibinfo {year} {2022})}\BibitemShut {NoStop}%
\bibitem [{\citenamefont {Albers}(2020)}]{Albers2020phd}%
  \BibitemOpen
  \bibfield  {author} {\bibinfo {author} {\bibfnamefont {H.}~\bibnamefont
  {Albers}},\ }\emph {\bibinfo {title} {Time-averaged optical potentials for
  creating and shaping Bose-Einstein condensates}},\ \href
  {https://doi.org/10.15488/10073} {Ph.D. thesis},\ \bibinfo  {school} {Leibniz
  Universit\"at Hannover} (\bibinfo {year} {2020})\BibitemShut {NoStop}%
\bibitem [{\citenamefont {Salomon}\ \emph {et~al.}(2013)\citenamefont
  {Salomon}, \citenamefont {Fouch{\'e}}, \citenamefont {Wang}, \citenamefont
  {Aspect}, \citenamefont {Bouyer},\ and\ \citenamefont
  {Bourdel}}]{Salomon13EPL}%
  \BibitemOpen
  \bibfield  {author} {\bibinfo {author} {\bibfnamefont {G.}~\bibnamefont
  {Salomon}}, \bibinfo {author} {\bibfnamefont {L.}~\bibnamefont {Fouch{\'e}}},
  \bibinfo {author} {\bibfnamefont {P.}~\bibnamefont {Wang}}, \bibinfo {author}
  {\bibfnamefont {A.}~\bibnamefont {Aspect}}, \bibinfo {author} {\bibfnamefont
  {P.}~\bibnamefont {Bouyer}},\ and\ \bibinfo {author} {\bibfnamefont
  {T.}~\bibnamefont {Bourdel}},\ }\bibfield  {title} {\bibinfo {title}
  {Gray-molasses cooling of $^{39}\mathrm{K}$ to a high phase-space density},\
  }\href {https://doi.org/10.1209/0295-5075/104/63002} {\bibfield  {journal}
  {\bibinfo  {journal} {EPL (Europhysics Letters)}\ }\textbf {\bibinfo {volume}
  {104}},\ \bibinfo {pages} {63002} (\bibinfo {year} {2013})}\BibitemShut
  {NoStop}%
\bibitem [{\citenamefont {D'Errico}\ \emph {et~al.}(2007)\citenamefont
  {D'Errico}, \citenamefont {Zaccanti}, \citenamefont {Fattori}, \citenamefont
  {Roati}, \citenamefont {Inguscio}, \citenamefont {Modugno},\ and\
  \citenamefont {Simoni}}]{DErrico07NJP}%
  \BibitemOpen
  \bibfield  {author} {\bibinfo {author} {\bibfnamefont {C.}~\bibnamefont
  {D'Errico}}, \bibinfo {author} {\bibfnamefont {M.}~\bibnamefont {Zaccanti}},
  \bibinfo {author} {\bibfnamefont {M.}~\bibnamefont {Fattori}}, \bibinfo
  {author} {\bibfnamefont {G.}~\bibnamefont {Roati}}, \bibinfo {author}
  {\bibfnamefont {M.}~\bibnamefont {Inguscio}}, \bibinfo {author}
  {\bibfnamefont {G.}~\bibnamefont {Modugno}},\ and\ \bibinfo {author}
  {\bibfnamefont {A.}~\bibnamefont {Simoni}},\ }\bibfield  {title} {\bibinfo
  {title} {Feshbach resonances in ultracold $^{39}\mathrm{K}$},\ }\href
  {https://doi.org/10.1088/1367-2630/9/7/223} {\bibfield  {journal} {\bibinfo
  {journal} {New Journal of Physics}\ }\textbf {\bibinfo {volume} {9}},\
  \bibinfo {pages} {223} (\bibinfo {year} {2007})}\BibitemShut {NoStop}%
\bibitem [{\citenamefont {Landini}\ \emph {et~al.}(2012)\citenamefont
  {Landini}, \citenamefont {Roy}, \citenamefont {Roati}, \citenamefont
  {Simoni}, \citenamefont {Inguscio}, \citenamefont {Modugno},\ and\
  \citenamefont {Fattori}}]{Landini12PRA}%
  \BibitemOpen
  \bibfield  {author} {\bibinfo {author} {\bibfnamefont {M.}~\bibnamefont
  {Landini}}, \bibinfo {author} {\bibfnamefont {S.}~\bibnamefont {Roy}},
  \bibinfo {author} {\bibfnamefont {G.}~\bibnamefont {Roati}}, \bibinfo
  {author} {\bibfnamefont {A.}~\bibnamefont {Simoni}}, \bibinfo {author}
  {\bibfnamefont {M.}~\bibnamefont {Inguscio}}, \bibinfo {author}
  {\bibfnamefont {G.}~\bibnamefont {Modugno}},\ and\ \bibinfo {author}
  {\bibfnamefont {M.}~\bibnamefont {Fattori}},\ }\bibfield  {title} {\bibinfo
  {title} {Direct evaporative cooling of $^{39}\mathrm{K}$ atoms to
  bose-einstein condensation},\ }\href
  {https://doi.org/10.1103/PhysRevA.86.033421} {\bibfield  {journal} {\bibinfo
  {journal} {Phys. Rev. A}\ }\textbf {\bibinfo {volume} {86}},\ \bibinfo
  {pages} {033421} (\bibinfo {year} {2012})}\BibitemShut {NoStop}%
\bibitem [{\citenamefont {Castin}\ and\ \citenamefont
  {Dum}(1996)}]{Castin1996PRL}%
  \BibitemOpen
  \bibfield  {author} {\bibinfo {author} {\bibfnamefont {Y.}~\bibnamefont
  {Castin}}\ and\ \bibinfo {author} {\bibfnamefont {R.}~\bibnamefont {Dum}},\
  }\bibfield  {title} {\bibinfo {title} {Bose-einstein condensates in time
  dependent traps},\ }\href {https://doi.org/10.1103/physrevlett.77.5315}
  {\bibfield  {journal} {\bibinfo  {journal} {Physical Review Letters}\
  }\textbf {\bibinfo {volume} {77}},\ \bibinfo {pages} {5315} (\bibinfo {year}
  {1996})}\BibitemShut {NoStop}%
\bibitem [{\citenamefont {Kagan}\ \emph {et~al.}(1997)\citenamefont {Kagan},
  \citenamefont {Surkov},\ and\ \citenamefont {Shlyapnikov}}]{Kagan1997PRA}%
  \BibitemOpen
  \bibfield  {author} {\bibinfo {author} {\bibfnamefont {Y.}~\bibnamefont
  {Kagan}}, \bibinfo {author} {\bibfnamefont {E.~L.}\ \bibnamefont {Surkov}},\
  and\ \bibinfo {author} {\bibfnamefont {G.~V.}\ \bibnamefont {Shlyapnikov}},\
  }\bibfield  {title} {\bibinfo {title} {Evolution of a bose gas in anisotropic
  time-dependent traps},\ }\href {https://doi.org/10.1103/PhysRevA.55.R18}
  {\bibfield  {journal} {\bibinfo  {journal} {Phys. Rev. A}\ }\textbf {\bibinfo
  {volume} {55}},\ \bibinfo {pages} {R18} (\bibinfo {year} {1997})}\BibitemShut
  {NoStop}%
\bibitem [{\citenamefont {Pethick}\ and\ \citenamefont
  {Smith}(2008)}]{Pethick2008}%
  \BibitemOpen
  \bibfield  {author} {\bibinfo {author} {\bibfnamefont {C.~J.}\ \bibnamefont
  {Pethick}}\ and\ \bibinfo {author} {\bibfnamefont {H.}~\bibnamefont
  {Smith}},\ }\href {https://doi.org/10.1017/CBO9780511802850.007} {\emph
  {\bibinfo {title} {Bose–Einstein Condensation in Dilute Gases}}},\ \bibinfo
  {edition} {2nd}\ ed.\ (\bibinfo  {publisher} {Cambridge University Press},\
  \bibinfo {year} {2008})\BibitemShut {NoStop}%
\bibitem [{\citenamefont {P\'erez-Garc\'{\i}a}\ \emph
  {et~al.}(1996)\citenamefont {P\'erez-Garc\'{\i}a}, \citenamefont {Michinel},
  \citenamefont {Cirac}, \citenamefont {Lewenstein},\ and\ \citenamefont
  {Zoller}}]{Perez1996PRL}%
  \BibitemOpen
  \bibfield  {author} {\bibinfo {author} {\bibfnamefont {V.~M.}\ \bibnamefont
  {P\'erez-Garc\'{\i}a}}, \bibinfo {author} {\bibfnamefont {H.}~\bibnamefont
  {Michinel}}, \bibinfo {author} {\bibfnamefont {J.~I.}\ \bibnamefont {Cirac}},
  \bibinfo {author} {\bibfnamefont {M.}~\bibnamefont {Lewenstein}},\ and\
  \bibinfo {author} {\bibfnamefont {P.}~\bibnamefont {Zoller}},\ }\bibfield
  {title} {\bibinfo {title} {Low energy excitations of a bose-einstein
  condensate: A time-dependent variational analysis},\ }\href
  {https://doi.org/10.1103/PhysRevLett.77.5320} {\bibfield  {journal} {\bibinfo
   {journal} {Phys. Rev. Lett.}\ }\textbf {\bibinfo {volume} {77}},\ \bibinfo
  {pages} {5320} (\bibinfo {year} {1996})}\BibitemShut {NoStop}%
\bibitem [{\citenamefont {P\'erez-Garc\'{\i}a}\ \emph
  {et~al.}(1997)\citenamefont {P\'erez-Garc\'{\i}a}, \citenamefont {Michinel},
  \citenamefont {Cirac}, \citenamefont {Lewenstein},\ and\ \citenamefont
  {Zoller}}]{Perez1997PRA}%
  \BibitemOpen
  \bibfield  {author} {\bibinfo {author} {\bibfnamefont {V.~M.}\ \bibnamefont
  {P\'erez-Garc\'{\i}a}}, \bibinfo {author} {\bibfnamefont {H.}~\bibnamefont
  {Michinel}}, \bibinfo {author} {\bibfnamefont {J.~I.}\ \bibnamefont {Cirac}},
  \bibinfo {author} {\bibfnamefont {M.}~\bibnamefont {Lewenstein}},\ and\
  \bibinfo {author} {\bibfnamefont {P.}~\bibnamefont {Zoller}},\ }\bibfield
  {title} {\bibinfo {title} {Dynamics of bose-einstein condensates: Variational
  solutions of the gross-pitaevskii equations},\ }\href
  {https://doi.org/10.1103/PhysRevA.56.1424} {\bibfield  {journal} {\bibinfo
  {journal} {Phys. Rev. A}\ }\textbf {\bibinfo {volume} {56}},\ \bibinfo
  {pages} {1424} (\bibinfo {year} {1997})}\BibitemShut {NoStop}%
\end{thebibliography}%
\section*{Acknowledgements}
We thank Dorothee Tell for thorough proof reading.
This work is funded by the German Space Agency (DLR) with funds provided by the Federal Ministry for Economic Affairs and Climate Action due to an enactment of the German Bundestag under Grant No. DLR 50WM2041 (PRIMUS-IV), 50WM2253A (AI-Quadrat) and supported by the ``ADI 2022'' project founded by the IDEX Paris-Saclay, ANR-11-IDEX-0003-02.
The authors further acknowledge support by the Federal Ministry of Education and Research (BMBF) through the funding program Photonics Research Germany under contract number 13N14875 and by the Deutsche Forschungsgemeinschaft (DFG, German Research Foundation)–Project-ID 274200144–the SFB 1227 DQ-mat within Project No.~A05 and ~B07 and under Germany’s Excellence Strategy—EXC-2123 QuantumFrontiers—Project-ID 390837967.
\section*{Author contributions}
A.H., H.A., S.B., E.M.R. and D.S. designed the experimental setup and the dipole trapping laser system.
A.H., H.A., S.B., K.S., E.M.R. and D.S. contributed to the design, operation, and maintenance of the overall setup.
T.E., R.C., E.C. and N.G. set the theoretical framework of this work.
A.H., T.E. and H.A. with support of R.C., E.C. and N.G. performed the analysis of the data presented in this manuscript.
A.H., T.E. and R.C. with support of D.S., E.C., E.M.R. and N.G. drafted the initial manuscript.
All authors discussed and evaluated the results and contributed to, reviewed, and approved of the manuscript.
\section*{Competing interests}
All authors declare no competing interests.
\end{document}